\documentclass[aps, prb, twocolumn, superscriptaddress,preprintnumbers]{revtex4-2}
\usepackage{bm, amsmath, amsfonts, amssymb, ascmac, mathtools, braket}
\usepackage{multirow}
\usepackage{graphicx}
\usepackage{float, color, xcolor}

\usepackage[whole]{bxcjkjatype} 
\usepackage{subfigure} 
\usepackage{amscd} 

\usepackage{bbm}
\usepackage{tabularx}

\usepackage{comment}

\definecolor{rred}{rgb}{0.8, 0.0, 0.0}
\definecolor{bblue}{rgb}{0.0, 0.0, 0.8}

\usepackage[
pagebackref=false,
colorlinks=true,
linkcolor=bblue,
urlcolor=bblue,
filecolor=black,
citecolor=rred,
pdfstartview=FitV,
pdftitle={},
pdfauthor={},
pdfsubject={},
pdfkeywords={},
pdfpagemode=None,
bookmarksopen=true
]{hyperref}

\newcommand{\ii}{\text{i}}
\newcommand{\sphere}{S}

\begin{document}

\title{Non-Hermitian Hopf insulators}

\author{Daichi Nakamura}
\email{daichi.nakamura@issp.u-tokyo.ac.jp}
\affiliation{Institute for Solid State Physics, University of Tokyo, Kashiwa, Chiba 277-8581, Japan}

\author{Kohei Kawabata}
\email{kawabata@issp.u-tokyo.ac.jp}
\affiliation{Institute for Solid State Physics, University of Tokyo, Kashiwa, Chiba 277-8581, Japan}

\date{\today}

\begin{abstract}
Hopf insulators represent a unique class of topological insulators that exist exclusively in two-band systems and are inherently unstable upon the inclusion of additional bands.
Meanwhile, recent studies have shown that non-Hermiticity gives rise to distinctive complex-energy gap structures, known as point gaps, and associated topological phases with no analogs in Hermitian systems.
However, non-Hermitian counterparts of Hopf insulators have remained largely elusive.
Here, we generally classify topological phases of two-band non-Hermitian systems based on the homotopy theory and uncover Hopf-type point-gap topology present only for two bands.
Specifically, we reveal such Hopf-type point-gap topology for three-dimensional systems with chiral symmetry (class AIII) and four-dimensional systems with no symmetry (class A).
Explicitly constructing prototypical models from the Hermitian Hopf insulator, we further demonstrate that these non-Hermitian topological phases lead to anomalous point-gapless boundary states spectrally detachable from the bulk bands.
\end{abstract}

\maketitle

\section{Introduction}

Topological insulators and superconductors constitute a cornerstone of condensed matter physics~\cite{HK-review, QZ-review}.
Topological phases of band insulators and Bogoliubov-de Gennes superconductors are characterized by topology of wave functions, giving rise to the emergence of anomalous gapless states at boundaries.
A rich variety of topological insulators and superconductors are generally classified by symmetry~\cite{AZ-97}, leading to the tenfold periodic table~\cite{Schnyder-08, *Ryu-10, Kitaev-09, CTSR-review}.
This classification is based on the stable equivalence in $K$-theory~\cite{Karoubi} and thus applicable to the case of a sufficiently large number of bands.

A distinctive class of topological insulators arises in two-band systems in three dimensions, known as the Hopf insulator~\cite{Moore-08}.
Unlike ordinary topological insulators in three dimensions, the Hopf insulator does not rely on symmetry protection.
Instead, it necessitates the exactly two bands and is unstable against the inclusion of additional bands.
From a mathematical perspective, the Hopf insulator is guaranteed by the homotopy formula~\cite{Nakahara-textbook}
\begin{equation}
    \pi_3 \left( \sphere^2 \right) = \mathbb{Z},
\end{equation} 
where $\sphere^2$ represents the (two-dimensional) sphere as the classifying space of two-band insulators.
This homotopy classification is only relevant to two-band insulators and inapplicable to generic insulators with an arbitrary number of bands.
Consequently, the Hopf insulator is not incorporated in the periodic table of topological insulators and superconductors.
Unique properties of Hopf insulators have been investigated~\cite{Deng-13, Kennedy-16, Liu-17, Unal-19, *Unal-20, Schuster-21, *Schuster-19, Alexandradinata-21, Zhu-21, Lapierre-21, Zhu-23, Wang-23, Lim-23, Jankowski-24}.
More recently, the notion of Hopf insulator has been further extended to delicate topological insulators~\cite{Nelson-21, *Nelson-22, Brouwer-23}.

Meanwhile, topological characterization of non-Hermitian systems has attracted widespread attention~\cite{BBK-review, Okuma-Sato-review} in both theory~\cite{Rudner-09, Sato-11, *Esaki-11, Hu-11, Schomerus-13, Longhi-15, Malzard-15, Lee-16, Leykam-17, Xu-17, Xiong-18, Shen-18, *Kozii-17, Takata-18, MartinezAlvarez-18, Gong-18, YW-18-SSH, *YSW-18-Chern, Kunst-18, McDonald-18, Lee-Thomale-19, Liu-19, Lee-Li-Gong-19, KSUS-19, ZL-19, Herviou-19, Zirnstein-19, Borgnia-19, KBS-19, Yokomizo-19, JYLee-19, Schomerus-20, Chang-20, Wanjura-20, Zhang-20, OKSS-20, Yang-20, Terrier-20, Xue-20, Bessho-21, Denner-21, *Denner-23JPhysMater, Okugawa-20, KSS-20, KSR-21, Zhang-22, Sun-21, Shiozaki-21, Franca-22, Nakamura-24, Denner-23, Wang-24, Nakai-24, Ma-24, Schindler-23, Nakamura-23, Hamanaka-24, Zhang-24, YifanWang-24, Tanaka-24} and experiments~\cite{Poli-15, Zeuner-15, Zhen-15, Weimann-17, Xiao-17, St-Jean-17, Parto-17, Bahari-17, Zhao-18, Zhou-18, Harari-18, *Bandres-18, Cerjan-19, Zhao-19, Brandenbourger-19-skin-exp, *Ghatak-19-skin-exp, Helbig-19-skin-exp, *Hofmann-19-skin-exp, Xiao-19-skin-exp, Weidemann-20-skin-exp, Gou-20, *Liang-22, Wang-21S, *Wang-21N, Wengang-21, Palacios-21, Hu-21, Zhang-21, WangWangMa-22, Wang-23mech, Liu-24, Ochkan-24, Zhao-25, Shen-25, Wu-24}.
Physically, non-Hermiticity arises from coupling with external environments and appears in various open classical and quantum systems~\cite{Konotop-review, Christodoulides-review}.
The distinctive properties of non-Hermitian topology originate from two types of energy gaps ensured by complex-valued spectra:
point and line gaps~\cite{Shen-18, Gong-18, KSUS-19}.
In the presence of a point (line) gap, complex-energy bands are defined not to cross a reference point (line) in the complex-energy plane.
Line-gap topology is continuously deformable into Hermitian (or anti-Hermitian) topology and hence describes the stability of Hermitian topology against non-Hermitian perturbations. 
In contrast, point-gap topology is not necessarily continuously deformable into Hermitian (or anti-Hermitian) topology and thereby constitutes unique features intrinsic to non-Hermitian systems.
Such intrinsic point-gap topology underlies the non-Hermitian skin effect~\cite{Zhang-20, OKSS-20} and the emergence of anomalous boundary states~\cite{Denner-21, Bessho-21, KSR-21, Nakamura-24}.
The interplay of point and line gaps enriches the topological classification of non-Hermitian systems~\cite{KSUS-19}. 

Despite the substantial progress in the research on non-Hermitian topological phases, non-Hermitian analogs of the Hopf insulator have been largely unexplored.
Since the existing classification~\cite{KSUS-19} relies on $K$-theory and is applicable to a sufficiently large number of bands like the Hermitian counterpart~\cite{Schnyder-08, *Ryu-10, Kitaev-09, CTSR-review}, it cannot detect topology unique to non-Hermitian systems with a specific number of bands.
Consequently, it has remained unclear whether Hopf-type point-gap topology can manifest in non-Hermitian systems.
If such a topological phase exists, its implications for boundary phenomena have yet to be elucidated.

In this work, we systematically classify topological phases of two-band non-Hermitian systems through the framework of homotopy theory, uncovering Hopf-type point-gap topology  (Table~\ref{tab: N2}).
This Hopf-type point-gap topology is a unique feature exclusive to two-band non-Hermitian systems and not captured in the previous classification for an arbitrary number of bands (compare Table~\ref{tab: N2} with Table~\ref{tab: Ktheory}).
Specifically, we identify such Hopf-type point-gap topology in three-dimensional non-Hermitian systems with chiral symmetry (i.e., class AIII) and four-dimensional systems with no symmetry (i.e., class A).
Explicitly constructing prototypical models from the Hermitian Hopf insulator, we further demonstrate that these topological phases give rise to the emergence of anomalous point-gapless boundary states with complex-valued spectra that can be spectrally detached from the bulk bands.

It is notable that some previous works studied non-Hermitian topology using the homotopy approach~\cite{Li-21, Wojcik-20, Hu-21knot, Ryu-24, Yang-24, Yang-25}.
However, the relevance of the Hopf invariant to point-gap topology has remained unexplored.
Additionally, the effects of non-Hermitian perturbations on Hermitian Hopf-type insulators and semimetals were also investigated~\cite{Yang-19, *Yang-20Jones, He-20, Kim-23, Pak-24}.
While these previous works mainly focused on line-gap topology and the robustness of the Hermitian Hopf topology under non-Hermitian perturbations, we rather explore point-gap topology governed by the Hopf invariant and the concomitant anomalous point-gapless boundary states, which can be intrinsic to non-Hermitian systems.

The remainder of this work is structured as follows.
In Sec.~\ref{sec: review}, we provide a review of point-gap and line-gap topology, and consequent topological classification based on $K$-theory.
In Sec.~\ref{eq: classification}, we develop the topological classification of two-band non-Hermitian systems, as summarized in Table~\ref{tab: N2}.
As prototypical examples realizing Hopf-type point-gap topology, we investigate three-dimensional non-Hermitian systems with chiral symmetry (class AIII) in Sec.~\ref{sec: 3D-AIII} and four-dimensional ones with no symmetry (class A) in Sec.~\ref{sec: 4D-A}.
In Sec.~\ref{sec: conclusion}, we conclude this work.

\section{Non-Hermitian topology}
    \label{sec: review}

We begin with reviewing the definitions of complex-energy gaps and the corresponding topological classification of non-Hermitian systems.
The complex-valued nature of the spectrum in non-Hermitian systems gives rise to two distinct types of energy gaps:
point and line gaps~\cite{KSUS-19}.
In the presence of a point gap, complex eigenenergies $E_n \left( \bm{k} \right)$'s $\in \mathbb{C}$ of a non-Hermitian Bloch Hamiltonian $H \left( \bm{k} \right)$ are defined to satisfy 
\begin{equation}
\forall\,n \quad E_n \left( \bm{k} \right) \neq E_{\rm P}
\end{equation}
with respect to a reference energy $E_{\rm P} \in \mathbb{C}$.
On the other hand, in the presence of a real (imaginary) line gap, $E_n \left( \bm{k} \right)$'s $\in \mathbb{C}$ are defined to satisfy
\begin{equation}
    \forall\,n \quad \mathrm{Re}\,E_n \left( \bm{k} \right) \neq E_{\rm L} \quad \left[ \mathrm{Im}\,E_n \left( \bm{k} \right) \neq E_{\rm L} \right]
\end{equation}
with respect to a reference energy $E_{\rm L} \in \mathbb{R}$.

The topological classification of non-Hermitian systems depends on the types of these complex-energy gaps~\cite{KSUS-19}.
In general, non-Hermitian systems with real (imaginary) line gaps can be continuously deformed into Hermitian (anti-Hermitian) systems.
Accordingly, line-gap topology characterizes the robustness of Hermitian topology against non-Hermitian perturbations.
By contrast, point-gapped non-Hermitian systems are not necessarily continuously deformable into Hermitian or anti-Hermitian systems.
Point-gap topology of non-Hermitian Hamiltonians $H \left( \bm{k} \right)$ with respect to a reference energy $E_{\rm P}$
can be analyzed through their Hermitized Hamiltonians ${\sf H} \left( {\bm k} \right)$:
\begin{equation}
{\sf H} \left( {\bm k} \right) \coloneqq 
    \begin{pmatrix}
        0 & H \left( \bm{k} \right) - E_{\rm P} \\
        H^{\dag} \left( \bm{k} \right) - E_{\rm P}^{*} & 0
    \end{pmatrix},
        \label{eq: Hermitization}
\end{equation}
which reduces point-gap topology of $H \left( \bm{k} \right)$ to Hermitian topology of ${\sf H} \left( {\bm k} \right)$.
Notably, by construction, this Hermitized Hamiltonian ${\sf H} \left( {\bm k} \right)$ respects additional chiral symmetry,
\begin{equation}
    {\sf \Sigma}\,{\sf H} \left( {\bm k} \right) {\sf \Sigma}^{-1} 
    = - {\sf H} \left( {\bm k} \right),\quad
    {\sf \Sigma} \coloneqq \begin{pmatrix}
        1 & 0 \\
	  0 & -1 \\
    \end{pmatrix}, 
        \label{eq: Hermitization-CS}
\end{equation}
changing the relevant symmetry classes and associated topological classification.

In Table~\ref{tab: Ktheory}, we summarize the topological classification of non-Hermitian systems in Ref.~\cite{KSUS-19}, both in the absence (class A) and presence (class AIII) of chiral symmetry.
Importantly, this classification is based on $K$-theory~\cite{Karoubi}, which assumes the stable equivalence.
Consequently, it only captures topological phases that remain robust under the inclusion of an arbitrary number of additional bands and thus fails to detect topological phases unique to a fixed number of bands.
In the following, we develop the topological classification of two-band non-Hermitian systems and find distinctive point-gap topology that eludes the $K$-theory classification in Table~\ref{tab: Ktheory}.

\begin{table*}[t]
	\centering
	\caption{Topological classification of non-Hermitian systems with two bands $N=2$.
    For each symmetry class, type of complex-energy gap (i.e., point or line gap), spatial dimensions $d$, the classifying space $\star = \mathrm{U} \left( 2 \right), \mathrm{U} \left( 1 \right), \sphere^2, \sphere^0 \times \sphere^0$, and its associated homotopy group $\pi_{d} \left( \star \right)$ are presented. Nontrivial topology that appears in the trivial $K$-theory classification (``$0$" in Table~\ref{tab: Ktheory}) is highlighted by $^{*}$.}
	\label{tab: N2}
     \begin{tabular}{ccccccccccc} \hline \hline
    ~~Class~~ & ~~Gap~~ & ~~Classifying space~~ 
    & ~~$d=1$~~ & ~~$d=2$~~ & ~~$d=3$~~ & ~~$d=4$~~ & ~~$d=5$~~ & ~~$d=6$~~ & ~~$d=7$~~ & ~~$d=8$~~ \\ \hline
    \multirow{2}{*}{A}
    & P & $\mathrm{U} \left( 2 \right)$ & $\mathbb{Z}$ & $0$ & $\mathbb{Z}$ & 
    $\mathbb{Z}_2^{*}$ & $\mathbb{Z}_2$ & $\mathbb{Z}_{12}^{*}$ & $\mathbb{Z}_2$ & $\mathbb{Z}_2^{*}$ \\
    & L & $\sphere^2$ & $0$ & $\mathbb{Z}$ & $\mathbb{Z}^{*}$ & $\mathbb{Z}_2$ & $\mathbb{Z}_2^{*}$ & $\mathbb{Z}_{12}$ & $\mathbb{Z}_2^{*}$ & $\mathbb{Z}_2$ \\ \hline
    \multirow{3}{*}{AIII}
    & P & $\sphere^2$ & $0$ & $\mathbb{Z}$ & $\mathbb{Z}^{*}$ & $\mathbb{Z}_2$ & $\mathbb{Z}_2^{*}$ & $\mathbb{Z}_{12}$ & $\mathbb{Z}_2^{*}$ & $\mathbb{Z}_2$ \\    
    & L$_{\text r}$ & $\mathrm{U} \left( 1 \right)$ & $\mathbb{Z}$ & $0$ & $0$ & $0$ & $0$ & $0$ & $0$ & $0$ \\ 
    & L$_{\text i}$ & $\sphere^0 \times \sphere^0$ & $0$ & $0$ & $0$ & $0$ & $0$ & $0$ & $0$ & $0$ \\ \hline \hline
  \end{tabular}
\end{table*}

\begin{table*}[t]
	\centering
	\caption{Topological classification of non-Hermitian systems based on $K$-theory.
    For each symmetry class, type of complex-energy gap (i.e., point or line gap), spatial dimensions $d$, the classifying space, and its stable topology are presented.
    Reproduced from Table~III in Ref.~\cite{KSUS-19}.}
	\label{tab: Ktheory}
     \begin{tabular}{ccccccccccc} \hline \hline
    ~~Class~~ & ~~Gap~~ & ~~Classifying space~~ & 
    ~~$d=1$~~ & ~~$d=2$~~ & ~~$d=3$~~ & ~~$d=4$~~ & ~~$d=5$~~ & ~~$d=6$~~ & ~~$d=7$~~ & ~~$d=8$~~ \\ \hline
    \multirow{2}{*}{A}
    & P & $\mathcal{C}_1$ & $\mathbb{Z}$ & $0$ & $\mathbb{Z}$ & $0$ & $\mathbb{Z}$ & $0$ & $\mathbb{Z}$ & $0$ \\
    & L & $\mathcal{C}_0$ & $0$ & $\mathbb{Z}$ & $0$ & $\mathbb{Z}$ & $0$ & $\mathbb{Z}$ & $0$ & $\mathbb{Z}$\\ \hline
    \multirow{3}{*}{AIII}
    & P & $\mathcal{C}_0$ & $0$ & 
    $\mathbb{Z}$ & $0$ & $\mathbb{Z}$ & $0$ & $\mathbb{Z}$ & $0$ & $\mathbb{Z}$ \\    
    & L$_{\text r}$ & $\mathcal{C}_1$ & $\mathbb{Z}$ & $0$ & $\mathbb{Z}$ & $0$ & $\mathbb{Z}$ & $0$ & $\mathbb{Z}$ & $0$ \\ 
    & L$_{\text i}$ & $\mathcal{C}_0 \times \mathcal{C}_0$ & $0$ & $\mathbb{Z} \oplus \mathbb{Z}$ & $0$ & $\mathbb{Z} \oplus \mathbb{Z}$ & $0$ & $\mathbb{Z} \oplus \mathbb{Z}$ & $0$ & $\mathbb{Z} \oplus \mathbb{Z}$ \\ \hline \hline
  \end{tabular}
\end{table*}

\section{Topological classification}
    \label{eq: classification}

We provide topological classification of two-band non-Hermitian systems, both with and without chiral symmetry, as summarized in Table~\ref{tab: N2}.
While generic non-Hermitian systems in class A are defined by the absence of any symmetry constraints, those in class AIII are defined to respect chiral symmetry,
\begin{equation}
    \Gamma H^{\dag} \left( {\bm k} \right) \Gamma^{-1} = - H \left( {\bm k} \right),    \quad   \Gamma^{2} = 1,
        \label{eq: CS}
\end{equation}
with a unitary matrix $\Gamma$~\cite{KSUS-19}.
For each symmetry class and type of complex-energy gaps, we identify the relevant classifying space and elucidate the topology through its associated homotopy group.

\subsection{Class A (no symmetry)}
    \label{sec: classification-A}

In the presence of point gaps, generic $N$-band non-Hermitian systems without symmetry form 
\begin{equation}
    \mathrm{GL} \left( N, \mathbb{C} \right) \simeq \mathrm{U} \left( N \right),
        \label{eq: classifying UN}
\end{equation}
where $\mathrm{GL} \left( N, \mathbb{C} \right)$ denotes the general linear group of $N \times N$ complex matrices, $\mathrm{U} \left( N \right)$ denotes the group of $N \times N$ unitary matrices, and $\simeq$ represents the homotopy equivalence.
In particular, two-band non-Hermitian Bloch Hamiltonians $H \left( \bm{k} \right)$ provide maps from the $d$-dimensional Brillouin zone to the classifying space $\mathrm{U} \left( 2 \right)$.

In the presence of line gaps, on the other hand, generic two-band non-Hermitian systems without symmetry are diagonalized through nonorthogonal eigenstates forming $\mathrm{GL} \left( 2, \mathbb{C} \right)$.
Given the gauge ambiguity, the classifying space in terms of line gaps is given as
\begin{align}
    &\mathrm{GL} \left( 2, \mathbb{C} \right)/\mathrm{GL} \left( 1, \mathbb{C} \right) \times \mathrm{GL} \left( 1, \mathbb{C} \right) \nonumber \\
    &\qquad\quad \simeq \mathrm{U} \left( 2 \right)/\mathrm{U} \left( 1 \right) \times \mathrm{U} \left( 1 \right) \simeq \sphere^2,
        \label{eq: classifying S2}
\end{align}
where $\sphere^2$ is the (two-dimensional) sphere.

The topological classification for each type of complex-energy gaps is determined by the homotopy group of the corresponding classifying space, i.e., $\pi_d \left( \mathrm{U} \left( 2 \right)\right)$ or $\pi_d \left( \sphere^2 \right)$~\cite{Nakahara-textbook}, as summarized in Table~\ref{tab: N2}.
Crucially, certain nontrivial topology found for $N=2$ is unstable for a generic number of bands and cannot be captured by $K$-theory;
compare Table~\ref{tab: N2} with Table~\ref{tab: Ktheory}.
The Hermitian Hopf insulators~\cite{Moore-08} correspond to the $\mathbb{Z}$-classified topological phases in three dimensions for line gaps.
In particular, two-band non-Hermitian systems in four dimensions support the $\mathbb{Z}_2$ point-gap topology that has no counterparts in generic $N$-band systems, which we clarify in Sec.~\ref{sec: 4D-A}.

\subsection{Class AIII (chiral symmetry)}
    \label{sec: classification-AIII}

In class AIII, point-gap topology of generic non-Hermitian systems $H \left( \bm{k} \right)$ reduces to Hermitian topology of 
$\ii \left( H \left( \bm{k} \right) - E_{\rm P} \right) \Gamma$,
where 
$E_{\rm P} \in \ii \mathbb{R}$ is a reference energy, and
$\Gamma$ denotes the chiral-symmetry operator as defined in Eq.~(\ref{eq: CS}) (see Appendix~\ref{appendix: chiral} for details~\cite{KSUS-19, KBS-19}).
Accordingly, the classifying space in the case of two bands is given as $\sphere^2$, as in Eq.~(\ref{eq: classifying S2}).

In the presence of chiral symmetry, real and imaginary line gaps are distinguished from each other.
For real line gaps, generic $N$-band non-Hermitian systems are continuously deformed into Hermitian systems while preserving the real line gaps and chiral symmetry.
In such a Hermitian limit, the Hamiltonians $H \left( \bm{k} \right)$ take the form
\begin{equation}
H \left( {\bm k} \right) \coloneqq 
    \begin{pmatrix}
        0 & D \left( \bm{k} \right) \\
        D^{\dag} \left( \bm{k} \right) & 0
    \end{pmatrix}, \quad D \in \mathrm{GL} \left( N/2, \mathbb{C} \right),
\end{equation}
where the chiral-symmetry operator $\Gamma$ is chosen as the diagonal matrix $\Gamma = \mathrm{diag} \left( 1, -1 \right)$.
Consequently, the classifying space is identified as $\mathrm{GL} \left( N/2, \mathbb{C} \right) \simeq \mathrm{U} \left( N/2 \right)$, as in Eq.~(\ref{eq: classifying UN}).
Here, $N$ is assumed to be even since real line gaps cannot be open for odd $N$.

For imaginary line gaps, by contrast, generic $N$-band non-Hermitian systems can be continuously deformed into anti-Hermitian systems while preserving both imaginary line gaps and chiral symmetry.
In this anti-Hermitian limit, $N$-band Hamiltonians $H \left( \bm{k} \right)$ are expressed as $H \left( \bm{k} \right) \eqqcolon \ii \bar{H} \left( \bm{k} \right)$, where $\bar{H} \left( \bm{k} \right)$ satisfies Hermiticity and 
\begin{equation}
    \Gamma \bar{H} \left( \bm{k} \right) \Gamma^{-1} = \bar{H} \left( \bm{k} \right), \quad \Gamma^2 = 1.
\end{equation}
Accordingly, $\bar{H} \left( \bm{k} \right)$ commutes with the unitary operator $\Gamma$.
In particular, in the case of $N=2$, $\bar{H} \left( \bm{k} \right)$ consists of two independent scalars, further implying the classifying space $\sphere^0 \times \sphere^0$ (note that the zero-dimensional sphere $\sphere^0$ is the pair of points).

The topological classification of two-band non-Hermitian systems in class AIII is also summarized in Table~\ref{tab: N2}.
In a similar manner to class A, certain nontrivial topological features observed for $N=2$ are inherently unstable for generic $N$-band systems and thus cannot be described within the framework of $K$-theory.
As a prime example, two-band chiral-symmetric non-Hermitian systems in three dimensions host the $\mathbb{Z}$ point-gap topology, which we clarify in Sec.~\ref{sec: 3D-AIII}.

\section{3D class AIII}
    \label{sec: 3D-AIII}

According to the classification in Table~\ref{tab: N2}, the minimal nontrivial manifestation of Hopf-type non-Hermitian topology arises in three dimensions for class AIII.
Specifically, we have the $\mathbb{Z}$ topological classification for two-band chiral-symmetric non-Hermitian systems with point gaps, which becomes unstable upon the inclusion of additional bands.
In the following, we elucidate such Hopf-type point-gap topology in three dimensions by explicitly constructing a prototypical model from a Hermitian Hopf insulator.

\subsection{Hermitian Hopf insulator}

We begin with reviewing a Hermitian Hopf insulator~\cite{Moore-08, Alexandradinata-21}:
\begin{equation}
    H_{\rm Hopf} \left( \bm{k} \right) = - \left( \vec{z}^{\dag} \bm{\sigma} \vec{z} \right) \cdot \bm{\sigma},
        \label{eq: Hermitian Hopf}
\end{equation}
where $\bm{\sigma} = \left( \sigma_x~\sigma_y~\sigma_z \right)^{T}$ denotes Pauli matrices, and $\vec{z} = \left( z_1~z_2 \right)^{T}$ is defined as
\begin{equation}
\begin{split}
    z_1 &\coloneqq \sin k_x + \ii \sin k_y, \\
    z_2 &\coloneqq \sin k_z + \ii \left( \phi + \cos k_x + \cos k_y + \cos k_z - 3 \right)
\end{split}
\end{equation}
with $\phi \in \mathbb{R}$.
From this Hermitian model $H_{\rm Hopf} \left( \bm{k} \right)$, we shortly construct our non-Hermitian Hopf insulators.
The Hamiltonian $H_{\rm Hopf} \left( \bm{k} \right)$ is rewritten as
\begin{equation}
    H_{\rm Hopf} \left( \bm{k} \right) = -\bm{h} \cdot \bm{\sigma} = - h_x \sigma_x - h_y \sigma_y - h_z \sigma_z
    \label{eq: Hermitian Hopf2}
\end{equation}
with 
\begin{equation}
\begin{split}
    h_x &= z_1 z_2^{*} + z_1^{*} z_2, \\ 
    h_y &= \ii \left( z_1 z_2^{*} - z_1^{*} z_2 \right), \\
    h_z &= z_1 z_1^{*} - z_2 z_2^{*}.
\end{split}
    \label{eq: hxyz}
\end{equation}
The energy gap $\left| \bm{h} \right| = \sqrt{h_x^2 + h_y^2 + h_z^2}$ is closed for $h_x = h_y = h_z = 0$, i.e., $z_1 = z_2 = 0$, i.e., 
\begin{equation}
    \phi = 0, 2, 4, 6.
\end{equation}

In the presence of an energy gap $\left| \bm{h} \right| \neq 0$, this Hamiltonian provides a Hopf map:
\begin{equation}
    \begin{array}{rccc}
    &\sphere^3                     &\longrightarrow& \sphere^2                     \\
        & \rotatebox{90}{$\in$}&               & \rotatebox{90}{$\in$} \\
        & \bm{k}                   & \longmapsto   & \bm{h}/\left| \bm{h} \right|
\end{array}
    \label{eq: Hopf map}
\end{equation}
From the homotopy formula~\cite{Nakahara-textbook}
\begin{equation}
    \pi_3 \left( \sphere^2 \right) = \mathbb{Z},
\end{equation}
it follows that the Hamiltonian can exhibit nontrivial topological phases.
As discussed previously, this corresponds to the $\mathbb{Z}$ topological classification for class A, $d=3$, and line gap in Table~\ref{tab: N2}.
The associated Hopf invariant for an eigenstate $\ket{u}$ with the lower eigenenergy is given by the (Abelian) Chern-Simons three-form~\cite{Wilczek-83},
\begin{equation}    
    \chi = - \oint \frac{d^3k}{4\pi^2} \bm{F} \cdot \bm{A},
        \label{eq: Hopf invariant}
\end{equation}
with the Berry connection $\bm{A} \coloneqq \ii \braket{u | \bm{\nabla}_{\bm k} | u}$ and the Berry curvature $\bm{F} \coloneqq \bm{\nabla}_{\bm k} \times \bm{A}$.
For the Hopf insulator in Eq.~(\ref{eq: Hermitian Hopf}), this topological invariant is given as
\begin{equation}
    \chi = \begin{cases}    
        0 & \left( \phi < 0 \right), \\
        1 & \left( 0 < \phi < 2 \right), \\
        -2 & \left( 2 < \phi < 4 \right), \\
        1 & \left( 4 < \phi < 6 \right), \\
        0 & \left( 6 < \phi \right). \\
    \end{cases}
        \label{eq: chi}
\end{equation}
It should be noted that while momenta $\bm{k} = \left( k_x, k_y, k_z \right)$ live in the three-dimensional Brillouin zone torus $T^3$, the difference between $\sphere^3$ and $T^3$ manifests itself only in the three weak Chern numbers, which are assumed to vanish here.

\begin{figure}[t]
    \centering
\includegraphics[width=\columnwidth]{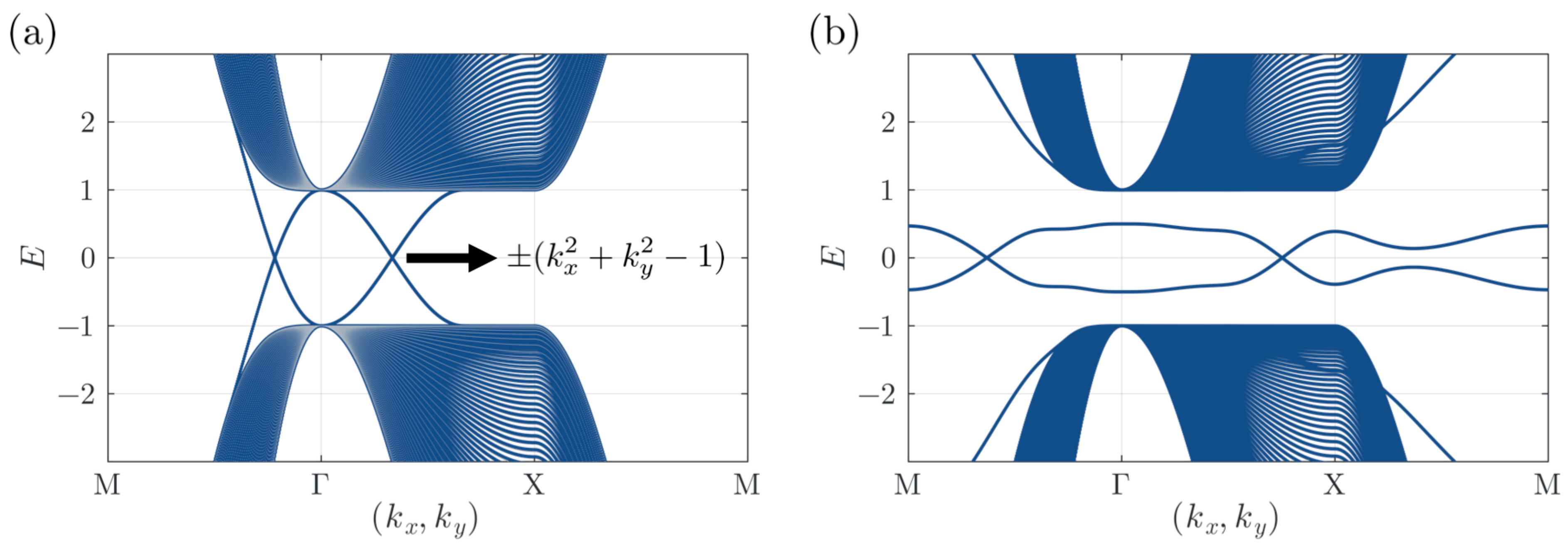}
\caption{(a)~Energy spectra of the Hermitian Hopf insulator in Eq.~(\ref{eq: Hermitian Hopf}) with $\phi=1$ under the open boundary conditions in the $z$ direction (blue) and the periodic boundary conditions (gray). 
Around the $\Gamma$ point, a gapless state with the quadratic energy dispersion appears at each surface. 
(b)~Detachment of the two surface (in-gap) bands for (a) in the presence of perturbations added at the top and bottom surfaces along the $z$ direction. 
Details of the surface perturbations are given in Appendix~\ref{appendix: Detachment}. 
The spectra are plotted along the high-symmetry line $M{\rm -}\Gamma{\rm -}X{\rm -}M$ of the two-dimensional Brillouin zone $(k_x,k_y)$, where $M=(\pi,\pi), \Gamma=(0,0)$, and $X=(\pi,0)$. 
The system size in the $z$ direction is $100$, and the momentum resolution along the high-symmetry line is $3\pi/899$.}
    \label{fig:3DclassA}
\end{figure}

As a consequence of the nontrivial Hopf invariant $\chi$, surface states emerge at boundaries. 
The number of such surface states is given by $\left| \chi \right|$.
In particular, for $\chi = 1$, the following effective surface Hamiltonian arises under the open boundary conditions along the $z$ direction (see Appendix~\ref{appendix: boundary} for a derivation):
\begin{equation}
    H \left( \bm{k} \right) = k_x^2 + k_y^2 - m \quad \left( m > 0 \right).
        \label{eq: Hermitian Hopf boundary}
\end{equation}
Consistently, the lattice model in Eq.~(\ref{eq: Hermitian Hopf}) supports surfaces states under the open boundary conditions, as shown by numerical calculations in Fig.~\ref{fig:3DclassA}\,(a).
Unlike ordinary topological insulators, this surface state is nonrelativistic, reflecting its unstable nature. 
This unique feature also manifests itself as the detachability of the surface states from the bulk bands~\cite{Alexandradinata-21}. 
In Fig.~\ref{fig:3DclassA}\,(b), we explicitly demonstrate the detachment of the surface bands by adding perturbations at surfaces (see Appendix~\ref{appendix: Detachment} for details).

\subsection{Non-Hermitian Hopf insulator}

As explained in Sec.~\ref{sec: classification-AIII} and Appendix~\ref{appendix: chiral}, point-gap topology of non-Hermitian systems with chiral symmetry in Eq.~(\ref{eq: CS}) is equivalent to Hermitian topology of $\ii H \left( \bm{k} \right) \Gamma$, where we choose a reference energy to be zero.
Using this correspondence, we construct a non-Hermitian Hopf insulator from the Hermitian Hopf insulator in Eq.~(\ref{eq: Hermitian Hopf}).
Specifically, we introduce the non-Hermitian Hopf insulator by
\begin{align}
    H \left( \bm{k} \right) &\coloneqq - \ii H_{\rm Hopf} \left( \bm{k} \right) \Gamma \nonumber \\
    &= \ii \left( \vec{z}^{\dag} \sigma_z \vec{z} \right) - \left( \vec{z}^{\dag} \sigma_y \vec{z} \right) \sigma_x + \left( \vec{z}^{\dag} \sigma_x \vec{z} \right) \sigma_y \nonumber \\
    &= \ii h_z - h_y \sigma_x + h_x \sigma_y,
        \label{eq: non-Hermitian Hopf}
\end{align}
where the chiral-symmetry operator is chosen as $\Gamma = \sigma_z$, and $h_x$, $h_y$, and $h_z$ are defined in Eq.~(\ref{eq: hxyz}).
Inheriting the nontrivial Hopf invariant in Eqs.~(\ref{eq: Hopf invariant}) and (\ref{eq: chi}), this non-Hermitian Hamiltonian $H \left( \bm{k} \right)$ exhibits nontrivial point-gap topology with respect to reference energy around $E = 0$.
In contrast to the Hermitian counterpart, this Hopf point-gap topology is protected by chiral symmetry. 

\begin{figure}[t]
    \centering
\includegraphics[width=\columnwidth]{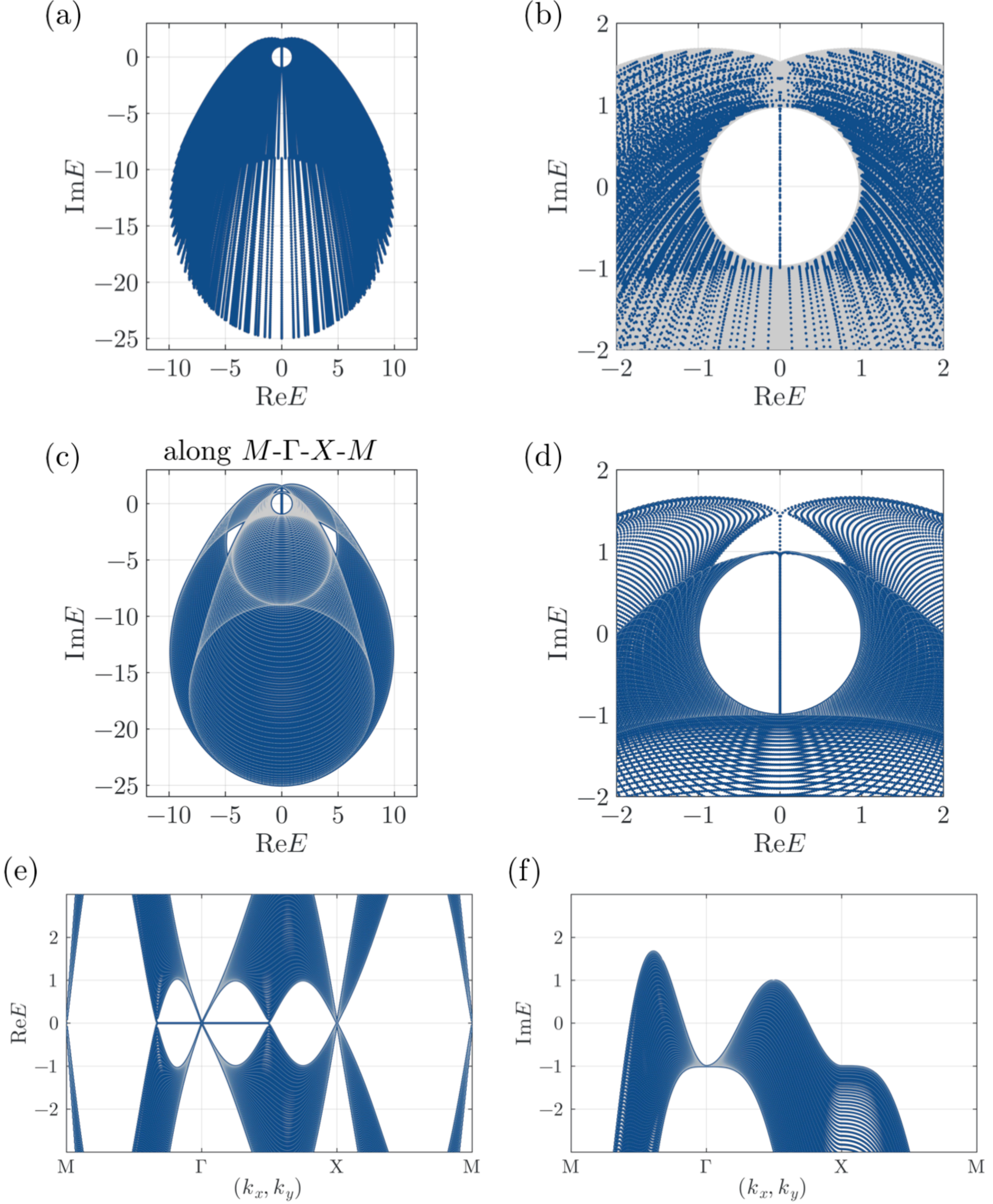}
\caption{Complex energy spectra of the non-Hermitian Hopf insulator in Eq.~(\ref{eq: non-Hermitian Hopf}) with $\phi=1$ under the open boundary conditions in the $z$ direction (blue) and the periodic boundary conditions (gray). 
(a)~Complex energy spectrum across the entire two-dimensional Brillouin zone $(k_x,k_y)$ under the open boundary conditions in the $z$ direction. 
The system size in the $z$ direction is $100$, and the momentum resolution is $(2\pi/59,2\pi/59)$. 
(b)~Complex energy spectrum of (a) around $E=0$, compared with the bulk spectrum under the periodic boundary conditions (gray). 
A point gap with a nontrivial Hopf invariant opens around $E=0$ under the periodic boundary conditions, leading to the in-gap states with the purely imaginary energy dispersion under the open boundary condition. 
(c-f)~Comparison of the complex energy spectra along the high-symmetry line $M{\rm -}\Gamma{\rm -}X{\rm -}M$ of the two-dimensional Brillouin zone $(k_x,k_y)$ under the open boundary conditions in the $z$ direction and the periodic boundary conditions. 
The system size in the $z$ direction is $100$, and the momentum resolution along the high-symmetry line is $3\pi/899$.}
    \label{fig:3DclassAIII}
\end{figure}

\begin{figure}[t]
    \centering
\includegraphics[width=\columnwidth]{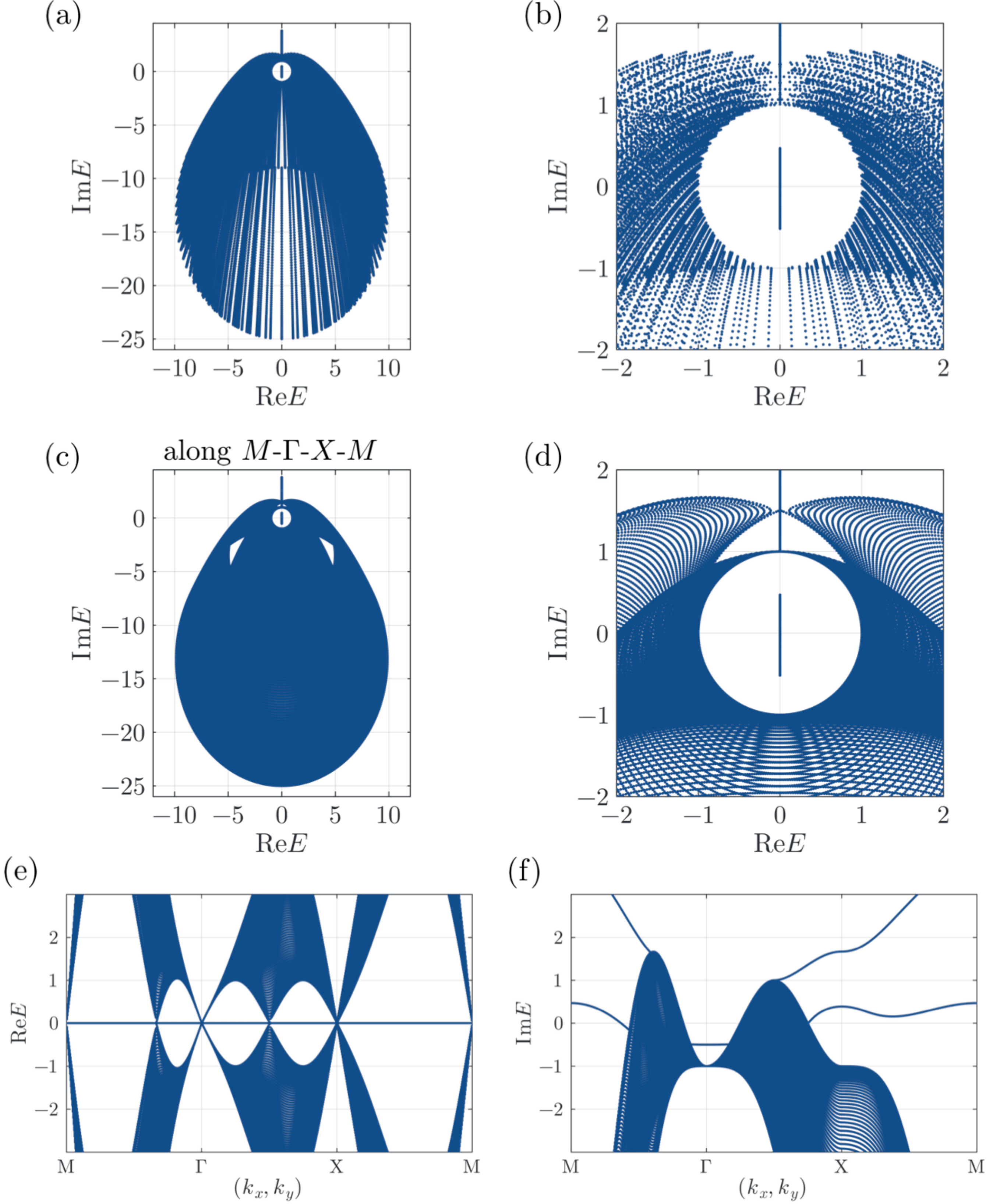}
\caption{Complex energy spectra of the perturbed non-Hermitian Hopf insulator constructed from the Hermitian model in Fig.~\ref{fig:3DclassA}\,(b) under the open boundary conditions in the $z$ direction. 
(a)~Complex energy spectrum across the entire two-dimensional Brillouin zone $(k_x,k_y)$. 
The system size in the $z$ direction is $100$, and the momentum resolution is $(2\pi/59,2\pi/59)$. 
(b)~Complex energy spectrum of (a) around $E=0$. 
In comparison with Fig.~\ref{fig:3DclassAIII}\,(b), the in-gap states with the purely imaginary energy dispersion are detached from the other bands. 
(c-f)~Complex energy spectra along the high-symmetry line $M{\rm -}\Gamma{\rm -}X{\rm -}M$ of the two-dimensional Brillouin zone $(k_x,k_y)$. 
The system size in the $z$ direction is $100$, and the momentum resolution along the high-symmetry line is $3\pi/899$.}
    \label{fig:3DclassAIII_Detach}
\end{figure}

\begin{figure}[t]
    \centering
\includegraphics[width=\columnwidth]{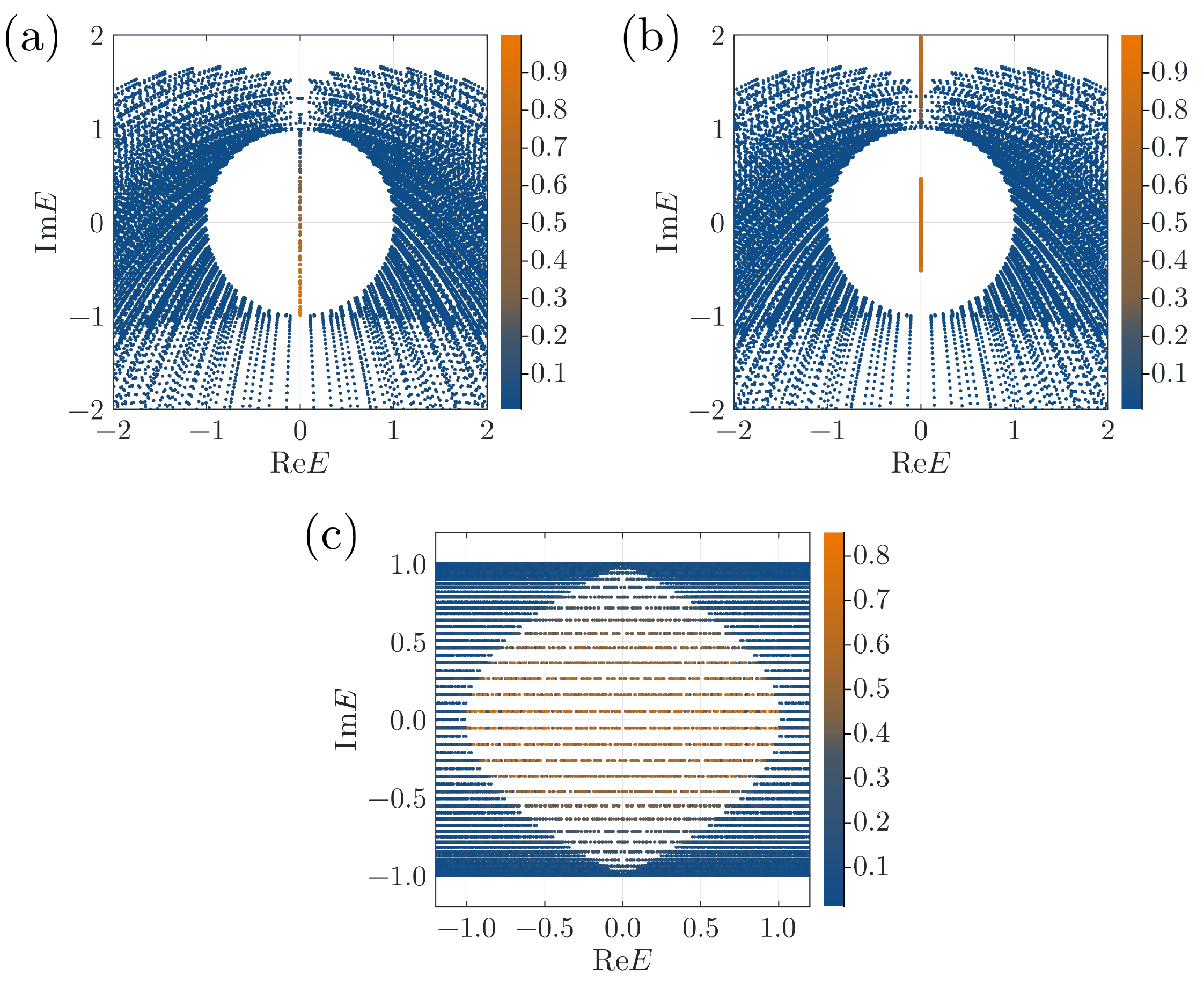}
\caption{Localization properties of surface states in Figs.~\ref{fig:3DclassAIII}\,(b), \ref{fig:3DclassAIII_Detach}\,(b), and \ref{fig:4DclassA}\,(b). 
The color represents the inverse participation ratio (IPR), where the larger values indicate the stronger localization toward the $z$ surfaces.}
    \label{fig:IPR}
\end{figure}

In Fig.~\ref{fig:3DclassAIII}, we provide the complex energy spectra of Eq.~(\ref{eq: non-Hermitian Hopf}) with $\phi=1$ under both periodic and open boundary conditions. 
As a result of the nontrivial Hopf point-gap topology $\chi=1$, the non-Hermitian Hopf insulator exhibits the point-gapless surface states with the purely imaginary energy dispersion. 
In general, the number of such point-gapless surface states is given by $\left| \chi \right|$.
Building upon the surface theory of the Hermitian Hopf insulator in Eq.~(\ref{eq: Hermitian Hopf boundary}), we obtain the corresponding surface Hamiltonian for the non-Hermitian Hopf insulator as
\begin{equation}
    H \left( \bm{k} \right) = \ii \left( k_x^2 + k_y^2 - m \right),
\end{equation}
which agrees with the numerical results in Fig.~\ref{fig:3DclassAIII}.
We also investigate the detachability of these surface states from the bulk bands under perturbations in Fig.~\ref{fig:3DclassAIII_Detach}. 
As expected from the detached surface states in the original Hermitian Hopf insulator [see Fig.~\ref{fig:3DclassA}\,(b)], we numerically demonstrate that the perturbed non-Hermitian Hopf insulator indeed supports the detached point-gapless surface states with the purely imaginary energy dispersion.
In Figs.~\ref{fig:IPR}\,(a) and\,(b), we further calculate the inverse participation ratios (IPRs) of eigenstates $\psi_{z,\alpha}$ along the $z$ direction,
\begin{align}
    \mathrm{IPR} \coloneqq \frac{\sum_{z,\alpha}|\psi_{z,\alpha}|^4}{\left( \sum_{z,\alpha}|\psi_{z,\alpha}|^2 \right)^2},
\end{align}
where $z$ and $\alpha$ represent spatial and internal degrees of freedom, respectively.
The surface states exhibit larger IPRs, showing their localization.

In contrast to the Hermitian counterparts, this surface state possesses the purely imaginary spectrum.
This implies that if this surface state, as well as the associated point-gap topology, is continuously deformable into a surface state ensured by certain line-gap topology, it should reduce to imaginary line gaps, in a similar manner to ordinary point-gap topology for two-dimensional non-Hermitian systems in class AIII~\cite{Nakamura-24}.
Nevertheless, the two-band classification in Table~\ref{tab: N2} indicates the absence of such imaginary-line-gap topology.
This further implies that the obtained Hopf point-gap topology should be irreducible to line-gap topology and hence intrinsic to non-Hermitian systems.
However, it should also be noted that it can be related to imaginary-line-gap topology for four-band non-Hermitian systems instead of two-band ones.
A comprehensive formulation of such $N$-band intrinsic and extrinsic point-gap topology is left for further research.

\section{4D class A}
    \label{sec: 4D-A}

As discussed in Sec.~\ref{sec: classification-A}, a prime example of nontrivial point-gap topology uniquely realized in two-band non-Hermitian systems can be found in four dimensions without symmetry (see Table~\ref{tab: N2}).
Indeed, the homotopy formula~\cite{Nakahara-textbook}
\begin{equation}
    \pi_4 \left( \mathrm{U} \left( 2 \right) \right) = \mathbb{Z}_2
        \label{eq: pi4}
\end{equation}
implies the existence of nontrivial $\mathbb{Z}_2$-classified point-gap topology, in contrast with its absence for larger bands, i.e.,~\cite{Nakahara-textbook}
\begin{equation}
    \pi_4 \left( \mathrm{U} \left( n \right) \right) = 0 \quad \left( n \geq 3 \right).
\end{equation}

\begin{figure}[t]
    \centering
\includegraphics[width=\columnwidth]{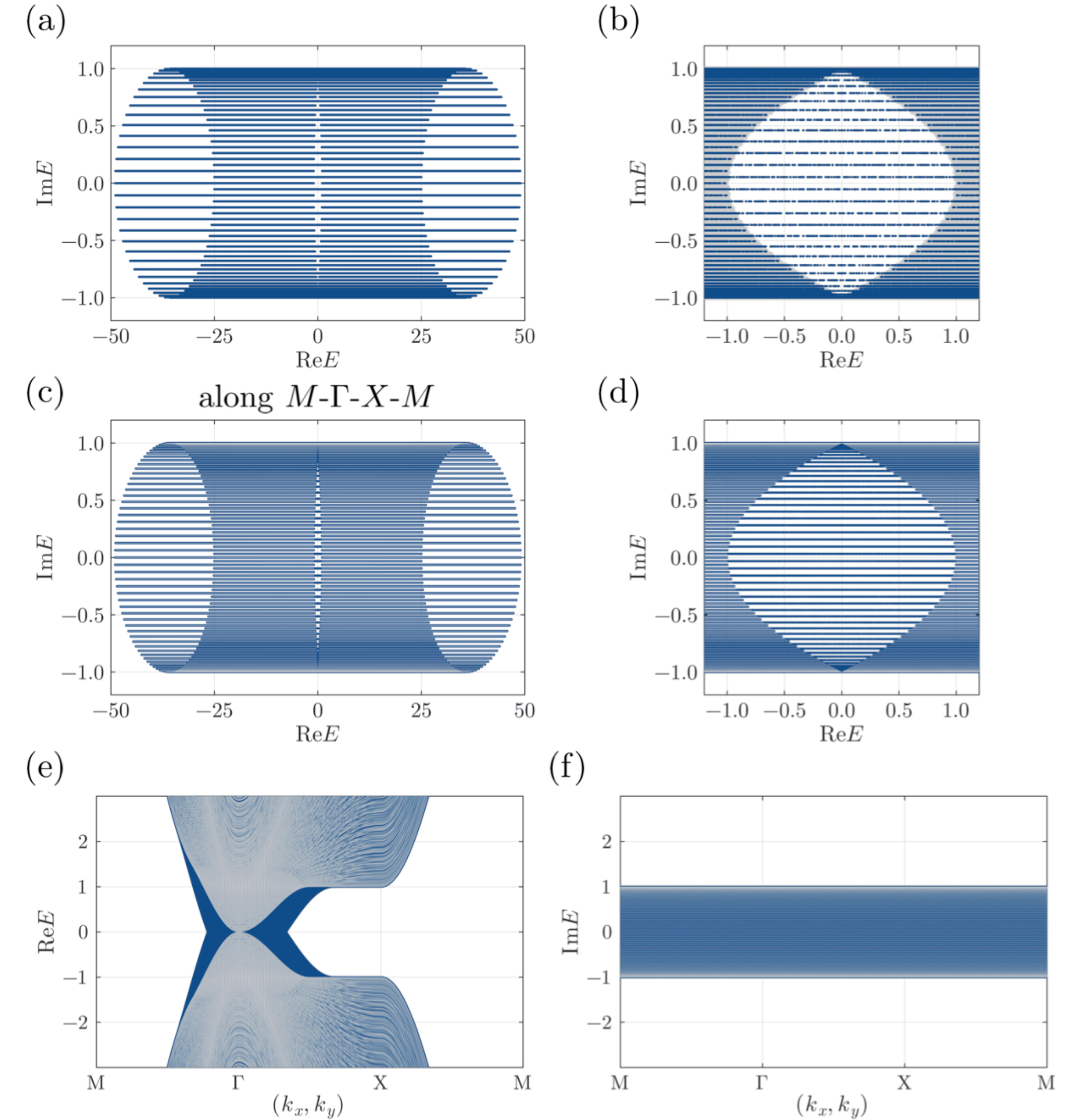}
\caption{Complex energy spectra of the four-dimensional non-Hermitian Hopf-type model in Eq.~(\ref{eq: non-Hermitian Hopf-type}) under the open boundary conditions in the $z$ direction (blue) and the periodic boundary conditions (gray).
(a)~Complex energy spectrum across the entire three-dimensional Brillouin zone $(k_x,k_y,k_w)$ under the open boundary conditions in the $z$ direction. 
The system size in the $z$ direction is $60$, and the momentum resolution is $(2\pi/59,2\pi/59,2\pi/59)$. 
(b)~Complex energy spectrum of (a) around $E=0$, compared with the bulk spectrum under the periodic boundary conditions (gray). 
A point gap with a nontrivial $\mathbb{Z}_2$ invariant opens around $E=0$ under the periodic boundary conditions. 
The consequent in-gap surface states cover the region around $E=0$ under the open boundary condition. 
(c-f)~Comparison of the complex energy spectra along the high-symmetry line $M{\rm -}\Gamma{\rm -}X{\rm -}M$ of the two-dimensional Brillouin zone $(k_x,k_y)$ with $k_w\in[-\pi,\pi]$ under the open boundary conditions in the $z$ direction and the periodic boundary conditions. 
The system size in the $z$ direction is $100$, and the momentum resolutions of $(k_x,k_y)$ along the high-symmetry line and $k_w$ are $3\pi/899$ and $2\pi/99$, respectively.}
    \label{fig:4DclassA}
\end{figure}

We explicitly construct a model realizing this non-Hermitian topological phase in four dimensions on the basis of the Hermitian Hopf insulator $H_{\rm Hopf} \left( \bm{k} \right)$ in three dimensions [i.e., Eq.~(\ref{eq: Hermitian Hopf})].
Specifically, we introduce a non-Hermitian Hopf-type model in four dimensions by
\begin{align}
    H \left( \bm{k} \right) = - \left( \vec{z}^{\dag} \bm{\sigma} \vec{z} \right) \cdot \bm{\sigma} + \ii \sin k_w,
    \label{eq: non-Hermitian Hopf-type}
\end{align}
where $\bm{k}$ represents a four-dimensional momentum $\bm{k} = \left( k_x, k_y, k_z, k_w \right)$, and $\vec{z} = \left( z_1~z_2 \right)^{T}$ is defined as
\begin{equation}
\begin{split}
    z_1 &\coloneqq \sin k_x + \ii \sin k_y, \\
    z_2 &\coloneqq \sin k_z + \ii \left( \cos k_x + \cos k_y \right. \\
    &\qquad\qquad\qquad\quad \left. + \cos k_z + \cos k_w - 3 \right).
\end{split}
\end{equation}
Notably, the parameter $\phi$ in the Hermitian Hopf insulator in Eq.~(\ref{eq: Hermitian Hopf}) is replaced by $\cos k_w$, and this non-Hermitian model $H \left( \bm{k} \right)$ continuously interpolates between the nontrivial and trivial Hermitian Hopf insulators for $k_w = 0$ and $k_w = \pi$, respectively.
Consequently, inheriting the nontrivial Hopf invariant in Eqs.~(\ref{eq: Hopf invariant}) and (\ref{eq: chi}), this non-Hermitian Hamiltonian $H \left( \bm{k} \right)$ exhibits nontrivial point-gap topology with respect to reference energy around $E = 0$.
Mathematically, this construction corresponds to the suspension of the Hopf map $\sphere^3 \to \sphere^2$ in Eq.~(\ref{eq: Hopf map}).
It is also notable that this construction is similar to the formulation of exceptional topological insulators (i.e., non-Hermitian point-gapped models in three dimensions for class A) through the non-Hermitian interpolation of Hermitian Chern insulators in two dimensions~\cite{Denner-21, Bessho-21, KSR-21, Nakamura-24}.
In passing, we note that the homotopy formula in Eq.~(\ref{eq: pi4}) is also relevant to the Witten anomaly~\cite{Witten-83}.

Extending the surface theory of the Hermitian Hopf insulator in Eq.~(\ref{eq: Hermitian Hopf boundary}), we obtain the corresponding non-Hermitian surface Hamiltonian under the open boundary conditions along the $z$ direction as
\begin{equation}
    H \left( \bm{k} \right) = k_x^2 + k_y^2 + \ii k_w - m.
    \label{eq: non-Hermitian Hopf-type boundary}
\end{equation}
In Fig.~\ref{fig:4DclassA}, we numerically demonstrate the emergence of such surface states with the complex spectrum. 
In contrast to the purely imaginary energy dispersion in Eq.~(\ref{eq: non-Hermitian Hopf}), the present surface states close all point gaps around $E=0$ [compare Fig.~\ref{fig:4DclassA}\,(b) with Fig.~\ref{fig:3DclassAIII}\,(b) for details].
This is also similar to the surface states in exceptional topological insulators~\cite{Denner-21, Bessho-21, KSR-21, Nakamura-24}. 
Given that the surface states in exceptional topological insulators can host a single exceptional point, it is worth further investigating the possible connection between Eq.~(\ref{eq: non-Hermitian Hopf-type boundary}) and exceptional points. 
In Fig.~\ref{fig:IPR}\,(c), we also provide the IPRs along the $z$ direction, showing the localization of the surface states.

\section{Discussion}
    \label{sec: conclusion}

While generic topological insulators are stable against the inclusion of additional bands, Hopf insulators require exactly two bands, making them a distinct class of topological insulators.
In this work, we have investigated analogs of Hermitian Hopf insulators within the framework of non-Hermitian point-gap topology.
Employing homotopy theory, we have systematically classified topological phases of two-band non-Hermitian systems.
Based on this classification, we have found distinctive point-gap topology that exists only in the two-band setting, reminiscent of Hermitian Hopf insulators.
As prototypical examples, we have elucidated such Hopf-type point-gap topology for three-dimensional non-Hermitian systems with chiral symmetry (class AIII) and four-dimensional ones without symmetry (class A), demonstrating the emergence of anomalous point-gapless boundary states spectrally detachable from the bulk bands.

The Hopf-type point-gap topology studied in this work gives rise to the emergence of point-gapless boundary states in a similar manner to Hermitian topological insulators.
By contrast, other variants of Hopf-type point-gap topology can potentially induce the non-Hermitian skin effect.
It is noteworthy that whereas intrinsic point-gap topology in one dimension constitutes the origin of the skin effect~\cite{Zhang-20, OKSS-20}, the higher-dimensional counterparts do not typically cause the skin effect~\cite{Nakamura-24}.
In this respect, the absence of the skin effect in our non-Hermitian Hopf insulators in three and four dimensions seems reasonable.

While we have focused on two-band non-Hermitian systems in this work, different types of point-gap topology can appear in non-Hermitian systems with generic $N$ bands.
Additionally, although our analysis has been restricted to classes A and AIII, other symmetry classes can also host point-gap topological phases unique to a fixed number of bands.
Such different types of point-gap topology are worth further investigation.
Moreover, it is notable that detachable boundary states in stable Hermitian topological insulators are related to extrinsic point-gap topology~\cite{Altland-24, Lapierre-24, NSSSK-24, *SNSSK-24}.
Accordingly, our classification of $N$-band point-gap topology can be relevant to detachable boundary states in Hermitian Hopf insulators [see Fig.~\ref{fig:3DclassA}\,(b)]~\cite{Alexandradinata-21}, which is left for further study.

\medskip
{\it Note added}.---After the completion of this work, we became aware of a recent related work~\cite{Yoshida-25}.

\medskip
\begingroup
\renewcommand{\addcontentsline}[3]{}
\begin{acknowledgments}
We thank Tokiro Numasawa, Ken Shiozaki, and Zhenyu Xiao for helpful discussion.
D.N. is supported by JSPS KAKENHI Grant No.~JP24K22857.
K.K. is supported by MEXT KAKENHI Grant-in-Aid for Transformative Research Areas A ``Extreme Universe" No.~JP24H00945.
\end{acknowledgments}
\endgroup

\appendix

\section{Point-gap topology protected by chiral symmetry}
    \label{appendix: chiral}

We clarify point-gap topology protected by chiral symmetry.
For non-Hermitian Hamiltonians $H \left( \bm{k} \right)$ respecting chiral symmetry in Eq.~(\ref{eq: CS}), their point-gap topology reduces to Hermitian topology of the associated Hermitian matrix 
$\ii \left( H \left( {\bm k} \right) -E_{\rm P} \right) \Gamma$~\cite{KSUS-19, KBS-19}. 
To show this, we focus on the Hermitized Hamiltonian ${\sf H} \left( {\bm k} \right)$ introduced in Eq.~(\ref{eq: Hermitization}).
The presence of a point gap for $H \left( \bm{k} \right)$ is equivalent to the presence of an energy gap for ${\sf H} \left( {\bm k} \right)$.
Henceforth, we choose a reference energy $E_{\rm P} \in \ii \mathbb{R}$ to be zero for the sake of simplicity.
Owing to chiral symmetry in Eq.~(\ref{eq: CS}), ${\sf H} \left( {\bm k} \right)$ also respects additional chiral symmetry
\begin{equation}
    {\sf \Gamma}\,{\sf H} \left( {\bm k} \right) {\sf \Gamma}^{-1} 
    = - {\sf H} \left( {\bm k} \right),\quad
    {\sf \Gamma} \coloneqq 
    \begin{pmatrix}
        0 & \Gamma \\
	   \Gamma & 0 \\
    \end{pmatrix}.
\end{equation}
Moreover, by construction, ${\sf H} \left( {\bm k} \right)$ satisfies additional chiral symmetry in Eq.~(\ref{eq: Hermitization-CS}).

From the two independent chiral symmetries, the following commutation relation holds,
\begin{equation}
\left[ {\sf H} \left( {\bm k} \right),\,\ii {\sf \Gamma \Sigma} \right] = 0.
\end{equation}
Thus, the two Hermitian matrices ${\sf H} \left( {\bm k} \right)$ and $\ii {\sf \Gamma \Sigma}$ can be simultaneously diagonalized by a unitary matrix ${\sf U}$, leading to
\begin{align}
    {\sf U}^{\dag} {\sf H} \left( {\bm k} \right) {\sf U} &= 
    \begin{pmatrix}
        \ii H \left( {\bm k} \right) \Gamma & 0 \\
	   0 & -\ii H \left( {\bm k} \right) \Gamma \\
    \end{pmatrix}, \\
    {\sf U}^{\dag} \left( \ii {\sf \Gamma \Sigma} \right) {\sf U} &= \begin{pmatrix}
        1 & 0 \\
	   0 & -1 \\
    \end{pmatrix}; \\
    {\sf U} &\coloneqq \frac{1}{\sqrt{2}} \begin{pmatrix}
        1 & -\ii \\
	\ii \Gamma & -\Gamma \\
    \end{pmatrix}.
\end{align}
Consequently, the topological properties of non-Hermitian Hamiltonians $H \left( {\bm k} \right)$ are characterized by the Hermitian matrices $\ii H \left( {\bm k} \right) \Gamma$ with no symmetry.

\section{Effective surface theory of Hermitian Hopf insulator}
    \label{appendix: boundary}

We derive the effective surface theory of the Hermitian Hopf insulator in Eq.~(\ref{eq: Hermitian Hopf}), following Ref.~\cite{Alexandradinata-21}.
To this end, we focus on the low-energy behavior and expand the lattice model around $\bm{k} = 0$, leading to $z_1 \simeq k_x + \ii k_y$ and $z_2 \simeq k_z + \ii \phi$, and hence
\begin{align}
    H_{\rm Hopf} \left( \bm{k} \right) &= - 2 \left( k_x k_z + \phi k_y \right) \sigma_x - 2 \left( \phi k_x - k_y k_z \right) \sigma_y \nonumber \\
    &\qquad\qquad\quad - \left( k_x^2 + k_y^2 - k_z^2 - \phi^2 \right) \sigma_z.
\end{align}
The corresponding Hopf invariant in Eq.~(\ref{eq: Hopf invariant}) reads
\begin{equation}
    \chi = \frac{1}{2} \mathrm{sgn}\,\phi.
\end{equation}

Now, we make a domain wall around $z = 0$ by choosing the mass term as $\phi = m z$ ($m \in \mathbb{R}$).
Under this configuration, the Hopf invariant takes $\chi = -\left( \mathrm{sgn}\,m \right)/2$ for $z < 0$ and $\chi = \left( \mathrm{sgn}\,m \right)/2$ for $z > 0$, between which the corresponding boundary state should appear.
Indeed, we obtain the following Gaussian boundary states around the domain wall $z = 0$:
\begin{align}
    \ket{\psi_{\uparrow}} \propto e^{m z^2/2} \ket{\uparrow}&, \quad E_{\uparrow} \left( \bm{k} \right) = - k_x^2 - k_y^2 - m; \label{aeq: bdy1} \\
    \ket{\psi_{\downarrow}} \propto e^{-m z^2/2} \ket{\downarrow}&, \quad E_{\downarrow} \left( \bm{k} \right) = k_x^2 + k_y^2 - m, \label{aeq: bdy2}
\end{align}
with $\ket{\uparrow} \coloneqq \left( 1~0 \right)^T$ and $\ket{\downarrow} \coloneqq \left( 0~1 \right)^T$.
The normalizability conditions for the Gaussian wave functions in Eqs.~(\ref{aeq: bdy1}) and (\ref{aeq: bdy2}) necessitate $m < 0$ and $m > 0$, respectively.

\begin{figure*}[t]
    \centering
    \includegraphics[width=\linewidth]{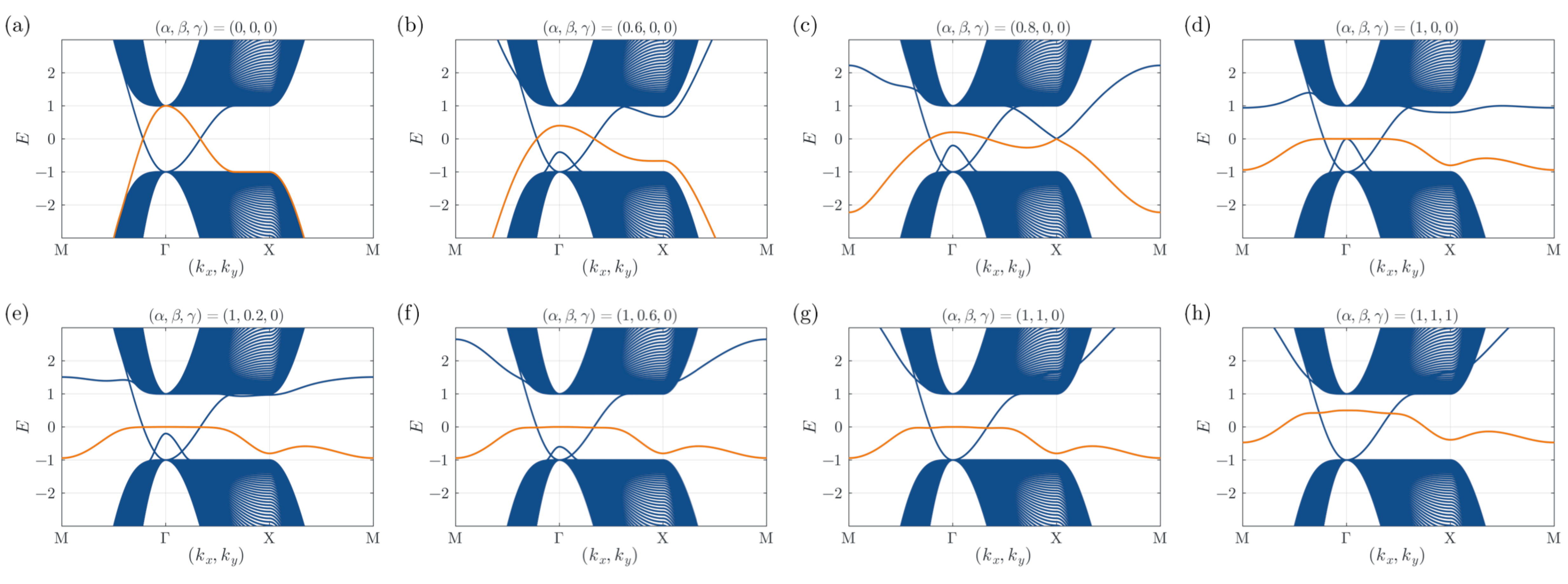} 
\caption{Detachment process of the surface state localized around $z=1$ surface (orange) from the bulk bands.
The Hermitian Hopf insulator in Eq.~(\ref{eq: Hermitian Hopf}) with $\phi=1$ is modified by replacing $H_{\rm Hopf}^{\rm slab}|_{z=1}$ with $\tilde{H}_{\rm Hopf}^{\rm slab}|_{z=1}$ in Eq.~(\ref{eq: modification at z=1}) under the open boundary conditions in the $z$ direction.
By increasing the parameters $(\alpha,\beta,\gamma)$ from $(0,0,0)$ to $(1,1,1)$, the surface state gradually becomes detached from the bulk bands. 
The spectra are plotted along the high-symmetry line $M{\rm -}\Gamma{\rm -}X{\rm -}M$ of the two-dimensional Brillouin zone $(k_x,k_y)$, where $M=(\pi,\pi), \Gamma=(0,0)$, and $X=(\pi,0)$. 
The system size in the $z$ direction is $100$, and the momentum resolution along the high-symmetry line is $3\pi/899$.}
        \label{fig:3dclassA_DetachProcess}
\end{figure*}

\section{Detachment of surface states in Hermitian Hopf insulator}
    \label{appendix: Detachment}
    
We provide the concrete perturbations that detach the surface states for $\phi = 1$ in Fig.~\ref{fig:3DclassA}\,(a) from the bulk bands, leading to Fig.~\ref{fig:3DclassA}\,(b). 
First of all, we replace the Hermitian Hopf insulator in Eq.~(\ref{eq: Hermitian Hopf2}) with the corresponding slab Hamiltonian $H_{\rm Hopf}^{\rm slab} \left( k_x,k_y \right)$ under the open boundary conditions in the $z$ direction, where $H_{\rm Hopf}^{\rm slab} \left( k_x,k_y \right)$ is defined as
\begin{align}
    H_{\rm Hopf}^{\rm slab}
    \coloneqq
    - h_x^{\rm slab}\otimes\sigma_x - h_y^{\rm slab}\otimes\sigma_y - h_z^{\rm slab}\otimes\sigma_z.
\end{align}
Here, $h_{i=x,y,z}^{\rm slab}=h_{i=x,y,z}^{\rm slab}\left( k_x,k_y \right)$ denotes the real-space representation (or equivalently the inverse Fourier transform) of $h_{i=x,y,z}$ along the $z$ direction. 
The Kronecker product for two matrices $A$ and $B$ is defined by $A\otimes B \coloneqq (A_{ij}B)$. 
Let $L_z$ be the system size in the $z$ direction, 
so that $h_{i=x,y,z}^{\rm slab}$ is an $L_z\times L_z$ matrix. 
Then, the first and last $2\times2$ diagonal submatrices in $H_{\rm Hopf}^{\rm slab}$ correspond to the onsite Hamiltonians at $z=1$ and $z=L_z$ surfaces with the internal degree of freedom.
We use $H_{\rm Hopf}^{\rm slab}|_{z=1}$ and $H_{\rm Hopf}^{\rm slab}|_{z=L_z}$ to represent these onsite Hamiltonians, respectively:
\begin{align}
    &H_{\rm Hopf}^{\rm slab}|_{z=1}
    \coloneqq
    \left[H_{\rm Hopf}^{\rm slab}\right]_{1:2,1:2}\nonumber\\
    &=
    - \left[h_x^{\rm slab}\right]_{1,1}\sigma_x - \left[h_y^{\rm slab}\right]_{1,1}\sigma_y - \left[h_z^{\rm slab}\right]_{1,1}\sigma_z,\\
    &H_{\rm Hopf}^{\rm slab}|_{z=L_z}
    \coloneqq
    \left[H_{\rm Hopf}^{\rm slab}\right]_{2L_z-1:2L_z,2L_z-1:2L_z}\nonumber\\
    &=
    - \left[h_x^{\rm slab}\right]_{L_z,L_z}\sigma_x - \left[h_y^{\rm slab}\right]_{L_z,L_z}\sigma_y - \left[h_z^{\rm slab}\right]_{L_z,L_z}\sigma_z,
\end{align}
where $\left[A\right]_{a:b,c:d}$ is a restricted matrix constructed from a matrix $A$ as $\left[A\right]_{a:b,c:d}\coloneqq (A_{ij})$ for $a\leq i\leq b$ and $c\leq j\leq d$.

To obtain Fig.~\ref{fig:3DclassA}\,(b), we replace $H_{\rm Hopf}^{\rm slab}|_{z=1,L_z}$ with $\tilde{H}_{\rm Hopf}^{\rm slab}|_{z=1,L_z}$, as follows:
\begin{align}
   &\tilde{H}_{\rm Hopf}^{\rm slab}|_{z=1} \coloneqq 
       (1-\alpha)H_{\rm Hopf}^{\rm slab} |_{z=1} \nonumber \\
       &\qquad\qquad\qquad\quad + \beta R\dfrac{\sigma_0-\sigma_z}{2} + \gamma\dfrac{\sigma_0+\sigma_z}{4},\label{eq: modification at z=1}\\
   &\tilde{H}_{\rm Hopf}^{\rm slab}|_{z=L_z} \coloneqq 
       (1-\alpha)H_{\rm Hopf}^{\rm slab} |_{z=L_z} \nonumber \\
       &\qquad\qquad\qquad\quad - \beta R\dfrac{\sigma_0+\sigma_z}{2} - \gamma\dfrac{\sigma_0-\sigma_z}{4},
\end{align}
with the $2\times2$ identity matrix $\sigma_0$. 
Notably, while $R \coloneqq 1 - \cos k_x - \cos k_y$ depends on momenta $(k_x,k_y)$, the parameters $\alpha,\beta,\gamma\in[0,1]$ are independent of $(k_x,k_y)$. 
The above perturbations are only valid for the following definition of the inverse Fourier transform under the open boundary conditions in the $z$ direction:
\begin{align}
    e^{+ \ii k_z}&\rightarrow(\delta_{i,j+1})_{1\leq i,j \leq L_z},\\
    e^{- \ii k_z}&\rightarrow(\delta_{i+1,j})_{1\leq i,j \leq L_z},
\end{align}
where $\delta_{i,j}$ is the Kronecker delta. For the other definition,
\begin{align}
    e^{+ \ii k_z}&\rightarrow(\delta_{i+1,j})_{1\leq i,j \leq L_z},\\
    e^{- \ii k_z}&\rightarrow(\delta_{i,j+1})_{1\leq i,j \leq L_z},
\end{align}
both the $\beta$ and $\gamma$ terms in $\tilde{H}_{\rm Hopf}^{\rm slab}|_{z=1}$ must be replaced by those in $\tilde{H}_{\rm Hopf}^{\rm slab}|_{z=L_z}$, and vice versa, since the different definition of the Fourier transform reverses the localization direction of surface states.

\begin{figure}[H]
    \centering
\includegraphics[width=\columnwidth]{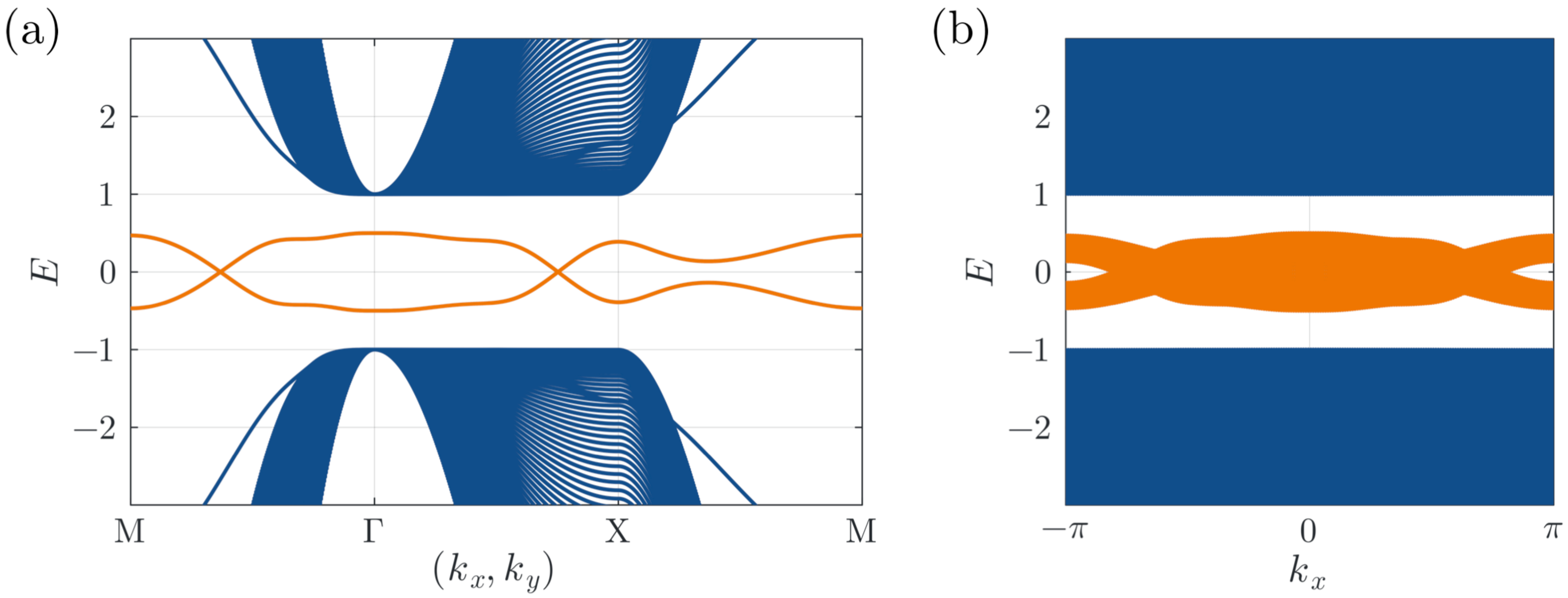}
\caption{Energy spectra of the Hermitian Hopf insulator in Eq.~(\ref{eq: Hermitian Hopf}) with $\phi=1$ modified by replacing $H_{\rm Hopf}^{\rm slab}|_{z=1,L_z}$ with $\tilde{H}_{\rm Hopf}^{\rm slab}|_{z=1,L_z}$ under the open boundary conditions in the $z$ direction. 
The bulk states and the detached surface states are shown in blue and orange, respectively. 
The parameters $(\alpha,\beta,\gamma)$ are set to $(1,1,1)$.
(a)~Energy spectrum along the high-symmetry line $M{\rm -}\Gamma{\rm -}X{\rm -}M$ of the two-dimensional Brillouin zone $(k_x,k_y)$, where $M=(\pi,\pi), \Gamma=(0,0)$, and $X=(\pi,0)$. 
The system size in the $z$ direction is $100$, and the momentum resolution along the high-symmetry line is $3\pi/899$. 
This figure is the same as Fig.~\ref{fig:3DclassA}\,(b).
(b)~Energy spectrum across the entire two-dimensional Brillouin zone $(k_x,k_y)$, where the $k_y$ direction is projected onto the $k_x$-$E$ plane.
The detached surface states in (a) are truly detached from the bulk bands over the whole two-dimensional Brillouin zone. 
The system size in the $z$ direction is $100$, and the momentum resolution is $(2\pi/199,2\pi/99)$.}
    \label{fig:3DclassA_DetachSpectralGap}
\end{figure}

In Fig.~\ref{fig:3dclassA_DetachProcess}, we demonstrate how the surface state localized around the $z=1$ surface (orange) in Fig.~\ref{fig:3dclassA_DetachProcess}\,(a) is detached from the bulk bands and evolves into that in Fig.~\ref{fig:3DclassA}\,(b), as the parameters $(\alpha,\beta,\gamma)$ are increased from $(0,0,0)$ to $(1,1,1)$. 
There, we only consider the modification of $H_{\rm Hopf}^{\rm slab}|_{z=1}$ just for simplicity. 
As can be seen in Figs.~\ref{fig:3dclassA_DetachProcess}\,(a-d), the term related to the parameter $\alpha$ detaches the surface state from the bulk bands and lifts 
it
into the bulk gap, as also discussed in Ref.~\cite{Alexandradinata-21}. 
It also makes some bulk bands moving towards the bulk gap.
The term containing the parameter $\beta$ pushes them back into the bulk [Figs.~\ref{fig:3dclassA_DetachProcess}\,(e-g)]. 
Finally, as shown in Fig.~\ref{fig:3dclassA_DetachProcess}\,(h), the $\gamma$ term adjusts the height of the surface spectrum, leading to the surface state localized around the $z=1$ surface in Fig.~\ref{fig:3DclassA}\,(b). 
The complete figure is obtained by simultaneously modifying $H_{\rm Hopf}^{\rm slab}|_{z=1}$ and $H_{\rm Hopf}^{\rm slab}|_{z=L_z}$, and setting $(\alpha,\beta,\gamma)$ to $(1,1,1)$, as shown in Fig.~\ref{fig:3DclassA_DetachSpectralGap}\,(a). 
We also confirm the surface states in Fig.~\ref{fig:3DclassA_DetachSpectralGap}\,(a) are fully detached from the bulk bands across the entire two-dimensional Brillouin zone $(k_x,k_y)$ [Fig.~\ref{fig:3DclassA_DetachSpectralGap}\,(b)]. 

\let\oldaddcontentsline\addcontentsline
\renewcommand{\addcontentsline}[3]{}
\bibliography{Hopf.bib}

\begin{thebibliography}{144}%
\makeatletter
\providecommand \@ifxundefined [1]{%
 \@ifx{#1\undefined}
}%
\providecommand \@ifnum [1]{%
 \ifnum #1\expandafter \@firstoftwo
 \else \expandafter \@secondoftwo
 \fi
}%
\providecommand \@ifx [1]{%
 \ifx #1\expandafter \@firstoftwo
 \else \expandafter \@secondoftwo
 \fi
}%
\providecommand \natexlab [1]{#1}%
\providecommand \enquote  [1]{``#1''}%
\providecommand \bibnamefont  [1]{#1}%
\providecommand \bibfnamefont [1]{#1}%
\providecommand \citenamefont [1]{#1}%
\providecommand \href@noop [0]{\@secondoftwo}%
\providecommand \href [0]{\begingroup \@sanitize@url \@href}%
\providecommand \@href[1]{\@@startlink{#1}\@@href}%
\providecommand \@@href[1]{\endgroup#1\@@endlink}%
\providecommand \@sanitize@url [0]{\catcode `\\12\catcode `\$12\catcode `\&12\catcode `\#12\catcode `\^12\catcode `\_12\catcode `\%12\relax}%
\providecommand \@@startlink[1]{}%
\providecommand \@@endlink[0]{}%
\providecommand \url  [0]{\begingroup\@sanitize@url \@url }%
\providecommand \@url [1]{\endgroup\@href {#1}{\urlprefix }}%
\providecommand \urlprefix  [0]{URL }%
\providecommand \Eprint [0]{\href }%
\providecommand \doibase [0]{https://doi.org/}%
\providecommand \selectlanguage [0]{\@gobble}%
\providecommand \bibinfo  [0]{\@secondoftwo}%
\providecommand \bibfield  [0]{\@secondoftwo}%
\providecommand \translation [1]{[#1]}%
\providecommand \BibitemOpen [0]{}%
\providecommand \bibitemStop [0]{}%
\providecommand \bibitemNoStop [0]{.\EOS\space}%
\providecommand \EOS [0]{\spacefactor3000\relax}%
\providecommand \BibitemShut  [1]{\csname bibitem#1\endcsname}%
\let\auto@bib@innerbib\@empty
\bibitem [{\citenamefont {Hasan}\ and\ \citenamefont {Kane}(2010)}]{HK-review}%
  \BibitemOpen
  \bibfield  {author} {\bibinfo {author} {\bibfnamefont {M.~Z.}\ \bibnamefont {Hasan}}\ and\ \bibinfo {author} {\bibfnamefont {C.~L.}\ \bibnamefont {Kane}},\ }\bibfield  {title} {\bibinfo {title} {{Colloquium: Topological insulators}},\ }\href {https://doi.org/10.1103/RevModPhys.82.3045} {\bibfield  {journal} {\bibinfo  {journal} {Rev. Mod. Phys.}\ }\textbf {\bibinfo {volume} {82}},\ \bibinfo {pages} {3045} (\bibinfo {year} {2010})}\BibitemShut {NoStop}%
\bibitem [{\citenamefont {Qi}\ and\ \citenamefont {Zhang}(2011)}]{QZ-review}%
  \BibitemOpen
  \bibfield  {author} {\bibinfo {author} {\bibfnamefont {X.-L.}\ \bibnamefont {Qi}}\ and\ \bibinfo {author} {\bibfnamefont {S.-C.}\ \bibnamefont {Zhang}},\ }\bibfield  {title} {\bibinfo {title} {{Topological insulators and superconductors}},\ }\href {https://doi.org/10.1103/RevModPhys.83.1057} {\bibfield  {journal} {\bibinfo  {journal} {{Rev. Mod. Phys.}}\ }\textbf {\bibinfo {volume} {83}},\ \bibinfo {pages} {1057} (\bibinfo {year} {2011})}\BibitemShut {NoStop}%
\bibitem [{\citenamefont {Altland}\ and\ \citenamefont {Zirnbauer}(1997)}]{AZ-97}%
  \BibitemOpen
  \bibfield  {author} {\bibinfo {author} {\bibfnamefont {A.}~\bibnamefont {Altland}}\ and\ \bibinfo {author} {\bibfnamefont {M.~R.}\ \bibnamefont {Zirnbauer}},\ }\bibfield  {title} {\bibinfo {title} {{Nonstandard symmetry classes in mesoscopic normal-superconducting hybrid structures}},\ }\href {https://doi.org/10.1103/PhysRevB.55.1142} {\bibfield  {journal} {\bibinfo  {journal} {Phys. Rev. B}\ }\textbf {\bibinfo {volume} {55}},\ \bibinfo {pages} {1142} (\bibinfo {year} {1997})}\BibitemShut {NoStop}%
\bibitem [{\citenamefont {Schnyder}\ \emph {et~al.}(2008)\citenamefont {Schnyder}, \citenamefont {Ryu}, \citenamefont {Furusaki},\ and\ \citenamefont {Ludwig}}]{Schnyder-08}%
  \BibitemOpen
  \bibfield  {author} {\bibinfo {author} {\bibfnamefont {A.~P.}\ \bibnamefont {Schnyder}}, \bibinfo {author} {\bibfnamefont {S.}~\bibnamefont {Ryu}}, \bibinfo {author} {\bibfnamefont {A.}~\bibnamefont {Furusaki}},\ and\ \bibinfo {author} {\bibfnamefont {A.~W.~W.}\ \bibnamefont {Ludwig}},\ }\bibfield  {title} {\bibinfo {title} {{Classification of topological insulators and superconductors in three spatial dimensions}},\ }\href {https://doi.org/10.1103/PhysRevB.78.195125} {\bibfield  {journal} {\bibinfo  {journal} {Phys. Rev. B}\ }\textbf {\bibinfo {volume} {78}},\ \bibinfo {pages} {195125} (\bibinfo {year} {2008})}\BibitemShut {NoStop}%
\bibitem [{\citenamefont {Ryu}\ \emph {et~al.}(2010)\citenamefont {Ryu}, \citenamefont {Schnyder}, \citenamefont {Furusaki},\ and\ \citenamefont {Ludwig}}]{Ryu-10}%
  \BibitemOpen
  \bibfield  {author} {\bibinfo {author} {\bibfnamefont {S.}~\bibnamefont {Ryu}}, \bibinfo {author} {\bibfnamefont {A.~P.}\ \bibnamefont {Schnyder}}, \bibinfo {author} {\bibfnamefont {A.}~\bibnamefont {Furusaki}},\ and\ \bibinfo {author} {\bibfnamefont {A.~W.~W.}\ \bibnamefont {Ludwig}},\ }\bibfield  {title} {\bibinfo {title} {{Topological insulators and superconductors: tenfold way and dimensional hierarchy}},\ }\href {https://doi.org/10.1088/1367-2630/12/6/065010} {\bibfield  {journal} {\bibinfo  {journal} {New J. Phys.}\ }\textbf {\bibinfo {volume} {12}},\ \bibinfo {pages} {065010} (\bibinfo {year} {2010})}\BibitemShut {NoStop}%
\bibitem [{\citenamefont {Kitaev}(2009)}]{Kitaev-09}%
  \BibitemOpen
  \bibfield  {author} {\bibinfo {author} {\bibfnamefont {A.}~\bibnamefont {Kitaev}},\ }\bibfield  {title} {\bibinfo {title} {{Periodic table for topological insulators and superconductors}},\ }\href {https://doi.org/10.1063/1.3149495} {\bibfield  {journal} {\bibinfo  {journal} {AIP Conf. Proc.}\ }\textbf {\bibinfo {volume} {1134}},\ \bibinfo {pages} {22} (\bibinfo {year} {2009})}\BibitemShut {NoStop}%
\bibitem [{\citenamefont {Chiu}\ \emph {et~al.}(2016)\citenamefont {Chiu}, \citenamefont {Teo}, \citenamefont {Schnyder},\ and\ \citenamefont {Ryu}}]{CTSR-review}%
  \BibitemOpen
  \bibfield  {author} {\bibinfo {author} {\bibfnamefont {C.-K.}\ \bibnamefont {Chiu}}, \bibinfo {author} {\bibfnamefont {J.~C.~Y.}\ \bibnamefont {Teo}}, \bibinfo {author} {\bibfnamefont {A.~P.}\ \bibnamefont {Schnyder}},\ and\ \bibinfo {author} {\bibfnamefont {S.}~\bibnamefont {Ryu}},\ }\bibfield  {title} {\bibinfo {title} {{Classification of topological quantum matter with symmetries}},\ }\href {https://doi.org/10.1103/RevModPhys.88.035005} {\bibfield  {journal} {\bibinfo  {journal} {Rev. Mod. Phys.}\ }\textbf {\bibinfo {volume} {88}},\ \bibinfo {pages} {035005} (\bibinfo {year} {2016})}\BibitemShut {NoStop}%
\bibitem [{\citenamefont {Karoubi}(1978)}]{Karoubi}%
  \BibitemOpen
  \bibfield  {author} {\bibinfo {author} {\bibfnamefont {M.}~\bibnamefont {Karoubi}},\ }\href {https://doi.org/10.1007/978-3-540-79890-3} {\emph {\bibinfo {title} {{K-Theory: An Introduction}}}}\ (\bibinfo  {publisher} {Springer},\ \bibinfo {address} {Berlin, Heidelberg},\ \bibinfo {year} {1978})\BibitemShut {NoStop}%
\bibitem [{\citenamefont {Moore}\ \emph {et~al.}(2008)\citenamefont {Moore}, \citenamefont {Ran},\ and\ \citenamefont {Wen}}]{Moore-08}%
  \BibitemOpen
  \bibfield  {author} {\bibinfo {author} {\bibfnamefont {J.~E.}\ \bibnamefont {Moore}}, \bibinfo {author} {\bibfnamefont {Y.}~\bibnamefont {Ran}},\ and\ \bibinfo {author} {\bibfnamefont {X.-G.}\ \bibnamefont {Wen}},\ }\bibfield  {title} {\bibinfo {title} {{Topological Surface States in Three-Dimensional Magnetic Insulators}},\ }\href {https://doi.org/10.1103/PhysRevLett.101.186805} {\bibfield  {journal} {\bibinfo  {journal} {Phys. Rev. Lett.}\ }\textbf {\bibinfo {volume} {101}},\ \bibinfo {pages} {186805} (\bibinfo {year} {2008})}\BibitemShut {NoStop}%
\bibitem [{\citenamefont {Nakahara}(2018)}]{Nakahara-textbook}%
  \BibitemOpen
  \bibfield  {author} {\bibinfo {author} {\bibfnamefont {M.}~\bibnamefont {Nakahara}},\ }\href {https://doi.org/https://doi.org/10.1201/9781315275826} {\emph {\bibinfo {title} {{Geometry, Topology and Physics}}}}\ (\bibinfo  {publisher} {CRC Press, Boca Raton},\ \bibinfo {year} {2018})\BibitemShut {NoStop}%
\bibitem [{\citenamefont {Deng}\ \emph {et~al.}(2013)\citenamefont {Deng}, \citenamefont {Wang}, \citenamefont {Shen},\ and\ \citenamefont {Duan}}]{Deng-13}%
  \BibitemOpen
  \bibfield  {author} {\bibinfo {author} {\bibfnamefont {D.-L.}\ \bibnamefont {Deng}}, \bibinfo {author} {\bibfnamefont {S.-T.}\ \bibnamefont {Wang}}, \bibinfo {author} {\bibfnamefont {C.}~\bibnamefont {Shen}},\ and\ \bibinfo {author} {\bibfnamefont {L.-M.}\ \bibnamefont {Duan}},\ }\bibfield  {title} {\bibinfo {title} {{Hopf insulators and their topologically protected surface states}},\ }\href {https://doi.org/10.1103/PhysRevB.88.201105} {\bibfield  {journal} {\bibinfo  {journal} {Phys. Rev. B}\ }\textbf {\bibinfo {volume} {88}},\ \bibinfo {pages} {201105} (\bibinfo {year} {2013})}\BibitemShut {NoStop}%
\bibitem [{\citenamefont {Kennedy}(2016)}]{Kennedy-16}%
  \BibitemOpen
  \bibfield  {author} {\bibinfo {author} {\bibfnamefont {R.}~\bibnamefont {Kennedy}},\ }\bibfield  {title} {\bibinfo {title} {{Topological Hopf-Chern insulators and the Hopf superconductor}},\ }\href {https://doi.org/10.1103/PhysRevB.94.035137} {\bibfield  {journal} {\bibinfo  {journal} {Phys. Rev. B}\ }\textbf {\bibinfo {volume} {94}},\ \bibinfo {pages} {035137} (\bibinfo {year} {2016})}\BibitemShut {NoStop}%
\bibitem [{\citenamefont {Liu}\ \emph {et~al.}(2017)\citenamefont {Liu}, \citenamefont {Vafa},\ and\ \citenamefont {Xu}}]{Liu-17}%
  \BibitemOpen
  \bibfield  {author} {\bibinfo {author} {\bibfnamefont {C.}~\bibnamefont {Liu}}, \bibinfo {author} {\bibfnamefont {F.}~\bibnamefont {Vafa}},\ and\ \bibinfo {author} {\bibfnamefont {C.}~\bibnamefont {Xu}},\ }\bibfield  {title} {\bibinfo {title} {{Symmetry-protected topological Hopf insulator and its generalizations}},\ }\href {https://doi.org/10.1103/PhysRevB.95.161116} {\bibfield  {journal} {\bibinfo  {journal} {Phys. Rev. B}\ }\textbf {\bibinfo {volume} {95}},\ \bibinfo {pages} {161116} (\bibinfo {year} {2017})}\BibitemShut {NoStop}%
\bibitem [{\citenamefont {\"Unal}\ \emph {et~al.}(2019)\citenamefont {\"Unal}, \citenamefont {Eckardt},\ and\ \citenamefont {Slager}}]{Unal-19}%
  \BibitemOpen
  \bibfield  {author} {\bibinfo {author} {\bibfnamefont {F.~N.}\ \bibnamefont {\"Unal}}, \bibinfo {author} {\bibfnamefont {A.}~\bibnamefont {Eckardt}},\ and\ \bibinfo {author} {\bibfnamefont {R.-J.}\ \bibnamefont {Slager}},\ }\bibfield  {title} {\bibinfo {title} {{Hopf characterization of two-dimensional Floquet topological insulators}},\ }\href {https://doi.org/10.1103/PhysRevResearch.1.022003} {\bibfield  {journal} {\bibinfo  {journal} {Phys. Rev. Research}\ }\textbf {\bibinfo {volume} {1}},\ \bibinfo {pages} {022003} (\bibinfo {year} {2019})}\BibitemShut {NoStop}%
\bibitem [{\citenamefont {\"Unal}\ \emph {et~al.}(2020)\citenamefont {\"Unal}, \citenamefont {Bouhon},\ and\ \citenamefont {Slager}}]{Unal-20}%
  \BibitemOpen
  \bibfield  {author} {\bibinfo {author} {\bibfnamefont {F.~N.}\ \bibnamefont {\"Unal}}, \bibinfo {author} {\bibfnamefont {A.}~\bibnamefont {Bouhon}},\ and\ \bibinfo {author} {\bibfnamefont {R.-J.}\ \bibnamefont {Slager}},\ }\bibfield  {title} {\bibinfo {title} {{Topological Euler Class as a Dynamical Observable in Optical Lattices}},\ }\href {https://doi.org/10.1103/PhysRevLett.125.053601} {\bibfield  {journal} {\bibinfo  {journal} {Phys. Rev. Lett.}\ }\textbf {\bibinfo {volume} {125}},\ \bibinfo {pages} {053601} (\bibinfo {year} {2020})}\BibitemShut {NoStop}%
\bibitem [{\citenamefont {Schuster}\ \emph {et~al.}(2021)\citenamefont {Schuster}, \citenamefont {Flicker}, \citenamefont {Li}, \citenamefont {Kotochigova}, \citenamefont {Moore}, \citenamefont {Ye},\ and\ \citenamefont {Yao}}]{Schuster-21}%
  \BibitemOpen
  \bibfield  {author} {\bibinfo {author} {\bibfnamefont {T.}~\bibnamefont {Schuster}}, \bibinfo {author} {\bibfnamefont {F.}~\bibnamefont {Flicker}}, \bibinfo {author} {\bibfnamefont {M.}~\bibnamefont {Li}}, \bibinfo {author} {\bibfnamefont {S.}~\bibnamefont {Kotochigova}}, \bibinfo {author} {\bibfnamefont {J.~E.}\ \bibnamefont {Moore}}, \bibinfo {author} {\bibfnamefont {J.}~\bibnamefont {Ye}},\ and\ \bibinfo {author} {\bibfnamefont {N.~Y.}\ \bibnamefont {Yao}},\ }\bibfield  {title} {\bibinfo {title} {{Realizing Hopf Insulators in Dipolar Spin Systems}},\ }\href {https://doi.org/10.1103/PhysRevLett.127.015301} {\bibfield  {journal} {\bibinfo  {journal} {Phys. Rev. Lett.}\ }\textbf {\bibinfo {volume} {127}},\ \bibinfo {pages} {015301} (\bibinfo {year} {2021})}\BibitemShut {NoStop}%
\bibitem [{\citenamefont {Schuster}\ \emph {et~al.}(2019)\citenamefont {Schuster}, \citenamefont {Gazit}, \citenamefont {Moore},\ and\ \citenamefont {Yao}}]{Schuster-19}%
  \BibitemOpen
  \bibfield  {author} {\bibinfo {author} {\bibfnamefont {T.}~\bibnamefont {Schuster}}, \bibinfo {author} {\bibfnamefont {S.}~\bibnamefont {Gazit}}, \bibinfo {author} {\bibfnamefont {J.~E.}\ \bibnamefont {Moore}},\ and\ \bibinfo {author} {\bibfnamefont {N.~Y.}\ \bibnamefont {Yao}},\ }\bibfield  {title} {\bibinfo {title} {{Floquet Hopf Insulators}},\ }\href {https://doi.org/10.1103/PhysRevLett.123.266803} {\bibfield  {journal} {\bibinfo  {journal} {Phys. Rev. Lett.}\ }\textbf {\bibinfo {volume} {123}},\ \bibinfo {pages} {266803} (\bibinfo {year} {2019})}\BibitemShut {NoStop}%
\bibitem [{\citenamefont {Alexandradinata}\ \emph {et~al.}(2021)\citenamefont {Alexandradinata}, \citenamefont {Nelson},\ and\ \citenamefont {Soluyanov}}]{Alexandradinata-21}%
  \BibitemOpen
  \bibfield  {author} {\bibinfo {author} {\bibfnamefont {A.}~\bibnamefont {Alexandradinata}}, \bibinfo {author} {\bibfnamefont {A.}~\bibnamefont {Nelson}},\ and\ \bibinfo {author} {\bibfnamefont {A.~A.}\ \bibnamefont {Soluyanov}},\ }\bibfield  {title} {\bibinfo {title} {{Teleportation of Berry curvature on the surface of a Hopf insulator}},\ }\href {https://doi.org/10.1103/PhysRevB.103.045107} {\bibfield  {journal} {\bibinfo  {journal} {Phys. Rev. B}\ }\textbf {\bibinfo {volume} {103}},\ \bibinfo {pages} {045107} (\bibinfo {year} {2021})}\BibitemShut {NoStop}%
\bibitem [{\citenamefont {Zhu}\ \emph {et~al.}(2021)\citenamefont {Zhu}, \citenamefont {Hughes},\ and\ \citenamefont {Alexandradinata}}]{Zhu-21}%
  \BibitemOpen
  \bibfield  {author} {\bibinfo {author} {\bibfnamefont {P.}~\bibnamefont {Zhu}}, \bibinfo {author} {\bibfnamefont {T.~L.}\ \bibnamefont {Hughes}},\ and\ \bibinfo {author} {\bibfnamefont {A.}~\bibnamefont {Alexandradinata}},\ }\bibfield  {title} {\bibinfo {title} {{Quantized surface magnetism and higher-order topology: Application to the Hopf insulator}},\ }\href {https://doi.org/10.1103/PhysRevB.103.014417} {\bibfield  {journal} {\bibinfo  {journal} {Phys. Rev. B}\ }\textbf {\bibinfo {volume} {103}},\ \bibinfo {pages} {014417} (\bibinfo {year} {2021})}\BibitemShut {NoStop}%
\bibitem [{\citenamefont {Lapierre}\ \emph {et~al.}(2021)\citenamefont {Lapierre}, \citenamefont {Neupert},\ and\ \citenamefont {Trifunovic}}]{Lapierre-21}%
  \BibitemOpen
  \bibfield  {author} {\bibinfo {author} {\bibfnamefont {B.}~\bibnamefont {Lapierre}}, \bibinfo {author} {\bibfnamefont {T.}~\bibnamefont {Neupert}},\ and\ \bibinfo {author} {\bibfnamefont {L.}~\bibnamefont {Trifunovic}},\ }\bibfield  {title} {\bibinfo {title} {{$N$-band Hopf insulator}},\ }\href {https://doi.org/10.1103/PhysRevResearch.3.033045} {\bibfield  {journal} {\bibinfo  {journal} {Phys. Rev. Research}\ }\textbf {\bibinfo {volume} {3}},\ \bibinfo {pages} {033045} (\bibinfo {year} {2021})}\BibitemShut {NoStop}%
\bibitem [{\citenamefont {Zhu}\ \emph {et~al.}(2023)\citenamefont {Zhu}, \citenamefont {Alexandradinata},\ and\ \citenamefont {Hughes}}]{Zhu-23}%
  \BibitemOpen
  \bibfield  {author} {\bibinfo {author} {\bibfnamefont {P.}~\bibnamefont {Zhu}}, \bibinfo {author} {\bibfnamefont {A.}~\bibnamefont {Alexandradinata}},\ and\ \bibinfo {author} {\bibfnamefont {T.~L.}\ \bibnamefont {Hughes}},\ }\bibfield  {title} {\bibinfo {title} {{${\mathbb{Z}}_{2}$ spin Hopf insulator: Helical hinge states and returning Thouless pump}},\ }\href {https://doi.org/10.1103/PhysRevB.107.115159} {\bibfield  {journal} {\bibinfo  {journal} {Phys. Rev. B}\ }\textbf {\bibinfo {volume} {107}},\ \bibinfo {pages} {115159} (\bibinfo {year} {2023})}\BibitemShut {NoStop}%
\bibitem [{\citenamefont {Wang}\ \emph {et~al.}(2023{\natexlab{a}})\citenamefont {Wang}, \citenamefont {Zeng}, \citenamefont {Biao}, \citenamefont {Yan},\ and\ \citenamefont {Yu}}]{Wang-23}%
  \BibitemOpen
  \bibfield  {author} {\bibinfo {author} {\bibfnamefont {Z.}~\bibnamefont {Wang}}, \bibinfo {author} {\bibfnamefont {X.-T.}\ \bibnamefont {Zeng}}, \bibinfo {author} {\bibfnamefont {Y.}~\bibnamefont {Biao}}, \bibinfo {author} {\bibfnamefont {Z.}~\bibnamefont {Yan}},\ and\ \bibinfo {author} {\bibfnamefont {R.}~\bibnamefont {Yu}},\ }\bibfield  {title} {\bibinfo {title} {{Realization of a Hopf Insulator in Circuit Systems}},\ }\href {https://doi.org/10.1103/PhysRevLett.130.057201} {\bibfield  {journal} {\bibinfo  {journal} {Phys. Rev. Lett.}\ }\textbf {\bibinfo {volume} {130}},\ \bibinfo {pages} {057201} (\bibinfo {year} {2023}{\natexlab{a}})}\BibitemShut {NoStop}%
\bibitem [{\citenamefont {Lim}\ \emph {et~al.}(2023)\citenamefont {Lim}, \citenamefont {Kim},\ and\ \citenamefont {Yang}}]{Lim-23}%
  \BibitemOpen
  \bibfield  {author} {\bibinfo {author} {\bibfnamefont {H.}~\bibnamefont {Lim}}, \bibinfo {author} {\bibfnamefont {S.}~\bibnamefont {Kim}},\ and\ \bibinfo {author} {\bibfnamefont {B.-J.}\ \bibnamefont {Yang}},\ }\bibfield  {title} {\bibinfo {title} {{Real Hopf insulator}},\ }\href {https://doi.org/10.1103/PhysRevB.108.125101} {\bibfield  {journal} {\bibinfo  {journal} {Phys. Rev. B}\ }\textbf {\bibinfo {volume} {108}},\ \bibinfo {pages} {125101} (\bibinfo {year} {2023})}\BibitemShut {NoStop}%
\bibitem [{\citenamefont {Jankowski}\ \emph {et~al.}(2024)\citenamefont {Jankowski}, \citenamefont {Morris}, \citenamefont {Davoyan}, \citenamefont {Bouhon}, \citenamefont {\"Unal},\ and\ \citenamefont {Slager}}]{Jankowski-24}%
  \BibitemOpen
  \bibfield  {author} {\bibinfo {author} {\bibfnamefont {W.~J.}\ \bibnamefont {Jankowski}}, \bibinfo {author} {\bibfnamefont {A.~S.}\ \bibnamefont {Morris}}, \bibinfo {author} {\bibfnamefont {Z.}~\bibnamefont {Davoyan}}, \bibinfo {author} {\bibfnamefont {A.}~\bibnamefont {Bouhon}}, \bibinfo {author} {\bibfnamefont {F.~N.}\ \bibnamefont {\"Unal}},\ and\ \bibinfo {author} {\bibfnamefont {R.-J.}\ \bibnamefont {Slager}},\ }\bibfield  {title} {\bibinfo {title} {{Non-Abelian Hopf-Euler insulators}},\ }\href {https://doi.org/10.1103/PhysRevB.110.075135} {\bibfield  {journal} {\bibinfo  {journal} {Phys. Rev. B}\ }\textbf {\bibinfo {volume} {110}},\ \bibinfo {pages} {075135} (\bibinfo {year} {2024})}\BibitemShut {NoStop}%
\bibitem [{\citenamefont {Nelson}\ \emph {et~al.}(2021)\citenamefont {Nelson}, \citenamefont {Neupert}, \citenamefont {Bzdu\ifmmode~\check{s}\else \v{s}\fi{}ek},\ and\ \citenamefont {Alexandradinata}}]{Nelson-21}%
  \BibitemOpen
  \bibfield  {author} {\bibinfo {author} {\bibfnamefont {A.}~\bibnamefont {Nelson}}, \bibinfo {author} {\bibfnamefont {T.}~\bibnamefont {Neupert}}, \bibinfo {author} {\bibfnamefont {T.}~\bibnamefont {Bzdu\ifmmode~\check{s}\else \v{s}\fi{}ek}},\ and\ \bibinfo {author} {\bibfnamefont {A.}~\bibnamefont {Alexandradinata}},\ }\bibfield  {title} {\bibinfo {title} {{Multicellularity of Delicate Topological Insulators}},\ }\href {https://doi.org/10.1103/PhysRevLett.126.216404} {\bibfield  {journal} {\bibinfo  {journal} {Phys. Rev. Lett.}\ }\textbf {\bibinfo {volume} {126}},\ \bibinfo {pages} {216404} (\bibinfo {year} {2021})}\BibitemShut {NoStop}%
\bibitem [{\citenamefont {Nelson}\ \emph {et~al.}(2022)\citenamefont {Nelson}, \citenamefont {Neupert}, \citenamefont {Alexandradinata},\ and\ \citenamefont {Bzdu\ifmmode~\check{s}\else \v{s}\fi{}ek}}]{Nelson-22}%
  \BibitemOpen
  \bibfield  {author} {\bibinfo {author} {\bibfnamefont {A.}~\bibnamefont {Nelson}}, \bibinfo {author} {\bibfnamefont {T.}~\bibnamefont {Neupert}}, \bibinfo {author} {\bibfnamefont {A.}~\bibnamefont {Alexandradinata}},\ and\ \bibinfo {author} {\bibfnamefont {T.}~\bibnamefont {Bzdu\ifmmode~\check{s}\else \v{s}\fi{}ek}},\ }\bibfield  {title} {\bibinfo {title} {{Delicate topology protected by rotation symmetry: Crystalline Hopf insulators and beyond}},\ }\href {https://doi.org/10.1103/PhysRevB.106.075124} {\bibfield  {journal} {\bibinfo  {journal} {Phys. Rev. B}\ }\textbf {\bibinfo {volume} {106}},\ \bibinfo {pages} {075124} (\bibinfo {year} {2022})}\BibitemShut {NoStop}%
\bibitem [{\citenamefont {Brouwer}\ and\ \citenamefont {Dwivedi}(2023)}]{Brouwer-23}%
  \BibitemOpen
  \bibfield  {author} {\bibinfo {author} {\bibfnamefont {P.~W.}\ \bibnamefont {Brouwer}}\ and\ \bibinfo {author} {\bibfnamefont {V.}~\bibnamefont {Dwivedi}},\ }\bibfield  {title} {\bibinfo {title} {{Homotopic classification of band structures: Stable, fragile, delicate, and stable representation-protected topology}},\ }\href {https://doi.org/10.1103/PhysRevB.108.155137} {\bibfield  {journal} {\bibinfo  {journal} {Phys. Rev. B}\ }\textbf {\bibinfo {volume} {108}},\ \bibinfo {pages} {155137} (\bibinfo {year} {2023})}\BibitemShut {NoStop}%
\bibitem [{\citenamefont {Bergholtz}\ \emph {et~al.}(2021)\citenamefont {Bergholtz}, \citenamefont {Budich},\ and\ \citenamefont {Kunst}}]{BBK-review}%
  \BibitemOpen
  \bibfield  {author} {\bibinfo {author} {\bibfnamefont {E.~J.}\ \bibnamefont {Bergholtz}}, \bibinfo {author} {\bibfnamefont {J.~C.}\ \bibnamefont {Budich}},\ and\ \bibinfo {author} {\bibfnamefont {F.~K.}\ \bibnamefont {Kunst}},\ }\bibfield  {title} {\bibinfo {title} {{Exceptional topology of non-Hermitian systems}},\ }\href {https://doi.org/10.1103/RevModPhys.93.015005} {\bibfield  {journal} {\bibinfo  {journal} {Rev. Mod. Phys.}\ }\textbf {\bibinfo {volume} {93}},\ \bibinfo {pages} {015005} (\bibinfo {year} {2021})}\BibitemShut {NoStop}%
\bibitem [{\citenamefont {Okuma}\ and\ \citenamefont {Sato}(2023)}]{Okuma-Sato-review}%
  \BibitemOpen
  \bibfield  {author} {\bibinfo {author} {\bibfnamefont {N.}~\bibnamefont {Okuma}}\ and\ \bibinfo {author} {\bibfnamefont {M.}~\bibnamefont {Sato}},\ }\bibfield  {title} {\bibinfo {title} {{Non-Hermitian Topological Phenomena: A Review}},\ }\href {https://doi.org/10.1146/annurev-conmatphys-040521-033133} {\bibfield  {journal} {\bibinfo  {journal} {Annu. Rev. Condens. Matter Phys.}\ }\textbf {\bibinfo {volume} {14}},\ \bibinfo {pages} {83} (\bibinfo {year} {2023})}\BibitemShut {NoStop}%
\bibitem [{\citenamefont {Rudner}\ and\ \citenamefont {Levitov}(2009)}]{Rudner-09}%
  \BibitemOpen
  \bibfield  {author} {\bibinfo {author} {\bibfnamefont {M.~S.}\ \bibnamefont {Rudner}}\ and\ \bibinfo {author} {\bibfnamefont {L.~S.}\ \bibnamefont {Levitov}},\ }\bibfield  {title} {\bibinfo {title} {{Topological Transition in a Non-Hermitian Quantum Walk}},\ }\href {https://doi.org/10.1103/PhysRevLett.102.065703} {\bibfield  {journal} {\bibinfo  {journal} {Phys. Rev. Lett.}\ }\textbf {\bibinfo {volume} {102}},\ \bibinfo {pages} {065703} (\bibinfo {year} {2009})}\BibitemShut {NoStop}%
\bibitem [{\citenamefont {Sato}\ \emph {et~al.}(2012)\citenamefont {Sato}, \citenamefont {Hasebe}, \citenamefont {Esaki},\ and\ \citenamefont {Kohmoto}}]{Sato-11}%
  \BibitemOpen
  \bibfield  {author} {\bibinfo {author} {\bibfnamefont {M.}~\bibnamefont {Sato}}, \bibinfo {author} {\bibfnamefont {K.}~\bibnamefont {Hasebe}}, \bibinfo {author} {\bibfnamefont {K.}~\bibnamefont {Esaki}},\ and\ \bibinfo {author} {\bibfnamefont {M.}~\bibnamefont {Kohmoto}},\ }\bibfield  {title} {\bibinfo {title} {{Time-Reversal Symmetry in Non-Hermitian Systems}},\ }\href {https://doi.org/10.1143/PTP.127.937} {\bibfield  {journal} {\bibinfo  {journal} {Prog. Theor. Phys.}\ }\textbf {\bibinfo {volume} {127}},\ \bibinfo {pages} {937} (\bibinfo {year} {2012})}\BibitemShut {NoStop}%
\bibitem [{\citenamefont {Esaki}\ \emph {et~al.}(2011)\citenamefont {Esaki}, \citenamefont {Sato}, \citenamefont {Hasebe},\ and\ \citenamefont {Kohmoto}}]{Esaki-11}%
  \BibitemOpen
  \bibfield  {author} {\bibinfo {author} {\bibfnamefont {K.}~\bibnamefont {Esaki}}, \bibinfo {author} {\bibfnamefont {M.}~\bibnamefont {Sato}}, \bibinfo {author} {\bibfnamefont {K.}~\bibnamefont {Hasebe}},\ and\ \bibinfo {author} {\bibfnamefont {M.}~\bibnamefont {Kohmoto}},\ }\bibfield  {title} {\bibinfo {title} {{Edge states and topological phases in non-Hermitian systems}},\ }\href {https://doi.org/10.1103/PhysRevB.84.205128} {\bibfield  {journal} {\bibinfo  {journal} {Phys. Rev. B}\ }\textbf {\bibinfo {volume} {84}},\ \bibinfo {pages} {205128} (\bibinfo {year} {2011})}\BibitemShut {NoStop}%
\bibitem [{\citenamefont {Hu}\ and\ \citenamefont {Hughes}(2011)}]{Hu-11}%
  \BibitemOpen
  \bibfield  {author} {\bibinfo {author} {\bibfnamefont {Y.~C.}\ \bibnamefont {Hu}}\ and\ \bibinfo {author} {\bibfnamefont {T.~L.}\ \bibnamefont {Hughes}},\ }\bibfield  {title} {\bibinfo {title} {{Absence of topological insulator phases in non-Hermitian $\textit{PT}$-symmetric Hamiltonians}},\ }\href {https://doi.org/10.1103/PhysRevB.84.153101} {\bibfield  {journal} {\bibinfo  {journal} {Phys. Rev. B}\ }\textbf {\bibinfo {volume} {84}},\ \bibinfo {pages} {153101} (\bibinfo {year} {2011})}\BibitemShut {NoStop}%
\bibitem [{\citenamefont {Schomerus}(2013)}]{Schomerus-13}%
  \BibitemOpen
  \bibfield  {author} {\bibinfo {author} {\bibfnamefont {H.}~\bibnamefont {Schomerus}},\ }\bibfield  {title} {\bibinfo {title} {{Topologically protected midgap states in complex photonic lattices}},\ }\href {https://doi.org/10.1364/OL.38.001912} {\bibfield  {journal} {\bibinfo  {journal} {Opt. Lett.}\ }\textbf {\bibinfo {volume} {38}},\ \bibinfo {pages} {1912} (\bibinfo {year} {2013})}\BibitemShut {NoStop}%
\bibitem [{\citenamefont {Longhi}\ \emph {et~al.}(2015)\citenamefont {Longhi}, \citenamefont {Gatti},\ and\ \citenamefont {Valle}}]{Longhi-15}%
  \BibitemOpen
  \bibfield  {author} {\bibinfo {author} {\bibfnamefont {S.}~\bibnamefont {Longhi}}, \bibinfo {author} {\bibfnamefont {D.}~\bibnamefont {Gatti}},\ and\ \bibinfo {author} {\bibfnamefont {G.~D.}\ \bibnamefont {Valle}},\ }\bibfield  {title} {\bibinfo {title} {{Robust light transport in non-Hermitian photonic lattices}},\ }\href {https://doi.org/10.1038/srep13376} {\bibfield  {journal} {\bibinfo  {journal} {Sci. Rep.}\ }\textbf {\bibinfo {volume} {5}},\ \bibinfo {pages} {13376} (\bibinfo {year} {2015})}\BibitemShut {NoStop}%
\bibitem [{\citenamefont {Malzard}\ \emph {et~al.}(2015)\citenamefont {Malzard}, \citenamefont {Poli},\ and\ \citenamefont {Schomerus}}]{Malzard-15}%
  \BibitemOpen
  \bibfield  {author} {\bibinfo {author} {\bibfnamefont {S.}~\bibnamefont {Malzard}}, \bibinfo {author} {\bibfnamefont {C.}~\bibnamefont {Poli}},\ and\ \bibinfo {author} {\bibfnamefont {H.}~\bibnamefont {Schomerus}},\ }\bibfield  {title} {\bibinfo {title} {{Topologically Protected Defect States in Open Photonic Systems with Non-Hermitian Charge-Conjugation and Parity-Time Symmetry}},\ }\href {https://doi.org/10.1103/PhysRevLett.115.200402} {\bibfield  {journal} {\bibinfo  {journal} {Phys. Rev. Lett.}\ }\textbf {\bibinfo {volume} {115}},\ \bibinfo {pages} {200402} (\bibinfo {year} {2015})}\BibitemShut {NoStop}%
\bibitem [{\citenamefont {Lee}(2016)}]{Lee-16}%
  \BibitemOpen
  \bibfield  {author} {\bibinfo {author} {\bibfnamefont {T.~E.}\ \bibnamefont {Lee}},\ }\bibfield  {title} {\bibinfo {title} {{Anomalous Edge State in a Non-Hermitian Lattice}},\ }\href {https://doi.org/10.1103/PhysRevLett.116.133903} {\bibfield  {journal} {\bibinfo  {journal} {Phys. Rev. Lett.}\ }\textbf {\bibinfo {volume} {116}},\ \bibinfo {pages} {133903} (\bibinfo {year} {2016})}\BibitemShut {NoStop}%
\bibitem [{\citenamefont {Leykam}\ \emph {et~al.}(2017)\citenamefont {Leykam}, \citenamefont {Bliokh}, \citenamefont {Huang}, \citenamefont {Chong},\ and\ \citenamefont {Nori}}]{Leykam-17}%
  \BibitemOpen
  \bibfield  {author} {\bibinfo {author} {\bibfnamefont {D.}~\bibnamefont {Leykam}}, \bibinfo {author} {\bibfnamefont {K.~Y.}\ \bibnamefont {Bliokh}}, \bibinfo {author} {\bibfnamefont {C.}~\bibnamefont {Huang}}, \bibinfo {author} {\bibfnamefont {Y.~D.}\ \bibnamefont {Chong}},\ and\ \bibinfo {author} {\bibfnamefont {F.}~\bibnamefont {Nori}},\ }\bibfield  {title} {\bibinfo {title} {{Edge Modes, Degeneracies, and Topological Numbers in Non-Hermitian Systems}},\ }\href {https://doi.org/10.1103/PhysRevLett.118.040401} {\bibfield  {journal} {\bibinfo  {journal} {Phys. Rev. Lett.}\ }\textbf {\bibinfo {volume} {118}},\ \bibinfo {pages} {040401} (\bibinfo {year} {2017})}\BibitemShut {NoStop}%
\bibitem [{\citenamefont {Xu}\ \emph {et~al.}(2017)\citenamefont {Xu}, \citenamefont {Wang},\ and\ \citenamefont {Duan}}]{Xu-17}%
  \BibitemOpen
  \bibfield  {author} {\bibinfo {author} {\bibfnamefont {Y.}~\bibnamefont {Xu}}, \bibinfo {author} {\bibfnamefont {S.-T.}\ \bibnamefont {Wang}},\ and\ \bibinfo {author} {\bibfnamefont {L.-M.}\ \bibnamefont {Duan}},\ }\bibfield  {title} {\bibinfo {title} {{Weyl Exceptional Rings in a Three-Dimensional Dissipative Cold Atomic Gas}},\ }\href {https://doi.org/10.1103/PhysRevLett.118.045701} {\bibfield  {journal} {\bibinfo  {journal} {Phys. Rev. Lett.}\ }\textbf {\bibinfo {volume} {118}},\ \bibinfo {pages} {045701} (\bibinfo {year} {2017})}\BibitemShut {NoStop}%
\bibitem [{\citenamefont {Xiong}(2018)}]{Xiong-18}%
  \BibitemOpen
  \bibfield  {author} {\bibinfo {author} {\bibfnamefont {Y.}~\bibnamefont {Xiong}},\ }\bibfield  {title} {\bibinfo {title} {{Why does bulk boundary correspondence fail in some non-hermitian topological models}},\ }\href {https://doi.org/10.1088/2399-6528/aab64a} {\bibfield  {journal} {\bibinfo  {journal} {J. Phys. Commun.}\ }\textbf {\bibinfo {volume} {2}},\ \bibinfo {pages} {035043} (\bibinfo {year} {2018})}\BibitemShut {NoStop}%
\bibitem [{\citenamefont {Shen}\ \emph {et~al.}(2018)\citenamefont {Shen}, \citenamefont {Zhen},\ and\ \citenamefont {Fu}}]{Shen-18}%
  \BibitemOpen
  \bibfield  {author} {\bibinfo {author} {\bibfnamefont {H.}~\bibnamefont {Shen}}, \bibinfo {author} {\bibfnamefont {B.}~\bibnamefont {Zhen}},\ and\ \bibinfo {author} {\bibfnamefont {L.}~\bibnamefont {Fu}},\ }\bibfield  {title} {\bibinfo {title} {{Topological Band Theory for Non-Hermitian Hamiltonians}},\ }\href {https://doi.org/10.1103/PhysRevLett.120.146402} {\bibfield  {journal} {\bibinfo  {journal} {Phys. Rev. Lett.}\ }\textbf {\bibinfo {volume} {120}},\ \bibinfo {pages} {146402} (\bibinfo {year} {2018})}\BibitemShut {NoStop}%
\bibitem [{\citenamefont {Kozii}\ and\ \citenamefont {Fu}(2024)}]{Kozii-17}%
  \BibitemOpen
  \bibfield  {author} {\bibinfo {author} {\bibfnamefont {V.}~\bibnamefont {Kozii}}\ and\ \bibinfo {author} {\bibfnamefont {L.}~\bibnamefont {Fu}},\ }\bibfield  {title} {\bibinfo {title} {{Non-Hermitian topological theory of finite-lifetime quasiparticles: Prediction of bulk Fermi arc due to exceptional point}},\ }\href {https://doi.org/10.1103/PhysRevB.109.235139} {\bibfield  {journal} {\bibinfo  {journal} {Phys. Rev. B}\ }\textbf {\bibinfo {volume} {109}},\ \bibinfo {pages} {235139} (\bibinfo {year} {2024})}\BibitemShut {NoStop}%
\bibitem [{\citenamefont {Takata}\ and\ \citenamefont {Notomi}(2018)}]{Takata-18}%
  \BibitemOpen
  \bibfield  {author} {\bibinfo {author} {\bibfnamefont {K.}~\bibnamefont {Takata}}\ and\ \bibinfo {author} {\bibfnamefont {M.}~\bibnamefont {Notomi}},\ }\bibfield  {title} {\bibinfo {title} {{Photonic Topological Insulating Phase Induced Solely by Gain and Loss}},\ }\href {https://doi.org/10.1103/PhysRevLett.121.213902} {\bibfield  {journal} {\bibinfo  {journal} {Phys. Rev. Lett.}\ }\textbf {\bibinfo {volume} {121}},\ \bibinfo {pages} {213902} (\bibinfo {year} {2018})}\BibitemShut {NoStop}%
\bibitem [{\citenamefont {Martinez~Alvarez}\ \emph {et~al.}(2018)\citenamefont {Martinez~Alvarez}, \citenamefont {Barrios~Vargas},\ and\ \citenamefont {Foa~Torres}}]{MartinezAlvarez-18}%
  \BibitemOpen
  \bibfield  {author} {\bibinfo {author} {\bibfnamefont {V.~M.}\ \bibnamefont {Martinez~Alvarez}}, \bibinfo {author} {\bibfnamefont {J.~E.}\ \bibnamefont {Barrios~Vargas}},\ and\ \bibinfo {author} {\bibfnamefont {L.~E.~F.}\ \bibnamefont {Foa~Torres}},\ }\bibfield  {title} {\bibinfo {title} {{Non-Hermitian robust edge states in one dimension: Anomalous localization and eigenspace condensation at exceptional points}},\ }\href {https://doi.org/10.1103/PhysRevB.97.121401} {\bibfield  {journal} {\bibinfo  {journal} {Phys. Rev. B}\ }\textbf {\bibinfo {volume} {97}},\ \bibinfo {pages} {121401(R)} (\bibinfo {year} {2018})}\BibitemShut {NoStop}%
\bibitem [{\citenamefont {Gong}\ \emph {et~al.}(2018)\citenamefont {Gong}, \citenamefont {Ashida}, \citenamefont {Kawabata}, \citenamefont {Takasan}, \citenamefont {Higashikawa},\ and\ \citenamefont {Ueda}}]{Gong-18}%
  \BibitemOpen
  \bibfield  {author} {\bibinfo {author} {\bibfnamefont {Z.}~\bibnamefont {Gong}}, \bibinfo {author} {\bibfnamefont {Y.}~\bibnamefont {Ashida}}, \bibinfo {author} {\bibfnamefont {K.}~\bibnamefont {Kawabata}}, \bibinfo {author} {\bibfnamefont {K.}~\bibnamefont {Takasan}}, \bibinfo {author} {\bibfnamefont {S.}~\bibnamefont {Higashikawa}},\ and\ \bibinfo {author} {\bibfnamefont {M.}~\bibnamefont {Ueda}},\ }\bibfield  {title} {\bibinfo {title} {{Topological Phases of Non-Hermitian Systems}},\ }\href {https://doi.org/10.1103/PhysRevX.8.031079} {\bibfield  {journal} {\bibinfo  {journal} {Phys. Rev. X}\ }\textbf {\bibinfo {volume} {8}},\ \bibinfo {pages} {031079} (\bibinfo {year} {2018})}\BibitemShut {NoStop}%
\bibitem [{\citenamefont {Yao}\ and\ \citenamefont {Wang}(2018)}]{YW-18-SSH}%
  \BibitemOpen
  \bibfield  {author} {\bibinfo {author} {\bibfnamefont {S.}~\bibnamefont {Yao}}\ and\ \bibinfo {author} {\bibfnamefont {Z.}~\bibnamefont {Wang}},\ }\bibfield  {title} {\bibinfo {title} {{Edge States and Topological Invariants of Non-Hermitian Systems}},\ }\href {https://doi.org/10.1103/PhysRevLett.121.086803} {\bibfield  {journal} {\bibinfo  {journal} {Phys. Rev. Lett.}\ }\textbf {\bibinfo {volume} {121}},\ \bibinfo {pages} {086803} (\bibinfo {year} {2018})}\BibitemShut {NoStop}%
\bibitem [{\citenamefont {Yao}\ \emph {et~al.}(2018)\citenamefont {Yao}, \citenamefont {Song},\ and\ \citenamefont {Wang}}]{YSW-18-Chern}%
  \BibitemOpen
  \bibfield  {author} {\bibinfo {author} {\bibfnamefont {S.}~\bibnamefont {Yao}}, \bibinfo {author} {\bibfnamefont {F.}~\bibnamefont {Song}},\ and\ \bibinfo {author} {\bibfnamefont {Z.}~\bibnamefont {Wang}},\ }\bibfield  {title} {\bibinfo {title} {{Non-Hermitian Chern Bands}},\ }\href {https://doi.org/10.1103/PhysRevLett.121.136802} {\bibfield  {journal} {\bibinfo  {journal} {{Phys. Rev. Lett.}}\ }\textbf {\bibinfo {volume} {121}},\ \bibinfo {pages} {136802} (\bibinfo {year} {2018})}\BibitemShut {NoStop}%
\bibitem [{\citenamefont {Kunst}\ \emph {et~al.}(2018)\citenamefont {Kunst}, \citenamefont {Edvardsson}, \citenamefont {Budich},\ and\ \citenamefont {Bergholtz}}]{Kunst-18}%
  \BibitemOpen
  \bibfield  {author} {\bibinfo {author} {\bibfnamefont {F.~K.}\ \bibnamefont {Kunst}}, \bibinfo {author} {\bibfnamefont {E.}~\bibnamefont {Edvardsson}}, \bibinfo {author} {\bibfnamefont {J.~C.}\ \bibnamefont {Budich}},\ and\ \bibinfo {author} {\bibfnamefont {E.~J.}\ \bibnamefont {Bergholtz}},\ }\bibfield  {title} {\bibinfo {title} {{Biorthogonal Bulk-Boundary Correspondence in Non-Hermitian Systems}},\ }\href {https://doi.org/10.1103/PhysRevLett.121.026808} {\bibfield  {journal} {\bibinfo  {journal} {Phys. Rev. Lett.}\ }\textbf {\bibinfo {volume} {121}},\ \bibinfo {pages} {026808} (\bibinfo {year} {2018})}\BibitemShut {NoStop}%
\bibitem [{\citenamefont {McDonald}\ \emph {et~al.}(2018)\citenamefont {McDonald}, \citenamefont {Pereg-Barnea},\ and\ \citenamefont {Clerk}}]{McDonald-18}%
  \BibitemOpen
  \bibfield  {author} {\bibinfo {author} {\bibfnamefont {A.}~\bibnamefont {McDonald}}, \bibinfo {author} {\bibfnamefont {T.}~\bibnamefont {Pereg-Barnea}},\ and\ \bibinfo {author} {\bibfnamefont {A.~A.}\ \bibnamefont {Clerk}},\ }\bibfield  {title} {\bibinfo {title} {{Phase-Dependent Chiral Transport and Effective Non-Hermitian Dynamics in a Bosonic Kitaev-Majorana Chain}},\ }\href {https://doi.org/10.1103/PhysRevX.8.041031} {\bibfield  {journal} {\bibinfo  {journal} {Phys. Rev. X}\ }\textbf {\bibinfo {volume} {8}},\ \bibinfo {pages} {041031} (\bibinfo {year} {2018})}\BibitemShut {NoStop}%
\bibitem [{\citenamefont {Lee}\ and\ \citenamefont {Thomale}(2019)}]{Lee-Thomale-19}%
  \BibitemOpen
  \bibfield  {author} {\bibinfo {author} {\bibfnamefont {C.~H.}\ \bibnamefont {Lee}}\ and\ \bibinfo {author} {\bibfnamefont {R.}~\bibnamefont {Thomale}},\ }\bibfield  {title} {\bibinfo {title} {{Anatomy of skin modes and topology in non-Hermitian systems}},\ }\href {https://doi.org/10.1103/PhysRevB.99.201103} {\bibfield  {journal} {\bibinfo  {journal} {Phys. Rev. B}\ }\textbf {\bibinfo {volume} {99}},\ \bibinfo {pages} {201103(R)} (\bibinfo {year} {2019})}\BibitemShut {NoStop}%
\bibitem [{\citenamefont {Liu}\ \emph {et~al.}(2019)\citenamefont {Liu}, \citenamefont {Zhang}, \citenamefont {Ai}, \citenamefont {Gong}, \citenamefont {Kawabata}, \citenamefont {Ueda},\ and\ \citenamefont {Nori}}]{Liu-19}%
  \BibitemOpen
  \bibfield  {author} {\bibinfo {author} {\bibfnamefont {T.}~\bibnamefont {Liu}}, \bibinfo {author} {\bibfnamefont {Y.-R.}\ \bibnamefont {Zhang}}, \bibinfo {author} {\bibfnamefont {Q.}~\bibnamefont {Ai}}, \bibinfo {author} {\bibfnamefont {Z.}~\bibnamefont {Gong}}, \bibinfo {author} {\bibfnamefont {K.}~\bibnamefont {Kawabata}}, \bibinfo {author} {\bibfnamefont {M.}~\bibnamefont {Ueda}},\ and\ \bibinfo {author} {\bibfnamefont {F.}~\bibnamefont {Nori}},\ }\bibfield  {title} {\bibinfo {title} {{Second-Order Topological Phases in Non-Hermitian Systems}},\ }\href {https://doi.org/10.1103/PhysRevLett.122.076801} {\bibfield  {journal} {\bibinfo  {journal} {Phys. Rev. Lett.}\ }\textbf {\bibinfo {volume} {122}},\ \bibinfo {pages} {076801} (\bibinfo {year} {2019})}\BibitemShut {NoStop}%
\bibitem [{\citenamefont {Lee}\ \emph {et~al.}(2019{\natexlab{a}})\citenamefont {Lee}, \citenamefont {Li},\ and\ \citenamefont {Gong}}]{Lee-Li-Gong-19}%
  \BibitemOpen
  \bibfield  {author} {\bibinfo {author} {\bibfnamefont {C.~H.}\ \bibnamefont {Lee}}, \bibinfo {author} {\bibfnamefont {L.}~\bibnamefont {Li}},\ and\ \bibinfo {author} {\bibfnamefont {J.}~\bibnamefont {Gong}},\ }\bibfield  {title} {\bibinfo {title} {{Hybrid Higher-Order Skin-Topological Modes in Nonreciprocal Systems}},\ }\href {https://doi.org/10.1103/PhysRevLett.123.016805} {\bibfield  {journal} {\bibinfo  {journal} {Phys. Rev. Lett.}\ }\textbf {\bibinfo {volume} {123}},\ \bibinfo {pages} {016805} (\bibinfo {year} {2019}{\natexlab{a}})}\BibitemShut {NoStop}%
\bibitem [{\citenamefont {Kawabata}\ \emph {et~al.}(2019{\natexlab{a}})\citenamefont {Kawabata}, \citenamefont {Shiozaki}, \citenamefont {Ueda},\ and\ \citenamefont {Sato}}]{KSUS-19}%
  \BibitemOpen
  \bibfield  {author} {\bibinfo {author} {\bibfnamefont {K.}~\bibnamefont {Kawabata}}, \bibinfo {author} {\bibfnamefont {K.}~\bibnamefont {Shiozaki}}, \bibinfo {author} {\bibfnamefont {M.}~\bibnamefont {Ueda}},\ and\ \bibinfo {author} {\bibfnamefont {M.}~\bibnamefont {Sato}},\ }\bibfield  {title} {\bibinfo {title} {{Symmetry and Topology in Non-Hermitian Physics}},\ }\href {https://doi.org/10.1103/PhysRevX.9.041015} {\bibfield  {journal} {\bibinfo  {journal} {Phys. Rev. X}\ }\textbf {\bibinfo {volume} {9}},\ \bibinfo {pages} {041015} (\bibinfo {year} {2019}{\natexlab{a}})}\BibitemShut {NoStop}%
\bibitem [{\citenamefont {Zhou}\ and\ \citenamefont {Lee}(2019)}]{ZL-19}%
  \BibitemOpen
  \bibfield  {author} {\bibinfo {author} {\bibfnamefont {H.}~\bibnamefont {Zhou}}\ and\ \bibinfo {author} {\bibfnamefont {J.~Y.}\ \bibnamefont {Lee}},\ }\bibfield  {title} {\bibinfo {title} {{Periodic table for topological bands with non-Hermitian symmetries}},\ }\href {https://doi.org/10.1103/PhysRevB.99.235112} {\bibfield  {journal} {\bibinfo  {journal} {Phys. Rev. B}\ }\textbf {\bibinfo {volume} {99}},\ \bibinfo {pages} {235112} (\bibinfo {year} {2019})}\BibitemShut {NoStop}%
\bibitem [{\citenamefont {Herviou}\ \emph {et~al.}(2019)\citenamefont {Herviou}, \citenamefont {Bardarson},\ and\ \citenamefont {Regnault}}]{Herviou-19}%
  \BibitemOpen
  \bibfield  {author} {\bibinfo {author} {\bibfnamefont {L.}~\bibnamefont {Herviou}}, \bibinfo {author} {\bibfnamefont {J.~H.}\ \bibnamefont {Bardarson}},\ and\ \bibinfo {author} {\bibfnamefont {N.}~\bibnamefont {Regnault}},\ }\bibfield  {title} {\bibinfo {title} {{Defining a bulk-edge correspondence for non-Hermitian Hamiltonians via singular-value decomposition}},\ }\href {https://doi.org/10.1103/PhysRevA.99.052118} {\bibfield  {journal} {\bibinfo  {journal} {Phys. Rev. A}\ }\textbf {\bibinfo {volume} {99}},\ \bibinfo {pages} {052118} (\bibinfo {year} {2019})}\BibitemShut {NoStop}%
\bibitem [{\citenamefont {Zirnstein}\ \emph {et~al.}(2021)\citenamefont {Zirnstein}, \citenamefont {Refael},\ and\ \citenamefont {Rosenow}}]{Zirnstein-19}%
  \BibitemOpen
  \bibfield  {author} {\bibinfo {author} {\bibfnamefont {H.-G.}\ \bibnamefont {Zirnstein}}, \bibinfo {author} {\bibfnamefont {G.}~\bibnamefont {Refael}},\ and\ \bibinfo {author} {\bibfnamefont {B.}~\bibnamefont {Rosenow}},\ }\bibfield  {title} {\bibinfo {title} {{Bulk-Boundary Correspondence for Non-Hermitian Hamiltonians via Green Functions}},\ }\href {https://doi.org/10.1103/PhysRevLett.126.216407} {\bibfield  {journal} {\bibinfo  {journal} {Phys. Rev. Lett.}\ }\textbf {\bibinfo {volume} {126}},\ \bibinfo {pages} {216407} (\bibinfo {year} {2021})}\BibitemShut {NoStop}%
\bibitem [{\citenamefont {Borgnia}\ \emph {et~al.}(2020)\citenamefont {Borgnia}, \citenamefont {Kruchkov},\ and\ \citenamefont {Slager}}]{Borgnia-19}%
  \BibitemOpen
  \bibfield  {author} {\bibinfo {author} {\bibfnamefont {D.~S.}\ \bibnamefont {Borgnia}}, \bibinfo {author} {\bibfnamefont {A.~J.}\ \bibnamefont {Kruchkov}},\ and\ \bibinfo {author} {\bibfnamefont {R.-J.}\ \bibnamefont {Slager}},\ }\bibfield  {title} {\bibinfo {title} {{Non-Hermitian Boundary Modes and Topology}},\ }\href {https://doi.org/10.1103/PhysRevLett.124.056802} {\bibfield  {journal} {\bibinfo  {journal} {Phys. Rev. Lett.}\ }\textbf {\bibinfo {volume} {124}},\ \bibinfo {pages} {056802} (\bibinfo {year} {2020})}\BibitemShut {NoStop}%
\bibitem [{\citenamefont {Kawabata}\ \emph {et~al.}(2019{\natexlab{b}})\citenamefont {Kawabata}, \citenamefont {Bessho},\ and\ \citenamefont {Sato}}]{KBS-19}%
  \BibitemOpen
  \bibfield  {author} {\bibinfo {author} {\bibfnamefont {K.}~\bibnamefont {Kawabata}}, \bibinfo {author} {\bibfnamefont {T.}~\bibnamefont {Bessho}},\ and\ \bibinfo {author} {\bibfnamefont {M.}~\bibnamefont {Sato}},\ }\bibfield  {title} {\bibinfo {title} {{Classification of Exceptional Points and Non-Hermitian Topological Semimetals}},\ }\href {https://doi.org/10.1103/PhysRevLett.123.066405} {\bibfield  {journal} {\bibinfo  {journal} {Phys. Rev. Lett.}\ }\textbf {\bibinfo {volume} {123}},\ \bibinfo {pages} {066405} (\bibinfo {year} {2019}{\natexlab{b}})}\BibitemShut {NoStop}%
\bibitem [{\citenamefont {Yokomizo}\ and\ \citenamefont {Murakami}(2019)}]{Yokomizo-19}%
  \BibitemOpen
  \bibfield  {author} {\bibinfo {author} {\bibfnamefont {K.}~\bibnamefont {Yokomizo}}\ and\ \bibinfo {author} {\bibfnamefont {S.}~\bibnamefont {Murakami}},\ }\bibfield  {title} {\bibinfo {title} {{Non-Bloch Band Theory of Non-Hermitian Systems}},\ }\href {https://doi.org/10.1103/PhysRevLett.123.066404} {\bibfield  {journal} {\bibinfo  {journal} {Phys. Rev. Lett.}\ }\textbf {\bibinfo {volume} {123}},\ \bibinfo {pages} {066404} (\bibinfo {year} {2019})}\BibitemShut {NoStop}%
\bibitem [{\citenamefont {Lee}\ \emph {et~al.}(2019{\natexlab{b}})\citenamefont {Lee}, \citenamefont {Ahn}, \citenamefont {Zhou},\ and\ \citenamefont {Vishwanath}}]{JYLee-19}%
  \BibitemOpen
  \bibfield  {author} {\bibinfo {author} {\bibfnamefont {J.~Y.}\ \bibnamefont {Lee}}, \bibinfo {author} {\bibfnamefont {J.}~\bibnamefont {Ahn}}, \bibinfo {author} {\bibfnamefont {H.}~\bibnamefont {Zhou}},\ and\ \bibinfo {author} {\bibfnamefont {A.}~\bibnamefont {Vishwanath}},\ }\bibfield  {title} {\bibinfo {title} {{Topological Correspondence between Hermitian and Non-Hermitian Systems: Anomalous Dynamics}},\ }\href {https://doi.org/10.1103/PhysRevLett.123.206404} {\bibfield  {journal} {\bibinfo  {journal} {Phys. Rev. Lett.}\ }\textbf {\bibinfo {volume} {123}},\ \bibinfo {pages} {206404} (\bibinfo {year} {2019}{\natexlab{b}})}\BibitemShut {NoStop}%
\bibitem [{\citenamefont {Schomerus}(2020)}]{Schomerus-20}%
  \BibitemOpen
  \bibfield  {author} {\bibinfo {author} {\bibfnamefont {H.}~\bibnamefont {Schomerus}},\ }\bibfield  {title} {\bibinfo {title} {{Nonreciprocal response theory of non-Hermitian mechanical metamaterials: Response phase transition from the skin effect of zero modes}},\ }\href {https://doi.org/10.1103/PhysRevResearch.2.013058} {\bibfield  {journal} {\bibinfo  {journal} {Phys. Rev. Research}\ }\textbf {\bibinfo {volume} {2}},\ \bibinfo {pages} {013058} (\bibinfo {year} {2020})}\BibitemShut {NoStop}%
\bibitem [{\citenamefont {Chang}\ \emph {et~al.}(2020)\citenamefont {Chang}, \citenamefont {You}, \citenamefont {Wen},\ and\ \citenamefont {Ryu}}]{Chang-20}%
  \BibitemOpen
  \bibfield  {author} {\bibinfo {author} {\bibfnamefont {P.-Y.}\ \bibnamefont {Chang}}, \bibinfo {author} {\bibfnamefont {J.-S.}\ \bibnamefont {You}}, \bibinfo {author} {\bibfnamefont {X.}~\bibnamefont {Wen}},\ and\ \bibinfo {author} {\bibfnamefont {S.}~\bibnamefont {Ryu}},\ }\bibfield  {title} {\bibinfo {title} {{Entanglement spectrum and entropy in topological non-Hermitian systems and nonunitary conformal field theory}},\ }\href {https://doi.org/10.1103/PhysRevResearch.2.033069} {\bibfield  {journal} {\bibinfo  {journal} {Phys. Rev. Research}\ }\textbf {\bibinfo {volume} {2}},\ \bibinfo {pages} {033069} (\bibinfo {year} {2020})}\BibitemShut {NoStop}%
\bibitem [{\citenamefont {Wanjura}\ \emph {et~al.}(2020)\citenamefont {Wanjura}, \citenamefont {Brunelli},\ and\ \citenamefont {Nunnenkamp}}]{Wanjura-20}%
  \BibitemOpen
  \bibfield  {author} {\bibinfo {author} {\bibfnamefont {C.~C.}\ \bibnamefont {Wanjura}}, \bibinfo {author} {\bibfnamefont {M.}~\bibnamefont {Brunelli}},\ and\ \bibinfo {author} {\bibfnamefont {A.}~\bibnamefont {Nunnenkamp}},\ }\bibfield  {title} {\bibinfo {title} {{Topological framework for directional amplification in driven-dissipative cavity arrays}},\ }\href {https://doi.org/10.1038/s41467-020-16863-9} {\bibfield  {journal} {\bibinfo  {journal} {Nat. Commun.}\ }\textbf {\bibinfo {volume} {11}},\ \bibinfo {pages} {3149} (\bibinfo {year} {2020})}\BibitemShut {NoStop}%
\bibitem [{\citenamefont {Zhang}\ \emph {et~al.}(2020)\citenamefont {Zhang}, \citenamefont {Yang},\ and\ \citenamefont {Fang}}]{Zhang-20}%
  \BibitemOpen
  \bibfield  {author} {\bibinfo {author} {\bibfnamefont {K.}~\bibnamefont {Zhang}}, \bibinfo {author} {\bibfnamefont {Z.}~\bibnamefont {Yang}},\ and\ \bibinfo {author} {\bibfnamefont {C.}~\bibnamefont {Fang}},\ }\bibfield  {title} {\bibinfo {title} {{Correspondence between Winding Numbers and Skin Modes in Non-Hermitian Systems}},\ }\href {https://doi.org/10.1103/PhysRevLett.125.126402} {\bibfield  {journal} {\bibinfo  {journal} {Phys. Rev. Lett.}\ }\textbf {\bibinfo {volume} {125}},\ \bibinfo {pages} {126402} (\bibinfo {year} {2020})}\BibitemShut {NoStop}%
\bibitem [{\citenamefont {Okuma}\ \emph {et~al.}(2020)\citenamefont {Okuma}, \citenamefont {Kawabata}, \citenamefont {Shiozaki},\ and\ \citenamefont {Sato}}]{OKSS-20}%
  \BibitemOpen
  \bibfield  {author} {\bibinfo {author} {\bibfnamefont {N.}~\bibnamefont {Okuma}}, \bibinfo {author} {\bibfnamefont {K.}~\bibnamefont {Kawabata}}, \bibinfo {author} {\bibfnamefont {K.}~\bibnamefont {Shiozaki}},\ and\ \bibinfo {author} {\bibfnamefont {M.}~\bibnamefont {Sato}},\ }\bibfield  {title} {\bibinfo {title} {{Topological Origin of Non-Hermitian Skin Effects}},\ }\href {https://doi.org/10.1103/PhysRevLett.124.086801} {\bibfield  {journal} {\bibinfo  {journal} {Phys. Rev. Lett.}\ }\textbf {\bibinfo {volume} {124}},\ \bibinfo {pages} {086801} (\bibinfo {year} {2020})}\BibitemShut {NoStop}%
\bibitem [{\citenamefont {Yang}\ \emph {et~al.}(2020{\natexlab{a}})\citenamefont {Yang}, \citenamefont {Zhang}, \citenamefont {Fang},\ and\ \citenamefont {Hu}}]{Yang-20}%
  \BibitemOpen
  \bibfield  {author} {\bibinfo {author} {\bibfnamefont {Z.}~\bibnamefont {Yang}}, \bibinfo {author} {\bibfnamefont {K.}~\bibnamefont {Zhang}}, \bibinfo {author} {\bibfnamefont {C.}~\bibnamefont {Fang}},\ and\ \bibinfo {author} {\bibfnamefont {J.}~\bibnamefont {Hu}},\ }\bibfield  {title} {\bibinfo {title} {{Non-Hermitian Bulk-Boundary Correspondence and Auxiliary Generalized Brillouin Zone Theory}},\ }\href {https://doi.org/10.1103/PhysRevLett.125.226402} {\bibfield  {journal} {\bibinfo  {journal} {Phys. Rev. Lett.}\ }\textbf {\bibinfo {volume} {125}},\ \bibinfo {pages} {226402} (\bibinfo {year} {2020}{\natexlab{a}})}\BibitemShut {NoStop}%
\bibitem [{\citenamefont {Terrier}\ and\ \citenamefont {Kunst}(2020)}]{Terrier-20}%
  \BibitemOpen
  \bibfield  {author} {\bibinfo {author} {\bibfnamefont {F.}~\bibnamefont {Terrier}}\ and\ \bibinfo {author} {\bibfnamefont {F.~K.}\ \bibnamefont {Kunst}},\ }\bibfield  {title} {\bibinfo {title} {{Dissipative analog of four-dimensional quantum Hall physics}},\ }\href {https://doi.org/10.1103/PhysRevResearch.2.023364} {\bibfield  {journal} {\bibinfo  {journal} {Phys. Rev. Research}\ }\textbf {\bibinfo {volume} {2}},\ \bibinfo {pages} {023364} (\bibinfo {year} {2020})}\BibitemShut {NoStop}%
\bibitem [{\citenamefont {Xue}\ \emph {et~al.}(2021)\citenamefont {Xue}, \citenamefont {Li}, \citenamefont {Hu}, \citenamefont {Song},\ and\ \citenamefont {Wang}}]{Xue-20}%
  \BibitemOpen
  \bibfield  {author} {\bibinfo {author} {\bibfnamefont {W.-T.}\ \bibnamefont {Xue}}, \bibinfo {author} {\bibfnamefont {M.-R.}\ \bibnamefont {Li}}, \bibinfo {author} {\bibfnamefont {Y.-M.}\ \bibnamefont {Hu}}, \bibinfo {author} {\bibfnamefont {F.}~\bibnamefont {Song}},\ and\ \bibinfo {author} {\bibfnamefont {Z.}~\bibnamefont {Wang}},\ }\bibfield  {title} {\bibinfo {title} {{Simple formulas of directional amplification from non-Bloch band theory}},\ }\href {https://doi.org/10.1103/PhysRevB.103.L241408} {\bibfield  {journal} {\bibinfo  {journal} {Phys. Rev. B}\ }\textbf {\bibinfo {volume} {103}},\ \bibinfo {pages} {L241408} (\bibinfo {year} {2021})}\BibitemShut {NoStop}%
\bibitem [{\citenamefont {Bessho}\ and\ \citenamefont {Sato}(2021)}]{Bessho-21}%
  \BibitemOpen
  \bibfield  {author} {\bibinfo {author} {\bibfnamefont {T.}~\bibnamefont {Bessho}}\ and\ \bibinfo {author} {\bibfnamefont {M.}~\bibnamefont {Sato}},\ }\bibfield  {title} {\bibinfo {title} {{Nielsen-Ninomiya Theorem with Bulk Topology: Duality in Floquet and Non-Hermitian Systems}},\ }\href {https://doi.org/10.1103/PhysRevLett.127.196404} {\bibfield  {journal} {\bibinfo  {journal} {Phys. Rev. Lett.}\ }\textbf {\bibinfo {volume} {127}},\ \bibinfo {pages} {196404} (\bibinfo {year} {2021})}\BibitemShut {NoStop}%
\bibitem [{\citenamefont {Denner}\ \emph {et~al.}(2021)\citenamefont {Denner}, \citenamefont {Skurativska}, \citenamefont {Schindler}, \citenamefont {Fischer}, \citenamefont {Thomale}, \citenamefont {Bzdu\v{s}ek},\ and\ \citenamefont {Neupert}}]{Denner-21}%
  \BibitemOpen
  \bibfield  {author} {\bibinfo {author} {\bibfnamefont {M.~M.}\ \bibnamefont {Denner}}, \bibinfo {author} {\bibfnamefont {A.}~\bibnamefont {Skurativska}}, \bibinfo {author} {\bibfnamefont {F.}~\bibnamefont {Schindler}}, \bibinfo {author} {\bibfnamefont {M.~H.}\ \bibnamefont {Fischer}}, \bibinfo {author} {\bibfnamefont {R.}~\bibnamefont {Thomale}}, \bibinfo {author} {\bibfnamefont {T.}~\bibnamefont {Bzdu\v{s}ek}},\ and\ \bibinfo {author} {\bibfnamefont {T.}~\bibnamefont {Neupert}},\ }\bibfield  {title} {\bibinfo {title} {{Exceptional topological insulators}},\ }\href {https://doi.org/10.1038/s41467-021-25947-z} {\bibfield  {journal} {\bibinfo  {journal} {Nat. Commun.}\ }\textbf {\bibinfo {volume} {12}},\ \bibinfo {pages} {5681} (\bibinfo {year} {2021})}\BibitemShut {NoStop}%
\bibitem [{\citenamefont {Denner}\ \emph {et~al.}(2023)\citenamefont {Denner}, \citenamefont {Neupert},\ and\ \citenamefont {Schindler}}]{Denner-23JPhysMater}%
  \BibitemOpen
  \bibfield  {author} {\bibinfo {author} {\bibfnamefont {M.~M.}\ \bibnamefont {Denner}}, \bibinfo {author} {\bibfnamefont {T.}~\bibnamefont {Neupert}},\ and\ \bibinfo {author} {\bibfnamefont {F.}~\bibnamefont {Schindler}},\ }\bibfield  {title} {\bibinfo {title} {{Infernal and exceptional edge modes: non-Hermitian topology beyond the skin effect}},\ }\href {https://doi.org/10.1088/2515-7639/acf2ca} {\bibfield  {journal} {\bibinfo  {journal} {J. Phys. Mater.}\ }\textbf {\bibinfo {volume} {6}},\ \bibinfo {pages} {045006} (\bibinfo {year} {2023})}\BibitemShut {NoStop}%
\bibitem [{\citenamefont {Okugawa}\ \emph {et~al.}(2020)\citenamefont {Okugawa}, \citenamefont {Takahashi},\ and\ \citenamefont {Yokomizo}}]{Okugawa-20}%
  \BibitemOpen
  \bibfield  {author} {\bibinfo {author} {\bibfnamefont {R.}~\bibnamefont {Okugawa}}, \bibinfo {author} {\bibfnamefont {R.}~\bibnamefont {Takahashi}},\ and\ \bibinfo {author} {\bibfnamefont {K.}~\bibnamefont {Yokomizo}},\ }\bibfield  {title} {\bibinfo {title} {{Second-order topological non-Hermitian skin effects}},\ }\href {https://doi.org/10.1103/PhysRevB.102.241202} {\bibfield  {journal} {\bibinfo  {journal} {Phys. Rev. B}\ }\textbf {\bibinfo {volume} {102}},\ \bibinfo {pages} {241202(R)} (\bibinfo {year} {2020})}\BibitemShut {NoStop}%
\bibitem [{\citenamefont {Kawabata}\ \emph {et~al.}(2020)\citenamefont {Kawabata}, \citenamefont {Sato},\ and\ \citenamefont {Shiozaki}}]{KSS-20}%
  \BibitemOpen
  \bibfield  {author} {\bibinfo {author} {\bibfnamefont {K.}~\bibnamefont {Kawabata}}, \bibinfo {author} {\bibfnamefont {M.}~\bibnamefont {Sato}},\ and\ \bibinfo {author} {\bibfnamefont {K.}~\bibnamefont {Shiozaki}},\ }\bibfield  {title} {\bibinfo {title} {{Higher-order non-Hermitian skin effect}},\ }\href {https://doi.org/10.1103/PhysRevB.102.205118} {\bibfield  {journal} {\bibinfo  {journal} {Phys. Rev. B}\ }\textbf {\bibinfo {volume} {102}},\ \bibinfo {pages} {205118} (\bibinfo {year} {2020})}\BibitemShut {NoStop}%
\bibitem [{\citenamefont {Kawabata}\ \emph {et~al.}(2021)\citenamefont {Kawabata}, \citenamefont {Shiozaki},\ and\ \citenamefont {Ryu}}]{KSR-21}%
  \BibitemOpen
  \bibfield  {author} {\bibinfo {author} {\bibfnamefont {K.}~\bibnamefont {Kawabata}}, \bibinfo {author} {\bibfnamefont {K.}~\bibnamefont {Shiozaki}},\ and\ \bibinfo {author} {\bibfnamefont {S.}~\bibnamefont {Ryu}},\ }\bibfield  {title} {\bibinfo {title} {{Topological Field Theory of Non-Hermitian Systems}},\ }\href {https://doi.org/10.1103/PhysRevLett.126.216405} {\bibfield  {journal} {\bibinfo  {journal} {Phys. Rev. Lett.}\ }\textbf {\bibinfo {volume} {126}},\ \bibinfo {pages} {216405} (\bibinfo {year} {2021})}\BibitemShut {NoStop}%
\bibitem [{\citenamefont {Zhang}\ \emph {et~al.}(2022)\citenamefont {Zhang}, \citenamefont {Yang},\ and\ \citenamefont {Fang}}]{Zhang-22}%
  \BibitemOpen
  \bibfield  {author} {\bibinfo {author} {\bibfnamefont {K.}~\bibnamefont {Zhang}}, \bibinfo {author} {\bibfnamefont {Z.}~\bibnamefont {Yang}},\ and\ \bibinfo {author} {\bibfnamefont {C.}~\bibnamefont {Fang}},\ }\bibfield  {title} {\bibinfo {title} {{Universal non-Hermitian skin effect in two and higher dimensions}},\ }\href {https://doi.org/10.1038/s41467-022-30161-6} {\bibfield  {journal} {\bibinfo  {journal} {Nat. Commun.}\ }\textbf {\bibinfo {volume} {13}},\ \bibinfo {pages} {2496} (\bibinfo {year} {2022})}\BibitemShut {NoStop}%
\bibitem [{\citenamefont {Sun}\ \emph {et~al.}(2021)\citenamefont {Sun}, \citenamefont {Zhu},\ and\ \citenamefont {Hughes}}]{Sun-21}%
  \BibitemOpen
  \bibfield  {author} {\bibinfo {author} {\bibfnamefont {X.-Q.}\ \bibnamefont {Sun}}, \bibinfo {author} {\bibfnamefont {P.}~\bibnamefont {Zhu}},\ and\ \bibinfo {author} {\bibfnamefont {T.~L.}\ \bibnamefont {Hughes}},\ }\bibfield  {title} {\bibinfo {title} {{Geometric Response and Disclination-Induced Skin Effects in Non-Hermitian Systems}},\ }\href {https://doi.org/10.1103/PhysRevLett.127.066401} {\bibfield  {journal} {\bibinfo  {journal} {Phys. Rev. Lett.}\ }\textbf {\bibinfo {volume} {127}},\ \bibinfo {pages} {066401} (\bibinfo {year} {2021})}\BibitemShut {NoStop}%
\bibitem [{\citenamefont {Shiozaki}\ and\ \citenamefont {Ono}(2021)}]{Shiozaki-21}%
  \BibitemOpen
  \bibfield  {author} {\bibinfo {author} {\bibfnamefont {K.}~\bibnamefont {Shiozaki}}\ and\ \bibinfo {author} {\bibfnamefont {S.}~\bibnamefont {Ono}},\ }\bibfield  {title} {\bibinfo {title} {{Symmetry indicator in non-Hermitian systems}},\ }\href {https://doi.org/10.1103/PhysRevB.104.035424} {\bibfield  {journal} {\bibinfo  {journal} {Phys. Rev. B}\ }\textbf {\bibinfo {volume} {104}},\ \bibinfo {pages} {035424} (\bibinfo {year} {2021})}\BibitemShut {NoStop}%
\bibitem [{\citenamefont {Franca}\ \emph {et~al.}(2022)\citenamefont {Franca}, \citenamefont {K\"onye}, \citenamefont {Hassler}, \citenamefont {van~den Brink},\ and\ \citenamefont {Fulga}}]{Franca-22}%
  \BibitemOpen
  \bibfield  {author} {\bibinfo {author} {\bibfnamefont {S.}~\bibnamefont {Franca}}, \bibinfo {author} {\bibfnamefont {V.}~\bibnamefont {K\"onye}}, \bibinfo {author} {\bibfnamefont {F.}~\bibnamefont {Hassler}}, \bibinfo {author} {\bibfnamefont {J.}~\bibnamefont {van~den Brink}},\ and\ \bibinfo {author} {\bibfnamefont {C.}~\bibnamefont {Fulga}},\ }\bibfield  {title} {\bibinfo {title} {{Non-Hermitian Physics without Gain or Loss: The Skin Effect of Reflected Waves}},\ }\href {https://doi.org/10.1103/PhysRevLett.129.086601} {\bibfield  {journal} {\bibinfo  {journal} {Phys. Rev. Lett.}\ }\textbf {\bibinfo {volume} {129}},\ \bibinfo {pages} {086601} (\bibinfo {year} {2022})}\BibitemShut {NoStop}%
\bibitem [{\citenamefont {Nakamura}\ \emph {et~al.}(2024)\citenamefont {Nakamura}, \citenamefont {Bessho},\ and\ \citenamefont {Sato}}]{Nakamura-24}%
  \BibitemOpen
  \bibfield  {author} {\bibinfo {author} {\bibfnamefont {D.}~\bibnamefont {Nakamura}}, \bibinfo {author} {\bibfnamefont {T.}~\bibnamefont {Bessho}},\ and\ \bibinfo {author} {\bibfnamefont {M.}~\bibnamefont {Sato}},\ }\bibfield  {title} {\bibinfo {title} {{Bulk-Boundary Correspondence in Point-Gap Topological Phases}},\ }\href {https://doi.org/10.1103/PhysRevLett.132.136401} {\bibfield  {journal} {\bibinfo  {journal} {Phys. Rev. Lett.}\ }\textbf {\bibinfo {volume} {132}},\ \bibinfo {pages} {136401} (\bibinfo {year} {2024})}\BibitemShut {NoStop}%
\bibitem [{\citenamefont {Denner}\ and\ \citenamefont {Schindler}(2023)}]{Denner-23}%
  \BibitemOpen
  \bibfield  {author} {\bibinfo {author} {\bibfnamefont {M.~M.}\ \bibnamefont {Denner}}\ and\ \bibinfo {author} {\bibfnamefont {F.}~\bibnamefont {Schindler}},\ }\bibfield  {title} {\bibinfo {title} {{Magnetic flux response of non-Hermitian topological phases}},\ }\href {https://doi.org/10.21468/SciPostPhys.14.5.107} {\bibfield  {journal} {\bibinfo  {journal} {SciPost Phys.}\ }\textbf {\bibinfo {volume} {14}},\ \bibinfo {pages} {107} (\bibinfo {year} {2023})}\BibitemShut {NoStop}%
\bibitem [{\citenamefont {Wang}\ \emph {et~al.}(2024)\citenamefont {Wang}, \citenamefont {Song},\ and\ \citenamefont {Wang}}]{Wang-24}%
  \BibitemOpen
  \bibfield  {author} {\bibinfo {author} {\bibfnamefont {H.-Y.}\ \bibnamefont {Wang}}, \bibinfo {author} {\bibfnamefont {F.}~\bibnamefont {Song}},\ and\ \bibinfo {author} {\bibfnamefont {Z.}~\bibnamefont {Wang}},\ }\bibfield  {title} {\bibinfo {title} {{Amoeba Formulation of Non-Bloch Band Theory in Arbitrary Dimensions}},\ }\href {https://doi.org/10.1103/PhysRevX.14.021011} {\bibfield  {journal} {\bibinfo  {journal} {Phys. Rev. X}\ }\textbf {\bibinfo {volume} {14}},\ \bibinfo {pages} {021011} (\bibinfo {year} {2024})}\BibitemShut {NoStop}%
\bibitem [{\citenamefont {Nakai}\ \emph {et~al.}(2024)\citenamefont {Nakai}, \citenamefont {Okuma}, \citenamefont {Nakamura}, \citenamefont {Shimomura},\ and\ \citenamefont {Sato}}]{Nakai-24}%
  \BibitemOpen
  \bibfield  {author} {\bibinfo {author} {\bibfnamefont {Y.~O.}\ \bibnamefont {Nakai}}, \bibinfo {author} {\bibfnamefont {N.}~\bibnamefont {Okuma}}, \bibinfo {author} {\bibfnamefont {D.}~\bibnamefont {Nakamura}}, \bibinfo {author} {\bibfnamefont {K.}~\bibnamefont {Shimomura}},\ and\ \bibinfo {author} {\bibfnamefont {M.}~\bibnamefont {Sato}},\ }\bibfield  {title} {\bibinfo {title} {{Topological enhancement of nonnormality in non-Hermitian skin effects}},\ }\href {https://doi.org/10.1103/PhysRevB.109.144203} {\bibfield  {journal} {\bibinfo  {journal} {Phys. Rev. B}\ }\textbf {\bibinfo {volume} {109}},\ \bibinfo {pages} {144203} (\bibinfo {year} {2024})}\BibitemShut {NoStop}%
\bibitem [{\citenamefont {Ma}\ \emph {et~al.}(2024)\citenamefont {Ma}, \citenamefont {Cao}, \citenamefont {Wang}, \citenamefont {Wei}, \citenamefont {Du},\ and\ \citenamefont {Kou}}]{Ma-24}%
  \BibitemOpen
  \bibfield  {author} {\bibinfo {author} {\bibfnamefont {X.-R.}\ \bibnamefont {Ma}}, \bibinfo {author} {\bibfnamefont {K.}~\bibnamefont {Cao}}, \bibinfo {author} {\bibfnamefont {X.-R.}\ \bibnamefont {Wang}}, \bibinfo {author} {\bibfnamefont {Z.}~\bibnamefont {Wei}}, \bibinfo {author} {\bibfnamefont {Q.}~\bibnamefont {Du}},\ and\ \bibinfo {author} {\bibfnamefont {S.-P.}\ \bibnamefont {Kou}},\ }\bibfield  {title} {\bibinfo {title} {{Non-Hermitian chiral skin effect}},\ }\href {https://doi.org/10.1103/PhysRevResearch.6.013213} {\bibfield  {journal} {\bibinfo  {journal} {Phys. Rev. Research}\ }\textbf {\bibinfo {volume} {6}},\ \bibinfo {pages} {013213} (\bibinfo {year} {2024})}\BibitemShut {NoStop}%
\bibitem [{\citenamefont {Schindler}\ \emph {et~al.}(2023)\citenamefont {Schindler}, \citenamefont {Gu}, \citenamefont {Lian},\ and\ \citenamefont {Kawabata}}]{Schindler-23}%
  \BibitemOpen
  \bibfield  {author} {\bibinfo {author} {\bibfnamefont {F.}~\bibnamefont {Schindler}}, \bibinfo {author} {\bibfnamefont {K.}~\bibnamefont {Gu}}, \bibinfo {author} {\bibfnamefont {B.}~\bibnamefont {Lian}},\ and\ \bibinfo {author} {\bibfnamefont {K.}~\bibnamefont {Kawabata}},\ }\bibfield  {title} {\bibinfo {title} {{Hermitian Bulk -- Non-Hermitian Boundary Correspondence}},\ }\href {https://doi.org/10.1103/PRXQuantum.4.030315} {\bibfield  {journal} {\bibinfo  {journal} {PRX Quantum}\ }\textbf {\bibinfo {volume} {4}},\ \bibinfo {pages} {030315} (\bibinfo {year} {2023})}\BibitemShut {NoStop}%
\bibitem [{\citenamefont {Nakamura}\ \emph {et~al.}(2023)\citenamefont {Nakamura}, \citenamefont {Inaka}, \citenamefont {Okuma},\ and\ \citenamefont {Sato}}]{Nakamura-23}%
  \BibitemOpen
  \bibfield  {author} {\bibinfo {author} {\bibfnamefont {D.}~\bibnamefont {Nakamura}}, \bibinfo {author} {\bibfnamefont {K.}~\bibnamefont {Inaka}}, \bibinfo {author} {\bibfnamefont {N.}~\bibnamefont {Okuma}},\ and\ \bibinfo {author} {\bibfnamefont {M.}~\bibnamefont {Sato}},\ }\bibfield  {title} {\bibinfo {title} {{Universal Platform of Point-Gap Topological Phases from Topological Materials}},\ }\href {https://doi.org/10.1103/PhysRevLett.131.256602} {\bibfield  {journal} {\bibinfo  {journal} {Phys. Rev. Lett.}\ }\textbf {\bibinfo {volume} {131}},\ \bibinfo {pages} {256602} (\bibinfo {year} {2023})}\BibitemShut {NoStop}%
\bibitem [{\citenamefont {Hamanaka}\ \emph {et~al.}(2024)\citenamefont {Hamanaka}, \citenamefont {Yoshida},\ and\ \citenamefont {Kawabata}}]{Hamanaka-24}%
  \BibitemOpen
  \bibfield  {author} {\bibinfo {author} {\bibfnamefont {S.}~\bibnamefont {Hamanaka}}, \bibinfo {author} {\bibfnamefont {T.}~\bibnamefont {Yoshida}},\ and\ \bibinfo {author} {\bibfnamefont {K.}~\bibnamefont {Kawabata}},\ }\bibfield  {title} {\bibinfo {title} {{Non-Hermitian Topology in Hermitian Topological Matter}},\ }\href {https://doi.org/10.1103/PhysRevLett.133.266604} {\bibfield  {journal} {\bibinfo  {journal} {Phys. Rev. Lett.}\ }\textbf {\bibinfo {volume} {133}},\ \bibinfo {pages} {266604} (\bibinfo {year} {2024})}\BibitemShut {NoStop}%
\bibitem [{\citenamefont {Zhang}\ \emph {et~al.}()\citenamefont {Zhang}, \citenamefont {Shu},\ and\ \citenamefont {Sun}}]{Zhang-24}%
  \BibitemOpen
  \bibfield  {author} {\bibinfo {author} {\bibfnamefont {K.}~\bibnamefont {Zhang}}, \bibinfo {author} {\bibfnamefont {C.}~\bibnamefont {Shu}},\ and\ \bibinfo {author} {\bibfnamefont {K.}~\bibnamefont {Sun}},\ }\bibfield  {title} {\bibinfo {title} {{Algebraic non-Hermitian skin effect and unified non-Bloch band theory in arbitrary dimensions}},\ }\Eprint {https://arxiv.org/abs/2406.06682} {arXiv:2406.06682} \BibitemShut {NoStop}%
\bibitem [{\citenamefont {Wang}()}]{YifanWang-24}%
  \BibitemOpen
  \bibfield  {author} {\bibinfo {author} {\bibfnamefont {Y.}~\bibnamefont {Wang}},\ }\bibfield  {title} {\bibinfo {title} {{Classifying Order-Two Spatial Symmetries in Non-Hermitian Hamiltonians: Point-gapped AZ and AZ$^{\dag}$ Classes}},\ }\Eprint {https://arxiv.org/abs/2411.03410} {arXiv:2411.03410} \BibitemShut {NoStop}%
\bibitem [{\citenamefont {Tanaka}\ \emph {et~al.}()\citenamefont {Tanaka}, \citenamefont {Nakamura}, \citenamefont {Okugawa},\ and\ \citenamefont {Kawabata}}]{Tanaka-24}%
  \BibitemOpen
  \bibfield  {author} {\bibinfo {author} {\bibfnamefont {Y.}~\bibnamefont {Tanaka}}, \bibinfo {author} {\bibfnamefont {D.}~\bibnamefont {Nakamura}}, \bibinfo {author} {\bibfnamefont {R.}~\bibnamefont {Okugawa}},\ and\ \bibinfo {author} {\bibfnamefont {K.}~\bibnamefont {Kawabata}},\ }\bibfield  {title} {\bibinfo {title} {{Exceptional Second-Order Topological Insulators}},\ }\Eprint {https://arxiv.org/abs/2411.06898} {arXiv:2411.06898} \BibitemShut {NoStop}%
\bibitem [{\citenamefont {Poli}\ \emph {et~al.}(2015)\citenamefont {Poli}, \citenamefont {Bellec}, \citenamefont {Kuhl}, \citenamefont {Mortessagne},\ and\ \citenamefont {Schomerus}}]{Poli-15}%
  \BibitemOpen
  \bibfield  {author} {\bibinfo {author} {\bibfnamefont {C.}~\bibnamefont {Poli}}, \bibinfo {author} {\bibfnamefont {M.}~\bibnamefont {Bellec}}, \bibinfo {author} {\bibfnamefont {U.}~\bibnamefont {Kuhl}}, \bibinfo {author} {\bibfnamefont {F.}~\bibnamefont {Mortessagne}},\ and\ \bibinfo {author} {\bibfnamefont {H.}~\bibnamefont {Schomerus}},\ }\bibfield  {title} {\bibinfo {title} {{Selective enhancement of topologically induced interface states in a dielectric resonator chain}},\ }\href {https://doi.org/10.1038/ncomms7710} {\bibfield  {journal} {\bibinfo  {journal} {Nat. Commun.}\ }\textbf {\bibinfo {volume} {6}},\ \bibinfo {pages} {6710} (\bibinfo {year} {2015})}\BibitemShut {NoStop}%
\bibitem [{\citenamefont {Zeuner}\ \emph {et~al.}(2015)\citenamefont {Zeuner}, \citenamefont {Rechtsman}, \citenamefont {Plotnik}, \citenamefont {Lumer}, \citenamefont {Nolte}, \citenamefont {Rudner}, \citenamefont {Segev},\ and\ \citenamefont {Szameit}}]{Zeuner-15}%
  \BibitemOpen
  \bibfield  {author} {\bibinfo {author} {\bibfnamefont {J.~M.}\ \bibnamefont {Zeuner}}, \bibinfo {author} {\bibfnamefont {M.~C.}\ \bibnamefont {Rechtsman}}, \bibinfo {author} {\bibfnamefont {Y.}~\bibnamefont {Plotnik}}, \bibinfo {author} {\bibfnamefont {Y.}~\bibnamefont {Lumer}}, \bibinfo {author} {\bibfnamefont {S.}~\bibnamefont {Nolte}}, \bibinfo {author} {\bibfnamefont {M.~S.}\ \bibnamefont {Rudner}}, \bibinfo {author} {\bibfnamefont {M.}~\bibnamefont {Segev}},\ and\ \bibinfo {author} {\bibfnamefont {A.}~\bibnamefont {Szameit}},\ }\bibfield  {title} {\bibinfo {title} {{Observation of a Topological Transition in the Bulk of a Non-Hermitian System}},\ }\href {https://doi.org/10.1103/PhysRevLett.115.040402} {\bibfield  {journal} {\bibinfo  {journal} {{Phys. Rev. Lett.}}\ }\textbf {\bibinfo {volume} {115}},\ \bibinfo {pages} {040402} (\bibinfo {year} {2015})}\BibitemShut {NoStop}%
\bibitem [{\citenamefont {Zhen}\ \emph {et~al.}(2015)\citenamefont {Zhen}, \citenamefont {Hsu}, \citenamefont {Igarashi}, \citenamefont {Lu}, \citenamefont {Kaminer}, \citenamefont {Pick}, \citenamefont {Chua}, \citenamefont {Joannopoulos},\ and\ \citenamefont {Solja\v{c}i\'c}}]{Zhen-15}%
  \BibitemOpen
  \bibfield  {author} {\bibinfo {author} {\bibfnamefont {B.}~\bibnamefont {Zhen}}, \bibinfo {author} {\bibfnamefont {C.~W.}\ \bibnamefont {Hsu}}, \bibinfo {author} {\bibfnamefont {Y.}~\bibnamefont {Igarashi}}, \bibinfo {author} {\bibfnamefont {L.}~\bibnamefont {Lu}}, \bibinfo {author} {\bibfnamefont {I.}~\bibnamefont {Kaminer}}, \bibinfo {author} {\bibfnamefont {A.}~\bibnamefont {Pick}}, \bibinfo {author} {\bibfnamefont {S.-L.}\ \bibnamefont {Chua}}, \bibinfo {author} {\bibfnamefont {J.~D.}\ \bibnamefont {Joannopoulos}},\ and\ \bibinfo {author} {\bibfnamefont {M.}~\bibnamefont {Solja\v{c}i\'c}},\ }\bibfield  {title} {\bibinfo {title} {{Spawning rings of exceptional points out of Dirac cones}},\ }\href {https://doi.org/10.1038/nature14889} {\bibfield  {journal} {\bibinfo  {journal} {Nature}\ }\textbf {\bibinfo {volume} {525}},\ \bibinfo {pages} {354} (\bibinfo {year} {2015})}\BibitemShut {NoStop}%
\bibitem [{\citenamefont {Weimann}\ \emph {et~al.}(2017)\citenamefont {Weimann}, \citenamefont {Kremer}, \citenamefont {Plotnik}, \citenamefont {Lumer}, \citenamefont {Nolte}, \citenamefont {Makris}, \citenamefont {Segev}, \citenamefont {Rechtsman},\ and\ \citenamefont {Szameit}}]{Weimann-17}%
  \BibitemOpen
  \bibfield  {author} {\bibinfo {author} {\bibfnamefont {S.}~\bibnamefont {Weimann}}, \bibinfo {author} {\bibfnamefont {M.}~\bibnamefont {Kremer}}, \bibinfo {author} {\bibfnamefont {Y.}~\bibnamefont {Plotnik}}, \bibinfo {author} {\bibfnamefont {Y.}~\bibnamefont {Lumer}}, \bibinfo {author} {\bibfnamefont {S.}~\bibnamefont {Nolte}}, \bibinfo {author} {\bibfnamefont {K.~G.}\ \bibnamefont {Makris}}, \bibinfo {author} {\bibfnamefont {M.}~\bibnamefont {Segev}}, \bibinfo {author} {\bibfnamefont {M.~C.}\ \bibnamefont {Rechtsman}},\ and\ \bibinfo {author} {\bibfnamefont {A.}~\bibnamefont {Szameit}},\ }\bibfield  {title} {\bibinfo {title} {{Topologically protected bound states in photonic parity-time-symmetric crystals}},\ }\href {https://doi.org/10.1038/nmat4811} {\bibfield  {journal} {\bibinfo  {journal} {Nat. Mater.}\ }\textbf {\bibinfo {volume} {16}},\ \bibinfo {pages} {433} (\bibinfo {year} {2017})}\BibitemShut {NoStop}%
\bibitem [{\citenamefont {Xiao}\ \emph {et~al.}(2017)\citenamefont {Xiao}, \citenamefont {Zhan}, \citenamefont {Bian}, \citenamefont {Wang}, \citenamefont {Zhang}, \citenamefont {Wang}, \citenamefont {Li}, \citenamefont {Mochizuki}, \citenamefont {Kim}, \citenamefont {Kawakami}, \citenamefont {Yi}, \citenamefont {Obuse}, \citenamefont {Sanders},\ and\ \citenamefont {Xue}}]{Xiao-17}%
  \BibitemOpen
  \bibfield  {author} {\bibinfo {author} {\bibfnamefont {L.}~\bibnamefont {Xiao}}, \bibinfo {author} {\bibfnamefont {X.}~\bibnamefont {Zhan}}, \bibinfo {author} {\bibfnamefont {Z.~H.}\ \bibnamefont {Bian}}, \bibinfo {author} {\bibfnamefont {K.~K.}\ \bibnamefont {Wang}}, \bibinfo {author} {\bibfnamefont {X.}~\bibnamefont {Zhang}}, \bibinfo {author} {\bibfnamefont {X.~P.}\ \bibnamefont {Wang}}, \bibinfo {author} {\bibfnamefont {J.}~\bibnamefont {Li}}, \bibinfo {author} {\bibfnamefont {K.}~\bibnamefont {Mochizuki}}, \bibinfo {author} {\bibfnamefont {D.}~\bibnamefont {Kim}}, \bibinfo {author} {\bibfnamefont {N.}~\bibnamefont {Kawakami}}, \bibinfo {author} {\bibfnamefont {W.}~\bibnamefont {Yi}}, \bibinfo {author} {\bibfnamefont {H.}~\bibnamefont {Obuse}}, \bibinfo {author} {\bibfnamefont {B.~C.}\ \bibnamefont {Sanders}},\ and\ \bibinfo {author} {\bibfnamefont {P.}~\bibnamefont {Xue}},\ }\bibfield  {title} {\bibinfo {title} {{Observation of topological edge states in parity-time-symmetric quantum walks}},\ }\href
  {https://doi.org/10.1038/nphys4204} {\bibfield  {journal} {\bibinfo  {journal} {Nat. Phys.}\ }\textbf {\bibinfo {volume} {13}},\ \bibinfo {pages} {1117} (\bibinfo {year} {2017})}\BibitemShut {NoStop}%
\bibitem [{\citenamefont {St-Jean}\ \emph {et~al.}(2017)\citenamefont {St-Jean}, \citenamefont {Goblot}, \citenamefont {Galopin}, \citenamefont {Lema\^itre}, \citenamefont {Ozawa}, \citenamefont {Gratiet}, \citenamefont {Sagnes}, \citenamefont {Bloch},\ and\ \citenamefont {Amo}}]{St-Jean-17}%
  \BibitemOpen
  \bibfield  {author} {\bibinfo {author} {\bibfnamefont {P.}~\bibnamefont {St-Jean}}, \bibinfo {author} {\bibfnamefont {V.}~\bibnamefont {Goblot}}, \bibinfo {author} {\bibfnamefont {E.}~\bibnamefont {Galopin}}, \bibinfo {author} {\bibfnamefont {A.}~\bibnamefont {Lema\^itre}}, \bibinfo {author} {\bibfnamefont {T.}~\bibnamefont {Ozawa}}, \bibinfo {author} {\bibfnamefont {L.~L.}\ \bibnamefont {Gratiet}}, \bibinfo {author} {\bibfnamefont {I.}~\bibnamefont {Sagnes}}, \bibinfo {author} {\bibfnamefont {J.}~\bibnamefont {Bloch}},\ and\ \bibinfo {author} {\bibfnamefont {A.}~\bibnamefont {Amo}},\ }\bibfield  {title} {\bibinfo {title} {{Lasing in topological edge states of a one-dimensional lattice}},\ }\href {https://doi.org/10.1038/s41566-017-0006-2} {\bibfield  {journal} {\bibinfo  {journal} {Nat. Photon.}\ }\textbf {\bibinfo {volume} {11}},\ \bibinfo {pages} {651} (\bibinfo {year} {2017})}\BibitemShut {NoStop}%
\bibitem [{\citenamefont {Parto}\ \emph {et~al.}(2018)\citenamefont {Parto}, \citenamefont {Wittek}, \citenamefont {Hodaei}, \citenamefont {Harari}, \citenamefont {Bandres}, \citenamefont {Ren}, \citenamefont {Rechtsman}, \citenamefont {Segev}, \citenamefont {Christodoulides},\ and\ \citenamefont {Khajavikhan}}]{Parto-17}%
  \BibitemOpen
  \bibfield  {author} {\bibinfo {author} {\bibfnamefont {M.}~\bibnamefont {Parto}}, \bibinfo {author} {\bibfnamefont {S.}~\bibnamefont {Wittek}}, \bibinfo {author} {\bibfnamefont {H.}~\bibnamefont {Hodaei}}, \bibinfo {author} {\bibfnamefont {G.}~\bibnamefont {Harari}}, \bibinfo {author} {\bibfnamefont {M.~A.}\ \bibnamefont {Bandres}}, \bibinfo {author} {\bibfnamefont {J.}~\bibnamefont {Ren}}, \bibinfo {author} {\bibfnamefont {M.~C.}\ \bibnamefont {Rechtsman}}, \bibinfo {author} {\bibfnamefont {M.}~\bibnamefont {Segev}}, \bibinfo {author} {\bibfnamefont {D.~N.}\ \bibnamefont {Christodoulides}},\ and\ \bibinfo {author} {\bibfnamefont {M.}~\bibnamefont {Khajavikhan}},\ }\bibfield  {title} {\bibinfo {title} {{Edge-Mode Lasing in 1D Topological Active Arrays}},\ }\href {https://doi.org/10.1103/PhysRevLett.120.113901} {\bibfield  {journal} {\bibinfo  {journal} {Phys. Rev. Lett.}\ }\textbf {\bibinfo {volume} {120}},\ \bibinfo {pages} {113901} (\bibinfo {year} {2018})}\BibitemShut {NoStop}%
\bibitem [{\citenamefont {Bahari}\ \emph {et~al.}(2017)\citenamefont {Bahari}, \citenamefont {Ndao}, \citenamefont {Vallini}, \citenamefont {Amili}, \citenamefont {Fainman},\ and\ \citenamefont {Kant\'e}}]{Bahari-17}%
  \BibitemOpen
  \bibfield  {author} {\bibinfo {author} {\bibfnamefont {B.}~\bibnamefont {Bahari}}, \bibinfo {author} {\bibfnamefont {A.}~\bibnamefont {Ndao}}, \bibinfo {author} {\bibfnamefont {F.}~\bibnamefont {Vallini}}, \bibinfo {author} {\bibfnamefont {A.~E.}\ \bibnamefont {Amili}}, \bibinfo {author} {\bibfnamefont {Y.}~\bibnamefont {Fainman}},\ and\ \bibinfo {author} {\bibfnamefont {B.}~\bibnamefont {Kant\'e}},\ }\bibfield  {title} {\bibinfo {title} {{Nonreciprocal lasing in topological cavities of arbitrary geometries}},\ }\href {https://doi.org/10.1126/science.aao4551} {\bibfield  {journal} {\bibinfo  {journal} {Science}\ }\textbf {\bibinfo {volume} {358}},\ \bibinfo {pages} {636} (\bibinfo {year} {2017})}\BibitemShut {NoStop}%
\bibitem [{\citenamefont {Zhao}\ \emph {et~al.}(2018)\citenamefont {Zhao}, \citenamefont {Miao}, \citenamefont {Teimourpour}, \citenamefont {Malzard}, \citenamefont {El-Ganainy}, \citenamefont {Schomerus},\ and\ \citenamefont {Feng}}]{Zhao-18}%
  \BibitemOpen
  \bibfield  {author} {\bibinfo {author} {\bibfnamefont {H.}~\bibnamefont {Zhao}}, \bibinfo {author} {\bibfnamefont {P.}~\bibnamefont {Miao}}, \bibinfo {author} {\bibfnamefont {M.~H.}\ \bibnamefont {Teimourpour}}, \bibinfo {author} {\bibfnamefont {S.}~\bibnamefont {Malzard}}, \bibinfo {author} {\bibfnamefont {R.}~\bibnamefont {El-Ganainy}}, \bibinfo {author} {\bibfnamefont {H.}~\bibnamefont {Schomerus}},\ and\ \bibinfo {author} {\bibfnamefont {L.}~\bibnamefont {Feng}},\ }\bibfield  {title} {\bibinfo {title} {{Topological hybrid silicon microlasers}},\ }\href {https://doi.org/10.1038/s41467-018-03434-2} {\bibfield  {journal} {\bibinfo  {journal} {Nat. Commun.}\ }\textbf {\bibinfo {volume} {9}},\ \bibinfo {pages} {981} (\bibinfo {year} {2018})}\BibitemShut {NoStop}%
\bibitem [{\citenamefont {Zhou}\ \emph {et~al.}(2018)\citenamefont {Zhou}, \citenamefont {Peng}, \citenamefont {Yoon}, \citenamefont {Hsu}, \citenamefont {Nelson}, \citenamefont {Fu}, \citenamefont {Joannopoulos}, \citenamefont {Solja\v{c}i\'c},\ and\ \citenamefont {Zhen}}]{Zhou-18}%
  \BibitemOpen
  \bibfield  {author} {\bibinfo {author} {\bibfnamefont {H.}~\bibnamefont {Zhou}}, \bibinfo {author} {\bibfnamefont {C.}~\bibnamefont {Peng}}, \bibinfo {author} {\bibfnamefont {Y.}~\bibnamefont {Yoon}}, \bibinfo {author} {\bibfnamefont {C.~W.}\ \bibnamefont {Hsu}}, \bibinfo {author} {\bibfnamefont {K.~A.}\ \bibnamefont {Nelson}}, \bibinfo {author} {\bibfnamefont {L.}~\bibnamefont {Fu}}, \bibinfo {author} {\bibfnamefont {J.~D.}\ \bibnamefont {Joannopoulos}}, \bibinfo {author} {\bibfnamefont {M.}~\bibnamefont {Solja\v{c}i\'c}},\ and\ \bibinfo {author} {\bibfnamefont {B.}~\bibnamefont {Zhen}},\ }\bibfield  {title} {\bibinfo {title} {{Observation of bulk Fermi arc and polarization half charge from paired exceptional points}},\ }\href {https://doi.org/10.1126/science.aap9859} {\bibfield  {journal} {\bibinfo  {journal} {Science}\ }\textbf {\bibinfo {volume} {359}},\ \bibinfo {pages} {1009} (\bibinfo {year} {2018})}\BibitemShut {NoStop}%
\bibitem [{\citenamefont {Harari}\ \emph {et~al.}(2018)\citenamefont {Harari}, \citenamefont {Bandres}, \citenamefont {Lumer}, \citenamefont {Rechtsman}, \citenamefont {Chong}, \citenamefont {Khajavikhan}, \citenamefont {Christodoulides},\ and\ \citenamefont {Segev}}]{Harari-18}%
  \BibitemOpen
  \bibfield  {author} {\bibinfo {author} {\bibfnamefont {G.}~\bibnamefont {Harari}}, \bibinfo {author} {\bibfnamefont {M.~A.}\ \bibnamefont {Bandres}}, \bibinfo {author} {\bibfnamefont {Y.}~\bibnamefont {Lumer}}, \bibinfo {author} {\bibfnamefont {M.~C.}\ \bibnamefont {Rechtsman}}, \bibinfo {author} {\bibfnamefont {Y.~D.}\ \bibnamefont {Chong}}, \bibinfo {author} {\bibfnamefont {M.}~\bibnamefont {Khajavikhan}}, \bibinfo {author} {\bibfnamefont {D.~N.}\ \bibnamefont {Christodoulides}},\ and\ \bibinfo {author} {\bibfnamefont {M.}~\bibnamefont {Segev}},\ }\bibfield  {title} {\bibinfo {title} {{Topological insulator laser: Theory}},\ }\href {https://doi.org/10.1126/science.aar4003} {\bibfield  {journal} {\bibinfo  {journal} {Science}\ }\textbf {\bibinfo {volume} {359}},\ \bibinfo {pages} {eaar4003} (\bibinfo {year} {2018})}\BibitemShut {NoStop}%
\bibitem [{\citenamefont {Bandres}\ \emph {et~al.}(2018)\citenamefont {Bandres}, \citenamefont {Wittek}, \citenamefont {Harari}, \citenamefont {Parto}, \citenamefont {Ren}, \citenamefont {Segev}, \citenamefont {Christodoulides},\ and\ \citenamefont {Khajavikhan}}]{Bandres-18}%
  \BibitemOpen
  \bibfield  {author} {\bibinfo {author} {\bibfnamefont {M.~A.}\ \bibnamefont {Bandres}}, \bibinfo {author} {\bibfnamefont {S.}~\bibnamefont {Wittek}}, \bibinfo {author} {\bibfnamefont {G.}~\bibnamefont {Harari}}, \bibinfo {author} {\bibfnamefont {M.}~\bibnamefont {Parto}}, \bibinfo {author} {\bibfnamefont {J.}~\bibnamefont {Ren}}, \bibinfo {author} {\bibfnamefont {M.}~\bibnamefont {Segev}}, \bibinfo {author} {\bibfnamefont {D.}~\bibnamefont {Christodoulides}},\ and\ \bibinfo {author} {\bibfnamefont {M.}~\bibnamefont {Khajavikhan}},\ }\bibfield  {title} {\bibinfo {title} {{Topological insulator laser: Experiments}},\ }\href {https://doi.org/10.1126/science.aar4005} {\bibfield  {journal} {\bibinfo  {journal} {{Science}}\ }\textbf {\bibinfo {volume} {359}},\ \bibinfo {pages} {eaar4005} (\bibinfo {year} {2018})}\BibitemShut {NoStop}%
\bibitem [{\citenamefont {Cerjan}\ \emph {et~al.}(2019)\citenamefont {Cerjan}, \citenamefont {Huang}, \citenamefont {Chen}, \citenamefont {Chong},\ and\ \citenamefont {Rechtsman}}]{Cerjan-19}%
  \BibitemOpen
  \bibfield  {author} {\bibinfo {author} {\bibfnamefont {A.}~\bibnamefont {Cerjan}}, \bibinfo {author} {\bibfnamefont {S.}~\bibnamefont {Huang}}, \bibinfo {author} {\bibfnamefont {K.~P.}\ \bibnamefont {Chen}}, \bibinfo {author} {\bibfnamefont {Y.}~\bibnamefont {Chong}},\ and\ \bibinfo {author} {\bibfnamefont {M.~C.}\ \bibnamefont {Rechtsman}},\ }\bibfield  {title} {\bibinfo {title} {{Experimental realization of a Weyl exceptional ring}},\ }\href {https://doi.org/10.1038/s41566-019-0453-z} {\bibfield  {journal} {\bibinfo  {journal} {Nat. Photon.}\ }\textbf {\bibinfo {volume} {13}},\ \bibinfo {pages} {623} (\bibinfo {year} {2019})}\BibitemShut {NoStop}%
\bibitem [{\citenamefont {Zhao}\ \emph {et~al.}(2019)\citenamefont {Zhao}, \citenamefont {Qiao}, \citenamefont {Wu}, \citenamefont {Midya}, \citenamefont {Longhi},\ and\ \citenamefont {Feng}}]{Zhao-19}%
  \BibitemOpen
  \bibfield  {author} {\bibinfo {author} {\bibfnamefont {H.}~\bibnamefont {Zhao}}, \bibinfo {author} {\bibfnamefont {X.}~\bibnamefont {Qiao}}, \bibinfo {author} {\bibfnamefont {T.}~\bibnamefont {Wu}}, \bibinfo {author} {\bibfnamefont {B.}~\bibnamefont {Midya}}, \bibinfo {author} {\bibfnamefont {S.}~\bibnamefont {Longhi}},\ and\ \bibinfo {author} {\bibfnamefont {L.}~\bibnamefont {Feng}},\ }\bibfield  {title} {\bibinfo {title} {{Non-Hermitian topological light steering}},\ }\href {https://doi.org/10.1126/science.aay1064} {\bibfield  {journal} {\bibinfo  {journal} {Science}\ }\textbf {\bibinfo {volume} {365}},\ \bibinfo {pages} {1163} (\bibinfo {year} {2019})}\BibitemShut {NoStop}%
\bibitem [{\citenamefont {Brandenbourger}\ \emph {et~al.}(2019)\citenamefont {Brandenbourger}, \citenamefont {Locsin}, \citenamefont {Lerner},\ and\ \citenamefont {Coulais}}]{Brandenbourger-19-skin-exp}%
  \BibitemOpen
  \bibfield  {author} {\bibinfo {author} {\bibfnamefont {M.}~\bibnamefont {Brandenbourger}}, \bibinfo {author} {\bibfnamefont {X.}~\bibnamefont {Locsin}}, \bibinfo {author} {\bibfnamefont {E.}~\bibnamefont {Lerner}},\ and\ \bibinfo {author} {\bibfnamefont {C.}~\bibnamefont {Coulais}},\ }\bibfield  {title} {\bibinfo {title} {{Non-reciprocal robotic metamaterials}},\ }\href {https://doi.org/10.1038/s41467-019-12599-3} {\bibfield  {journal} {\bibinfo  {journal} {Nat. Commun.}\ }\textbf {\bibinfo {volume} {10}},\ \bibinfo {pages} {4608} (\bibinfo {year} {2019})}\BibitemShut {NoStop}%
\bibitem [{\citenamefont {Ghatak}\ \emph {et~al.}(2020)\citenamefont {Ghatak}, \citenamefont {Brandenbourger}, \citenamefont {van Wezel},\ and\ \citenamefont {Coulais}}]{Ghatak-19-skin-exp}%
  \BibitemOpen
  \bibfield  {author} {\bibinfo {author} {\bibfnamefont {A.}~\bibnamefont {Ghatak}}, \bibinfo {author} {\bibfnamefont {M.}~\bibnamefont {Brandenbourger}}, \bibinfo {author} {\bibfnamefont {J.}~\bibnamefont {van Wezel}},\ and\ \bibinfo {author} {\bibfnamefont {C.}~\bibnamefont {Coulais}},\ }\bibfield  {title} {\bibinfo {title} {{Observation of non-Hermitian topology and its bulk-edge correspondence in an active mechanical metamaterial}},\ }\href {https://doi.org/10.1073/pnas.2010580117} {\bibfield  {journal} {\bibinfo  {journal} {Proc. Natl. Acad. Sci. USA}\ }\textbf {\bibinfo {volume} {117}},\ \bibinfo {pages} {29561} (\bibinfo {year} {2020})}\BibitemShut {NoStop}%
\bibitem [{\citenamefont {Helbig}\ \emph {et~al.}(2020)\citenamefont {Helbig}, \citenamefont {Hofmann}, \citenamefont {Imhof}, \citenamefont {Abdelghany}, \citenamefont {Kiessling}, \citenamefont {Molenkamp}, \citenamefont {Lee}, \citenamefont {Szameit}, \citenamefont {Greiter},\ and\ \citenamefont {Thomale}}]{Helbig-19-skin-exp}%
  \BibitemOpen
  \bibfield  {author} {\bibinfo {author} {\bibfnamefont {T.}~\bibnamefont {Helbig}}, \bibinfo {author} {\bibfnamefont {T.}~\bibnamefont {Hofmann}}, \bibinfo {author} {\bibfnamefont {S.}~\bibnamefont {Imhof}}, \bibinfo {author} {\bibfnamefont {M.}~\bibnamefont {Abdelghany}}, \bibinfo {author} {\bibfnamefont {T.}~\bibnamefont {Kiessling}}, \bibinfo {author} {\bibfnamefont {L.~W.}\ \bibnamefont {Molenkamp}}, \bibinfo {author} {\bibfnamefont {C.~H.}\ \bibnamefont {Lee}}, \bibinfo {author} {\bibfnamefont {A.}~\bibnamefont {Szameit}}, \bibinfo {author} {\bibfnamefont {M.}~\bibnamefont {Greiter}},\ and\ \bibinfo {author} {\bibfnamefont {R.}~\bibnamefont {Thomale}},\ }\bibfield  {title} {\bibinfo {title} {{Generalized bulk-boundary correspondence in non-Hermitian topolectrical circuits}},\ }\href {https://doi.org/10.1038/s41567-020-0922-9} {\bibfield  {journal} {\bibinfo  {journal} {Nat. Phys.}\ }\textbf {\bibinfo {volume} {16}},\ \bibinfo {pages} {747} (\bibinfo {year} {2020})}\BibitemShut {NoStop}%
\bibitem [{\citenamefont {Hofmann}\ \emph {et~al.}(2020)\citenamefont {Hofmann}, \citenamefont {Helbig}, \citenamefont {Schindler}, \citenamefont {Salgo}, \citenamefont {Brzezi\'nska}, \citenamefont {Greiter}, \citenamefont {Kiessling}, \citenamefont {Wolf}, \citenamefont {Vollhardt}, \citenamefont {Kaba\v{s}i}, \citenamefont {Lee}, \citenamefont {Bilu\v{s}i\'c}, \citenamefont {Thomale},\ and\ \citenamefont {Neupert}}]{Hofmann-19-skin-exp}%
  \BibitemOpen
  \bibfield  {author} {\bibinfo {author} {\bibfnamefont {T.}~\bibnamefont {Hofmann}}, \bibinfo {author} {\bibfnamefont {T.}~\bibnamefont {Helbig}}, \bibinfo {author} {\bibfnamefont {F.}~\bibnamefont {Schindler}}, \bibinfo {author} {\bibfnamefont {N.}~\bibnamefont {Salgo}}, \bibinfo {author} {\bibfnamefont {M.}~\bibnamefont {Brzezi\'nska}}, \bibinfo {author} {\bibfnamefont {M.}~\bibnamefont {Greiter}}, \bibinfo {author} {\bibfnamefont {T.}~\bibnamefont {Kiessling}}, \bibinfo {author} {\bibfnamefont {D.}~\bibnamefont {Wolf}}, \bibinfo {author} {\bibfnamefont {A.}~\bibnamefont {Vollhardt}}, \bibinfo {author} {\bibfnamefont {A.}~\bibnamefont {Kaba\v{s}i}}, \bibinfo {author} {\bibfnamefont {C.~H.}\ \bibnamefont {Lee}}, \bibinfo {author} {\bibfnamefont {A.}~\bibnamefont {Bilu\v{s}i\'c}}, \bibinfo {author} {\bibfnamefont {R.}~\bibnamefont {Thomale}},\ and\ \bibinfo {author} {\bibfnamefont {T.}~\bibnamefont {Neupert}},\ }\bibfield  {title} {\bibinfo {title} {{Reciprocal skin effect and its realization in a
  topolectrical circuit}},\ }\href {https://doi.org/10.1103/PhysRevResearch.2.023265} {\bibfield  {journal} {\bibinfo  {journal} {Phys. Rev. Research}\ }\textbf {\bibinfo {volume} {2}},\ \bibinfo {pages} {023265} (\bibinfo {year} {2020})}\BibitemShut {NoStop}%
\bibitem [{\citenamefont {Xiao}\ \emph {et~al.}(2020)\citenamefont {Xiao}, \citenamefont {Deng}, \citenamefont {Wang}, \citenamefont {Zhu}, \citenamefont {Wang}, \citenamefont {Yi},\ and\ \citenamefont {Xue}}]{Xiao-19-skin-exp}%
  \BibitemOpen
  \bibfield  {author} {\bibinfo {author} {\bibfnamefont {L.}~\bibnamefont {Xiao}}, \bibinfo {author} {\bibfnamefont {T.}~\bibnamefont {Deng}}, \bibinfo {author} {\bibfnamefont {K.}~\bibnamefont {Wang}}, \bibinfo {author} {\bibfnamefont {G.}~\bibnamefont {Zhu}}, \bibinfo {author} {\bibfnamefont {Z.}~\bibnamefont {Wang}}, \bibinfo {author} {\bibfnamefont {W.}~\bibnamefont {Yi}},\ and\ \bibinfo {author} {\bibfnamefont {P.}~\bibnamefont {Xue}},\ }\bibfield  {title} {\bibinfo {title} {{Non-Hermitian bulk-boundary correspondence in quantum dynamics}},\ }\href {https://doi.org/10.1038/s41567-020-0836-6} {\bibfield  {journal} {\bibinfo  {journal} {Nat. Phys.}\ }\textbf {\bibinfo {volume} {16}},\ \bibinfo {pages} {761} (\bibinfo {year} {2020})}\BibitemShut {NoStop}%
\bibitem [{\citenamefont {Weidemann}\ \emph {et~al.}(2020)\citenamefont {Weidemann}, \citenamefont {Kremer}, \citenamefont {Helbig}, \citenamefont {Hofmann}, \citenamefont {Stegmaier}, \citenamefont {Greiter}, \citenamefont {Thomale},\ and\ \citenamefont {Szameit}}]{Weidemann-20-skin-exp}%
  \BibitemOpen
  \bibfield  {author} {\bibinfo {author} {\bibfnamefont {S.}~\bibnamefont {Weidemann}}, \bibinfo {author} {\bibfnamefont {M.}~\bibnamefont {Kremer}}, \bibinfo {author} {\bibfnamefont {T.}~\bibnamefont {Helbig}}, \bibinfo {author} {\bibfnamefont {T.}~\bibnamefont {Hofmann}}, \bibinfo {author} {\bibfnamefont {A.}~\bibnamefont {Stegmaier}}, \bibinfo {author} {\bibfnamefont {M.}~\bibnamefont {Greiter}}, \bibinfo {author} {\bibfnamefont {R.}~\bibnamefont {Thomale}},\ and\ \bibinfo {author} {\bibfnamefont {A.}~\bibnamefont {Szameit}},\ }\bibfield  {title} {\bibinfo {title} {{Topological funneling of light}},\ }\href {https://doi.org/10.1126/science.aaz8727} {\bibfield  {journal} {\bibinfo  {journal} {Science}\ }\textbf {\bibinfo {volume} {368}},\ \bibinfo {pages} {311} (\bibinfo {year} {2020})}\BibitemShut {NoStop}%
\bibitem [{\citenamefont {Gou}\ \emph {et~al.}(2020)\citenamefont {Gou}, \citenamefont {Chen}, \citenamefont {Xie}, \citenamefont {Xiao}, \citenamefont {Deng}, \citenamefont {Gadway}, \citenamefont {Yi},\ and\ \citenamefont {Yan}}]{Gou-20}%
  \BibitemOpen
  \bibfield  {author} {\bibinfo {author} {\bibfnamefont {W.}~\bibnamefont {Gou}}, \bibinfo {author} {\bibfnamefont {T.}~\bibnamefont {Chen}}, \bibinfo {author} {\bibfnamefont {D.}~\bibnamefont {Xie}}, \bibinfo {author} {\bibfnamefont {T.}~\bibnamefont {Xiao}}, \bibinfo {author} {\bibfnamefont {T.-S.}\ \bibnamefont {Deng}}, \bibinfo {author} {\bibfnamefont {B.}~\bibnamefont {Gadway}}, \bibinfo {author} {\bibfnamefont {W.}~\bibnamefont {Yi}},\ and\ \bibinfo {author} {\bibfnamefont {B.}~\bibnamefont {Yan}},\ }\bibfield  {title} {\bibinfo {title} {{Tunable Nonreciprocal Quantum Transport through a Dissipative Aharonov-Bohm Ring in Ultracold Atoms}},\ }\href {https://doi.org/10.1103/PhysRevLett.124.070402} {\bibfield  {journal} {\bibinfo  {journal} {Phys. Rev. Lett.}\ }\textbf {\bibinfo {volume} {124}},\ \bibinfo {pages} {070402} (\bibinfo {year} {2020})}\BibitemShut {NoStop}%
\bibitem [{\citenamefont {Liang}\ \emph {et~al.}(2022)\citenamefont {Liang}, \citenamefont {Xie}, \citenamefont {Dong}, \citenamefont {Li}, \citenamefont {Li}, \citenamefont {Gadway}, \citenamefont {Yi},\ and\ \citenamefont {Yan}}]{Liang-22}%
  \BibitemOpen
  \bibfield  {author} {\bibinfo {author} {\bibfnamefont {Q.}~\bibnamefont {Liang}}, \bibinfo {author} {\bibfnamefont {D.}~\bibnamefont {Xie}}, \bibinfo {author} {\bibfnamefont {Z.}~\bibnamefont {Dong}}, \bibinfo {author} {\bibfnamefont {H.}~\bibnamefont {Li}}, \bibinfo {author} {\bibfnamefont {H.}~\bibnamefont {Li}}, \bibinfo {author} {\bibfnamefont {B.}~\bibnamefont {Gadway}}, \bibinfo {author} {\bibfnamefont {W.}~\bibnamefont {Yi}},\ and\ \bibinfo {author} {\bibfnamefont {B.}~\bibnamefont {Yan}},\ }\bibfield  {title} {\bibinfo {title} {{Dynamic Signatures of Non-Hermitian Skin Effect and Topology in Ultracold Atoms}},\ }\href {https://doi.org/10.1103/PhysRevLett.129.070401} {\bibfield  {journal} {\bibinfo  {journal} {Phys. Rev. Lett.}\ }\textbf {\bibinfo {volume} {129}},\ \bibinfo {pages} {070401} (\bibinfo {year} {2022})}\BibitemShut {NoStop}%
\bibitem [{\citenamefont {Wang}\ \emph {et~al.}(2021{\natexlab{a}})\citenamefont {Wang}, \citenamefont {Dutt}, \citenamefont {Yang}, \citenamefont {Wojcik}, \citenamefont {Vu\v{c}kovi\'{c}},\ and\ \citenamefont {Fan}}]{Wang-21S}%
  \BibitemOpen
  \bibfield  {author} {\bibinfo {author} {\bibfnamefont {K.}~\bibnamefont {Wang}}, \bibinfo {author} {\bibfnamefont {A.}~\bibnamefont {Dutt}}, \bibinfo {author} {\bibfnamefont {K.~Y.}\ \bibnamefont {Yang}}, \bibinfo {author} {\bibfnamefont {C.~C.}\ \bibnamefont {Wojcik}}, \bibinfo {author} {\bibfnamefont {J.}~\bibnamefont {Vu\v{c}kovi\'{c}}},\ and\ \bibinfo {author} {\bibfnamefont {S.}~\bibnamefont {Fan}},\ }\bibfield  {title} {\bibinfo {title} {{Generating arbitrary topological windings of a non-Hermitian band}},\ }\href {https://doi.org/10.1126/science.abf6568} {\bibfield  {journal} {\bibinfo  {journal} {Science}\ }\textbf {\bibinfo {volume} {371}},\ \bibinfo {pages} {1240} (\bibinfo {year} {2021}{\natexlab{a}})}\BibitemShut {NoStop}%
\bibitem [{\citenamefont {Wang}\ \emph {et~al.}(2021{\natexlab{b}})\citenamefont {Wang}, \citenamefont {Dutt}, \citenamefont {Wojcik},\ and\ \citenamefont {Fan}}]{Wang-21N}%
  \BibitemOpen
  \bibfield  {author} {\bibinfo {author} {\bibfnamefont {K.}~\bibnamefont {Wang}}, \bibinfo {author} {\bibfnamefont {A.}~\bibnamefont {Dutt}}, \bibinfo {author} {\bibfnamefont {C.~C.}\ \bibnamefont {Wojcik}},\ and\ \bibinfo {author} {\bibfnamefont {S.}~\bibnamefont {Fan}},\ }\bibfield  {title} {\bibinfo {title} {{Topological complex-energy braiding of non-Hermitian bands}},\ }\href {https://doi.org/10.1038/s41586-021-03848-x} {\bibfield  {journal} {\bibinfo  {journal} {Nature}\ }\textbf {\bibinfo {volume} {598}},\ \bibinfo {pages} {59} (\bibinfo {year} {2021}{\natexlab{b}})}\BibitemShut {NoStop}%
\bibitem [{\citenamefont {Zhang}\ \emph {et~al.}(2021{\natexlab{a}})\citenamefont {Zhang}, \citenamefont {Ouyang}, \citenamefont {Huang}, \citenamefont {Wang}, \citenamefont {Zhang}, \citenamefont {Yu}, \citenamefont {Chang}, \citenamefont {Liu}, \citenamefont {Deng},\ and\ \citenamefont {Duan}}]{Wengang-21}%
  \BibitemOpen
  \bibfield  {author} {\bibinfo {author} {\bibfnamefont {W.}~\bibnamefont {Zhang}}, \bibinfo {author} {\bibfnamefont {X.}~\bibnamefont {Ouyang}}, \bibinfo {author} {\bibfnamefont {X.}~\bibnamefont {Huang}}, \bibinfo {author} {\bibfnamefont {X.}~\bibnamefont {Wang}}, \bibinfo {author} {\bibfnamefont {H.}~\bibnamefont {Zhang}}, \bibinfo {author} {\bibfnamefont {Y.}~\bibnamefont {Yu}}, \bibinfo {author} {\bibfnamefont {X.}~\bibnamefont {Chang}}, \bibinfo {author} {\bibfnamefont {Y.}~\bibnamefont {Liu}}, \bibinfo {author} {\bibfnamefont {D.-L.}\ \bibnamefont {Deng}},\ and\ \bibinfo {author} {\bibfnamefont {L.-M.}\ \bibnamefont {Duan}},\ }\bibfield  {title} {\bibinfo {title} {{Observation of Non-Hermitian Topology with Nonunitary Dynamics of Solid-State Spins}},\ }\href {https://doi.org/10.1103/PhysRevLett.127.090501} {\bibfield  {journal} {\bibinfo  {journal} {Phys. Rev. Lett.}\ }\textbf {\bibinfo {volume} {127}},\ \bibinfo {pages} {090501} (\bibinfo {year} {2021}{\natexlab{a}})}\BibitemShut {NoStop}%
\bibitem [{\citenamefont {Palacios}\ \emph {et~al.}(2020)\citenamefont {Palacios}, \citenamefont {Tchoumakov}, \citenamefont {Guix}, \citenamefont {S\'anchez},\ and\ \citenamefont {Grushin}}]{Palacios-21}%
  \BibitemOpen
  \bibfield  {author} {\bibinfo {author} {\bibfnamefont {L.~S.}\ \bibnamefont {Palacios}}, \bibinfo {author} {\bibfnamefont {S.}~\bibnamefont {Tchoumakov}}, \bibinfo {author} {\bibfnamefont {M.}~\bibnamefont {Guix}}, \bibinfo {author} {\bibfnamefont {I.~P.~S.}\ \bibnamefont {S\'anchez}},\ and\ \bibinfo {author} {\bibfnamefont {A.~G.}\ \bibnamefont {Grushin}},\ }\bibfield  {title} {\bibinfo {title} {{Guided accumulation of active particles by topological design of a second-order skin effect}},\ }\href {https://doi.org/10.1038/s41467-021-24948-2} {\bibfield  {journal} {\bibinfo  {journal} {Nat. Commun.}\ }\textbf {\bibinfo {volume} {12}},\ \bibinfo {pages} {4691} (\bibinfo {year} {2020})}\BibitemShut {NoStop}%
\bibitem [{\citenamefont {Hu}\ \emph {et~al.}(2021)\citenamefont {Hu}, \citenamefont {Zhang}, \citenamefont {Zhang}, \citenamefont {Zheng}, \citenamefont {Xiong}, \citenamefont {Yue}, \citenamefont {Wang}, \citenamefont {Xu}, \citenamefont {Cheng}, \citenamefont {Liu},\ and\ \citenamefont {Christensen}}]{Hu-21}%
  \BibitemOpen
  \bibfield  {author} {\bibinfo {author} {\bibfnamefont {B.}~\bibnamefont {Hu}}, \bibinfo {author} {\bibfnamefont {Z.}~\bibnamefont {Zhang}}, \bibinfo {author} {\bibfnamefont {H.}~\bibnamefont {Zhang}}, \bibinfo {author} {\bibfnamefont {L.}~\bibnamefont {Zheng}}, \bibinfo {author} {\bibfnamefont {W.}~\bibnamefont {Xiong}}, \bibinfo {author} {\bibfnamefont {Z.}~\bibnamefont {Yue}}, \bibinfo {author} {\bibfnamefont {X.}~\bibnamefont {Wang}}, \bibinfo {author} {\bibfnamefont {J.}~\bibnamefont {Xu}}, \bibinfo {author} {\bibfnamefont {Y.}~\bibnamefont {Cheng}}, \bibinfo {author} {\bibfnamefont {X.}~\bibnamefont {Liu}},\ and\ \bibinfo {author} {\bibfnamefont {J.}~\bibnamefont {Christensen}},\ }\bibfield  {title} {\bibinfo {title} {{Non-Hermitian topological whispering gallery}},\ }\href {https://doi.org/10.1038/s41586-021-03833-4} {\bibfield  {journal} {\bibinfo  {journal} {Nature}\ }\textbf {\bibinfo {volume} {597}},\ \bibinfo {pages} {655} (\bibinfo {year} {2021})}\BibitemShut {NoStop}%
\bibitem [{\citenamefont {Zhang}\ \emph {et~al.}(2021{\natexlab{b}})\citenamefont {Zhang}, \citenamefont {Tian}, \citenamefont {Jiang}, \citenamefont {Lu},\ and\ \citenamefont {Chen}}]{Zhang-21}%
  \BibitemOpen
  \bibfield  {author} {\bibinfo {author} {\bibfnamefont {X.}~\bibnamefont {Zhang}}, \bibinfo {author} {\bibfnamefont {Y.}~\bibnamefont {Tian}}, \bibinfo {author} {\bibfnamefont {J.-H.}\ \bibnamefont {Jiang}}, \bibinfo {author} {\bibfnamefont {M.-H.}\ \bibnamefont {Lu}},\ and\ \bibinfo {author} {\bibfnamefont {Y.-F.}\ \bibnamefont {Chen}},\ }\bibfield  {title} {\bibinfo {title} {{Observation of higher-order non-Hermitian skin effect}},\ }\href {https://doi.org/10.1038/s41467-021-25716-y} {\bibfield  {journal} {\bibinfo  {journal} {Nat. Commun.}\ }\textbf {\bibinfo {volume} {12}},\ \bibinfo {pages} {5377} (\bibinfo {year} {2021}{\natexlab{b}})}\BibitemShut {NoStop}%
\bibitem [{\citenamefont {Wang}\ \emph {et~al.}(2022)\citenamefont {Wang}, \citenamefont {Wang},\ and\ \citenamefont {Ma}}]{WangWangMa-22}%
  \BibitemOpen
  \bibfield  {author} {\bibinfo {author} {\bibfnamefont {W.}~\bibnamefont {Wang}}, \bibinfo {author} {\bibfnamefont {X.}~\bibnamefont {Wang}},\ and\ \bibinfo {author} {\bibfnamefont {G.}~\bibnamefont {Ma}},\ }\bibfield  {title} {\bibinfo {title} {{Non-Hermitian morphing of topological modes}},\ }\href {https://doi.org/https://doi.org/10.1038/s41586-022-04929-1} {\bibfield  {journal} {\bibinfo  {journal} {Nature}\ }\textbf {\bibinfo {volume} {608}},\ \bibinfo {pages} {50} (\bibinfo {year} {2022})}\BibitemShut {NoStop}%
\bibitem [{\citenamefont {Wang}\ \emph {et~al.}(2023{\natexlab{b}})\citenamefont {Wang}, \citenamefont {Meng},\ and\ \citenamefont {Chen}}]{Wang-23mech}%
  \BibitemOpen
  \bibfield  {author} {\bibinfo {author} {\bibfnamefont {A.}~\bibnamefont {Wang}}, \bibinfo {author} {\bibfnamefont {Z.}~\bibnamefont {Meng}},\ and\ \bibinfo {author} {\bibfnamefont {C.~Q.}\ \bibnamefont {Chen}},\ }\bibfield  {title} {\bibinfo {title} {{Non-Hermitian topology in static mechanical metamaterials}},\ }\href {https://doi.org/10.1126/sciadv.adf7299} {\bibfield  {journal} {\bibinfo  {journal} {Sci. Adv.}\ }\textbf {\bibinfo {volume} {9}},\ \bibinfo {pages} {eadf7299} (\bibinfo {year} {2023}{\natexlab{b}})}\BibitemShut {NoStop}%
\bibitem [{\citenamefont {Liu}\ \emph {et~al.}(2024)\citenamefont {Liu}, \citenamefont {Mandal}, \citenamefont {Zhou}, \citenamefont {Xi}, \citenamefont {Banerjee}, \citenamefont {Hu}, \citenamefont {Wei}, \citenamefont {Wang}, \citenamefont {Wang}, \citenamefont {Gao}, \citenamefont {Chen}, \citenamefont {Yang}, \citenamefont {Chong},\ and\ \citenamefont {Zhang}}]{Liu-24}%
  \BibitemOpen
  \bibfield  {author} {\bibinfo {author} {\bibfnamefont {G.-G.}\ \bibnamefont {Liu}}, \bibinfo {author} {\bibfnamefont {S.}~\bibnamefont {Mandal}}, \bibinfo {author} {\bibfnamefont {P.}~\bibnamefont {Zhou}}, \bibinfo {author} {\bibfnamefont {X.}~\bibnamefont {Xi}}, \bibinfo {author} {\bibfnamefont {R.}~\bibnamefont {Banerjee}}, \bibinfo {author} {\bibfnamefont {Y.-H.}\ \bibnamefont {Hu}}, \bibinfo {author} {\bibfnamefont {M.}~\bibnamefont {Wei}}, \bibinfo {author} {\bibfnamefont {M.}~\bibnamefont {Wang}}, \bibinfo {author} {\bibfnamefont {Q.}~\bibnamefont {Wang}}, \bibinfo {author} {\bibfnamefont {Z.}~\bibnamefont {Gao}}, \bibinfo {author} {\bibfnamefont {H.}~\bibnamefont {Chen}}, \bibinfo {author} {\bibfnamefont {Y.}~\bibnamefont {Yang}}, \bibinfo {author} {\bibfnamefont {Y.}~\bibnamefont {Chong}},\ and\ \bibinfo {author} {\bibfnamefont {B.}~\bibnamefont {Zhang}},\ }\bibfield  {title} {\bibinfo {title} {{Localization of Chiral Edge States by the Non-Hermitian Skin Effect}},\ }\href
  {https://doi.org/10.1103/PhysRevLett.132.113802} {\bibfield  {journal} {\bibinfo  {journal} {Phys. Rev. Lett.}\ }\textbf {\bibinfo {volume} {132}},\ \bibinfo {pages} {113802} (\bibinfo {year} {2024})}\BibitemShut {NoStop}%
\bibitem [{\citenamefont {Ochkan}\ \emph {et~al.}(2024)\citenamefont {Ochkan}, \citenamefont {Chaturvedi}, \citenamefont {K\"onye}, \citenamefont {Veyrat}, \citenamefont {Giraud}, \citenamefont {Mailly}, \citenamefont {Cavanna}, \citenamefont {Gennser}, \citenamefont {Hankiewicz}, \citenamefont {B\"uchner}, \citenamefont {van~den Brink}, \citenamefont {Dufouleur},\ and\ \citenamefont {Fulga}}]{Ochkan-24}%
  \BibitemOpen
  \bibfield  {author} {\bibinfo {author} {\bibfnamefont {K.}~\bibnamefont {Ochkan}}, \bibinfo {author} {\bibfnamefont {R.}~\bibnamefont {Chaturvedi}}, \bibinfo {author} {\bibfnamefont {V.}~\bibnamefont {K\"onye}}, \bibinfo {author} {\bibfnamefont {L.}~\bibnamefont {Veyrat}}, \bibinfo {author} {\bibfnamefont {R.}~\bibnamefont {Giraud}}, \bibinfo {author} {\bibfnamefont {D.}~\bibnamefont {Mailly}}, \bibinfo {author} {\bibfnamefont {A.}~\bibnamefont {Cavanna}}, \bibinfo {author} {\bibfnamefont {U.}~\bibnamefont {Gennser}}, \bibinfo {author} {\bibfnamefont {E.~M.}\ \bibnamefont {Hankiewicz}}, \bibinfo {author} {\bibfnamefont {B.}~\bibnamefont {B\"uchner}}, \bibinfo {author} {\bibfnamefont {J.}~\bibnamefont {van~den Brink}}, \bibinfo {author} {\bibfnamefont {J.}~\bibnamefont {Dufouleur}},\ and\ \bibinfo {author} {\bibfnamefont {I.~C.}\ \bibnamefont {Fulga}},\ }\bibfield  {title} {\bibinfo {title} {{Non-Hermitian topology in a multi-terminal quantum Hall device}},\ }\href
  {https://doi.org/10.1038/s41567-023-02337-4} {\bibfield  {journal} {\bibinfo  {journal} {Nat. Phys.}\ }\textbf {\bibinfo {volume} {20}},\ \bibinfo {pages} {395} (\bibinfo {year} {2024})}\BibitemShut {NoStop}%
\bibitem [{\citenamefont {Zhao}\ \emph {et~al.}(2025)\citenamefont {Zhao}, \citenamefont {Wang}, \citenamefont {He}, \citenamefont {Poon}, \citenamefont {Pak}, \citenamefont {Liu}, \citenamefont {Ren}, \citenamefont {Liu},\ and\ \citenamefont {Jo}}]{Zhao-25}%
  \BibitemOpen
  \bibfield  {author} {\bibinfo {author} {\bibfnamefont {E.}~\bibnamefont {Zhao}}, \bibinfo {author} {\bibfnamefont {Z.}~\bibnamefont {Wang}}, \bibinfo {author} {\bibfnamefont {C.}~\bibnamefont {He}}, \bibinfo {author} {\bibfnamefont {T.~F.~J.}\ \bibnamefont {Poon}}, \bibinfo {author} {\bibfnamefont {K.~K.}\ \bibnamefont {Pak}}, \bibinfo {author} {\bibfnamefont {Y.-J.}\ \bibnamefont {Liu}}, \bibinfo {author} {\bibfnamefont {P.}~\bibnamefont {Ren}}, \bibinfo {author} {\bibfnamefont {X.-J.}\ \bibnamefont {Liu}},\ and\ \bibinfo {author} {\bibfnamefont {G.-B.}\ \bibnamefont {Jo}},\ }\bibfield  {title} {\bibinfo {title} {{Two-dimensional non-Hermitian skin effect in an ultracold Fermi gas}},\ }\href {https://doi.org/https://doi.org/10.1038/s41586-024-08347-3} {\bibfield  {journal} {\bibinfo  {journal} {Nature}\ }\textbf {\bibinfo {volume} {637}},\ \bibinfo {pages} {565} (\bibinfo {year} {2025})}\BibitemShut {NoStop}%
\bibitem [{\citenamefont {Shen}\ \emph {et~al.}(2025)\citenamefont {Shen}, \citenamefont {Chen}, \citenamefont {Yang},\ and\ \citenamefont {Lee}}]{Shen-25}%
  \BibitemOpen
  \bibfield  {author} {\bibinfo {author} {\bibfnamefont {R.}~\bibnamefont {Shen}}, \bibinfo {author} {\bibfnamefont {T.}~\bibnamefont {Chen}}, \bibinfo {author} {\bibfnamefont {B.}~\bibnamefont {Yang}},\ and\ \bibinfo {author} {\bibfnamefont {C.~H.}\ \bibnamefont {Lee}},\ }\bibfield  {title} {\bibinfo {title} {{Observation of the non-Hermitian skin effect and Fermi skin on a digital quantum computer}},\ }\href {https://doi.org/https://doi.org/10.1038/s41467-025-55953-4} {\bibfield  {journal} {\bibinfo  {journal} {Nat. Commun.}\ }\textbf {\bibinfo {volume} {16}},\ \bibinfo {pages} {1340} (\bibinfo {year} {2025})}\BibitemShut {NoStop}%
\bibitem [{\citenamefont {Wu}\ \emph {et~al.}(2024)\citenamefont {Wu}, \citenamefont {Zheng}, \citenamefont {Liang}, \citenamefont {Ke}, \citenamefont {Lu}, \citenamefont {Deng}, \citenamefont {Huang},\ and\ \citenamefont {Liu}}]{Wu-24}%
  \BibitemOpen
  \bibfield  {author} {\bibinfo {author} {\bibfnamefont {J.}~\bibnamefont {Wu}}, \bibinfo {author} {\bibfnamefont {R.}~\bibnamefont {Zheng}}, \bibinfo {author} {\bibfnamefont {J.}~\bibnamefont {Liang}}, \bibinfo {author} {\bibfnamefont {M.}~\bibnamefont {Ke}}, \bibinfo {author} {\bibfnamefont {J.}~\bibnamefont {Lu}}, \bibinfo {author} {\bibfnamefont {W.}~\bibnamefont {Deng}}, \bibinfo {author} {\bibfnamefont {X.}~\bibnamefont {Huang}},\ and\ \bibinfo {author} {\bibfnamefont {Z.}~\bibnamefont {Liu}},\ }\bibfield  {title} {\bibinfo {title} {{Spin-Dependent Localization of Helical Edge States in a Non-Hermitian Phononic Crystal}},\ }\href {https://doi.org/10.1103/PhysRevLett.133.126601} {\bibfield  {journal} {\bibinfo  {journal} {Phys. Rev. Lett.}\ }\textbf {\bibinfo {volume} {133}},\ \bibinfo {pages} {126601} (\bibinfo {year} {2024})}\BibitemShut {NoStop}%
\bibitem [{\citenamefont {Konotop}\ \emph {et~al.}(2016)\citenamefont {Konotop}, \citenamefont {Yang},\ and\ \citenamefont {Zezyulin}}]{Konotop-review}%
  \BibitemOpen
  \bibfield  {author} {\bibinfo {author} {\bibfnamefont {V.~V.}\ \bibnamefont {Konotop}}, \bibinfo {author} {\bibfnamefont {J.}~\bibnamefont {Yang}},\ and\ \bibinfo {author} {\bibfnamefont {D.~A.}\ \bibnamefont {Zezyulin}},\ }\bibfield  {title} {\bibinfo {title} {{Nonlinear waves in $\mathcal{PT}$-symmetric systems}},\ }\href {https://doi.org/10.1103/RevModPhys.88.035002} {\bibfield  {journal} {\bibinfo  {journal} {Rev. Mod. Phys.}\ }\textbf {\bibinfo {volume} {88}},\ \bibinfo {pages} {035002} (\bibinfo {year} {2016})}\BibitemShut {NoStop}%
\bibitem [{\citenamefont {El-Ganainy}\ \emph {et~al.}(2018)\citenamefont {El-Ganainy}, \citenamefont {Makris}, \citenamefont {Khajavikhan}, \citenamefont {Musslimani}, \citenamefont {Rotter},\ and\ \citenamefont {Christodoulides}}]{Christodoulides-review}%
  \BibitemOpen
  \bibfield  {author} {\bibinfo {author} {\bibfnamefont {R.}~\bibnamefont {El-Ganainy}}, \bibinfo {author} {\bibfnamefont {K.~G.}\ \bibnamefont {Makris}}, \bibinfo {author} {\bibfnamefont {M.}~\bibnamefont {Khajavikhan}}, \bibinfo {author} {\bibfnamefont {Z.~H.}\ \bibnamefont {Musslimani}}, \bibinfo {author} {\bibfnamefont {S.}~\bibnamefont {Rotter}},\ and\ \bibinfo {author} {\bibfnamefont {D.~N.}\ \bibnamefont {Christodoulides}},\ }\bibfield  {title} {\bibinfo {title} {{Non-Hermitian physics and PT symmetry}},\ }\href {https://doi.org/10.1038/nphys4323} {\bibfield  {journal} {\bibinfo  {journal} {Nat. Phys.}\ }\textbf {\bibinfo {volume} {14}},\ \bibinfo {pages} {11} (\bibinfo {year} {2018})}\BibitemShut {NoStop}%
\bibitem [{\citenamefont {Li}\ and\ \citenamefont {Mong}(2021)}]{Li-21}%
  \BibitemOpen
  \bibfield  {author} {\bibinfo {author} {\bibfnamefont {Z.}~\bibnamefont {Li}}\ and\ \bibinfo {author} {\bibfnamefont {R.~S.~K.}\ \bibnamefont {Mong}},\ }\bibfield  {title} {\bibinfo {title} {{Homotopical characterization of non-Hermitian band structures}},\ }\href {https://doi.org/10.1103/PhysRevB.103.155129} {\bibfield  {journal} {\bibinfo  {journal} {Phys. Rev. B}\ }\textbf {\bibinfo {volume} {103}},\ \bibinfo {pages} {155129} (\bibinfo {year} {2021})}\BibitemShut {NoStop}%
\bibitem [{\citenamefont {Wojcik}\ \emph {et~al.}(2020)\citenamefont {Wojcik}, \citenamefont {Sun}, \citenamefont {Bzdu\ifmmode~\check{s}\else \v{s}\fi{}ek},\ and\ \citenamefont {Fan}}]{Wojcik-20}%
  \BibitemOpen
  \bibfield  {author} {\bibinfo {author} {\bibfnamefont {C.~C.}\ \bibnamefont {Wojcik}}, \bibinfo {author} {\bibfnamefont {X.-Q.}\ \bibnamefont {Sun}}, \bibinfo {author} {\bibfnamefont {T.}~\bibnamefont {Bzdu\ifmmode~\check{s}\else \v{s}\fi{}ek}},\ and\ \bibinfo {author} {\bibfnamefont {S.}~\bibnamefont {Fan}},\ }\bibfield  {title} {\bibinfo {title} {{Homotopy characterization of non-Hermitian Hamiltonians}},\ }\href {https://doi.org/10.1103/PhysRevB.101.205417} {\bibfield  {journal} {\bibinfo  {journal} {Phys. Rev. B}\ }\textbf {\bibinfo {volume} {101}},\ \bibinfo {pages} {205417} (\bibinfo {year} {2020})}\BibitemShut {NoStop}%
\bibitem [{\citenamefont {Hu}\ and\ \citenamefont {Zhao}(2021)}]{Hu-21knot}%
  \BibitemOpen
  \bibfield  {author} {\bibinfo {author} {\bibfnamefont {H.}~\bibnamefont {Hu}}\ and\ \bibinfo {author} {\bibfnamefont {E.}~\bibnamefont {Zhao}},\ }\bibfield  {title} {\bibinfo {title} {{Knots and Non-Hermitian Bloch Bands}},\ }\href {https://doi.org/10.1103/PhysRevLett.126.010401} {\bibfield  {journal} {\bibinfo  {journal} {Phys. Rev. Lett.}\ }\textbf {\bibinfo {volume} {126}},\ \bibinfo {pages} {010401} (\bibinfo {year} {2021})}\BibitemShut {NoStop}%
\bibitem [{\citenamefont {Ryu}\ \emph {et~al.}(2024)\citenamefont {Ryu}, \citenamefont {Han}, \citenamefont {Yi}, \citenamefont {Park},\ and\ \citenamefont {Park}}]{Ryu-24}%
  \BibitemOpen
  \bibfield  {author} {\bibinfo {author} {\bibfnamefont {J.-W.}\ \bibnamefont {Ryu}}, \bibinfo {author} {\bibfnamefont {J.-H.}\ \bibnamefont {Han}}, \bibinfo {author} {\bibfnamefont {C.-H.}\ \bibnamefont {Yi}}, \bibinfo {author} {\bibfnamefont {M.~J.}\ \bibnamefont {Park}},\ and\ \bibinfo {author} {\bibfnamefont {H.~C.}\ \bibnamefont {Park}},\ }\bibfield  {title} {\bibinfo {title} {{Exceptional classifications of non-Hermitian systems}},\ }\href {https://doi.org/https://doi.org/10.1038/s42005-024-01595-9} {\bibfield  {journal} {\bibinfo  {journal} {Commun. Phys.}\ }\textbf {\bibinfo {volume} {7}},\ \bibinfo {pages} {109} (\bibinfo {year} {2024})}\BibitemShut {NoStop}%
\bibitem [{\citenamefont {Yang}\ \emph {et~al.}(2024)\citenamefont {Yang}, \citenamefont {Li}, \citenamefont {K\"onig}, \citenamefont {R{\o}dland}, \citenamefont {St{\aa}lhammar},\ and\ \citenamefont {Bergholtz}}]{Yang-24}%
  \BibitemOpen
  \bibfield  {author} {\bibinfo {author} {\bibfnamefont {K.}~\bibnamefont {Yang}}, \bibinfo {author} {\bibfnamefont {Z.}~\bibnamefont {Li}}, \bibinfo {author} {\bibfnamefont {J.~L.~K.}\ \bibnamefont {K\"onig}}, \bibinfo {author} {\bibfnamefont {L.}~\bibnamefont {R{\o}dland}}, \bibinfo {author} {\bibfnamefont {M.}~\bibnamefont {St{\aa}lhammar}},\ and\ \bibinfo {author} {\bibfnamefont {E.~J.}\ \bibnamefont {Bergholtz}},\ }\bibfield  {title} {\bibinfo {title} {{Homotopy, symmetry, and non-Hermitian band topology}},\ }\href {https://doi.org/10.1088/1361-6633/ad4e64} {\bibfield  {journal} {\bibinfo  {journal} {Rep. Prog. Phys.}\ }\textbf {\bibinfo {volume} {87}},\ \bibinfo {pages} {078002} (\bibinfo {year} {2024})}\BibitemShut {NoStop}%
\bibitem [{\citenamefont {Yang}\ \emph {et~al.}()\citenamefont {Yang}, \citenamefont {Li}, \citenamefont {Xue}, \citenamefont {Bergholtz},\ and\ \citenamefont {Brouwer}}]{Yang-25}%
  \BibitemOpen
  \bibfield  {author} {\bibinfo {author} {\bibfnamefont {K.}~\bibnamefont {Yang}}, \bibinfo {author} {\bibfnamefont {Z.}~\bibnamefont {Li}}, \bibinfo {author} {\bibfnamefont {P.}~\bibnamefont {Xue}}, \bibinfo {author} {\bibfnamefont {E.~J.}\ \bibnamefont {Bergholtz}},\ and\ \bibinfo {author} {\bibfnamefont {P.~W.}\ \bibnamefont {Brouwer}},\ }\bibfield  {title} {\bibinfo {title} {{Spontaneous Chern-Euler Duality Transitions}},\ }\Eprint {https://arxiv.org/abs/2503.21861} {arXiv:2503.21861} \BibitemShut {NoStop}%
\bibitem [{\citenamefont {Yang}\ and\ \citenamefont {Hu}(2019)}]{Yang-19}%
  \BibitemOpen
  \bibfield  {author} {\bibinfo {author} {\bibfnamefont {Z.}~\bibnamefont {Yang}}\ and\ \bibinfo {author} {\bibfnamefont {J.}~\bibnamefont {Hu}},\ }\bibfield  {title} {\bibinfo {title} {{Non-Hermitian Hopf-link exceptional line semimetals}},\ }\href {https://doi.org/10.1103/PhysRevB.99.081102} {\bibfield  {journal} {\bibinfo  {journal} {Phys. Rev. B}\ }\textbf {\bibinfo {volume} {99}},\ \bibinfo {pages} {081102} (\bibinfo {year} {2019})}\BibitemShut {NoStop}%
\bibitem [{\citenamefont {Yang}\ \emph {et~al.}(2020{\natexlab{b}})\citenamefont {Yang}, \citenamefont {Chiu}, \citenamefont {Fang},\ and\ \citenamefont {Hu}}]{Yang-20Jones}%
  \BibitemOpen
  \bibfield  {author} {\bibinfo {author} {\bibfnamefont {Z.}~\bibnamefont {Yang}}, \bibinfo {author} {\bibfnamefont {C.-K.}\ \bibnamefont {Chiu}}, \bibinfo {author} {\bibfnamefont {C.}~\bibnamefont {Fang}},\ and\ \bibinfo {author} {\bibfnamefont {J.}~\bibnamefont {Hu}},\ }\bibfield  {title} {\bibinfo {title} {{Jones Polynomial and Knot Transitions in Hermitian and non-Hermitian Topological Semimetals}},\ }\href {https://doi.org/10.1103/PhysRevLett.124.186402} {\bibfield  {journal} {\bibinfo  {journal} {Phys. Rev. Lett.}\ }\textbf {\bibinfo {volume} {124}},\ \bibinfo {pages} {186402} (\bibinfo {year} {2020}{\natexlab{b}})}\BibitemShut {NoStop}%
\bibitem [{\citenamefont {He}\ and\ \citenamefont {Chien}(2020)}]{He-20}%
  \BibitemOpen
  \bibfield  {author} {\bibinfo {author} {\bibfnamefont {Y.}~\bibnamefont {He}}\ and\ \bibinfo {author} {\bibfnamefont {C.-C.}\ \bibnamefont {Chien}},\ }\bibfield  {title} {\bibinfo {title} {{Non-Hermitian three-dimensional two-band Hopf insulator}},\ }\href {https://doi.org/10.1103/PhysRevB.102.035101} {\bibfield  {journal} {\bibinfo  {journal} {Phys. Rev. B}\ }\textbf {\bibinfo {volume} {102}},\ \bibinfo {pages} {035101} (\bibinfo {year} {2020})}\BibitemShut {NoStop}%
\bibitem [{\citenamefont {Kim}\ \emph {et~al.}(2023)\citenamefont {Kim}, \citenamefont {Park}, \citenamefont {Kyung}, \citenamefont {Lee}, \citenamefont {Ryu}, \citenamefont {You}, \citenamefont {Zhang}, \citenamefont {Min},\ and\ \citenamefont {Park}}]{Kim-23}%
  \BibitemOpen
  \bibfield  {author} {\bibinfo {author} {\bibfnamefont {Y.}~\bibnamefont {Kim}}, \bibinfo {author} {\bibfnamefont {H.~C.}\ \bibnamefont {Park}}, \bibinfo {author} {\bibfnamefont {M.}~\bibnamefont {Kyung}}, \bibinfo {author} {\bibfnamefont {K.}~\bibnamefont {Lee}}, \bibinfo {author} {\bibfnamefont {J.-W.}\ \bibnamefont {Ryu}}, \bibinfo {author} {\bibfnamefont {O.}~\bibnamefont {You}}, \bibinfo {author} {\bibfnamefont {S.}~\bibnamefont {Zhang}}, \bibinfo {author} {\bibfnamefont {B.}~\bibnamefont {Min}},\ and\ \bibinfo {author} {\bibfnamefont {M.~J.}\ \bibnamefont {Park}},\ }\bibfield  {title} {\bibinfo {title} {{Realization of non-Hermitian Hopf bundle matter}},\ }\href {https://doi.org/https://doi.org/10.1038/s42005-023-01381-z} {\bibfield  {journal} {\bibinfo  {journal} {Commun. Phys.}\ }\textbf {\bibinfo {volume} {6}},\ \bibinfo {pages} {273} (\bibinfo {year} {2023})}\BibitemShut {NoStop}%
\bibitem [{\citenamefont {Pak}\ \emph {et~al.}(2024)\citenamefont {Pak}, \citenamefont {Yeom}, \citenamefont {Verma},\ and\ \citenamefont {Park}}]{Pak-24}%
  \BibitemOpen
  \bibfield  {author} {\bibinfo {author} {\bibfnamefont {S.}~\bibnamefont {Pak}}, \bibinfo {author} {\bibfnamefont {C.~H.}\ \bibnamefont {Yeom}}, \bibinfo {author} {\bibfnamefont {S.}~\bibnamefont {Verma}},\ and\ \bibinfo {author} {\bibfnamefont {M.~J.}\ \bibnamefont {Park}},\ }\bibfield  {title} {\bibinfo {title} {{$\mathcal{PT}$-symmetric non-Hermitian Hopf metal}},\ }\href {https://doi.org/10.1103/PhysRevResearch.6.L012053} {\bibfield  {journal} {\bibinfo  {journal} {Phys. Rev. Research}\ }\textbf {\bibinfo {volume} {6}},\ \bibinfo {pages} {L012053} (\bibinfo {year} {2024})}\BibitemShut {NoStop}%
\bibitem [{\citenamefont {Wilczek}\ and\ \citenamefont {Zee}(1983)}]{Wilczek-83}%
  \BibitemOpen
  \bibfield  {author} {\bibinfo {author} {\bibfnamefont {F.}~\bibnamefont {Wilczek}}\ and\ \bibinfo {author} {\bibfnamefont {A.}~\bibnamefont {Zee}},\ }\bibfield  {title} {\bibinfo {title} {{Linking Numbers, Spin, and Statistics of Solitons}},\ }\href {https://doi.org/10.1103/PhysRevLett.51.2250} {\bibfield  {journal} {\bibinfo  {journal} {Phys. Rev. Lett.}\ }\textbf {\bibinfo {volume} {51}},\ \bibinfo {pages} {2250} (\bibinfo {year} {1983})}\BibitemShut {NoStop}%
\bibitem [{\citenamefont {Witten}(1983)}]{Witten-83}%
  \BibitemOpen
  \bibfield  {author} {\bibinfo {author} {\bibfnamefont {E.}~\bibnamefont {Witten}},\ }\bibfield  {title} {\bibinfo {title} {{Current algebra, baryons, and quark confinement}},\ }\href {https://doi.org/https://doi.org/10.1016/0550-3213(83)90064-0} {\bibfield  {journal} {\bibinfo  {journal} {Nucl. Phys. B}\ }\textbf {\bibinfo {volume} {223}},\ \bibinfo {pages} {433} (\bibinfo {year} {1983})}\BibitemShut {NoStop}%
\bibitem [{\citenamefont {Altland}\ \emph {et~al.}(2024)\citenamefont {Altland}, \citenamefont {Brouwer}, \citenamefont {Dieplinger}, \citenamefont {Foster}, \citenamefont {Moreno-Gonzalez},\ and\ \citenamefont {Trifunovic}}]{Altland-24}%
  \BibitemOpen
  \bibfield  {author} {\bibinfo {author} {\bibfnamefont {A.}~\bibnamefont {Altland}}, \bibinfo {author} {\bibfnamefont {P.~W.}\ \bibnamefont {Brouwer}}, \bibinfo {author} {\bibfnamefont {J.}~\bibnamefont {Dieplinger}}, \bibinfo {author} {\bibfnamefont {M.~S.}\ \bibnamefont {Foster}}, \bibinfo {author} {\bibfnamefont {M.}~\bibnamefont {Moreno-Gonzalez}},\ and\ \bibinfo {author} {\bibfnamefont {L.}~\bibnamefont {Trifunovic}},\ }\bibfield  {title} {\bibinfo {title} {{Fragility of Surface States in Non-Wigner-Dyson Topological Insulators}},\ }\href {https://doi.org/10.1103/PhysRevX.14.011057} {\bibfield  {journal} {\bibinfo  {journal} {Phys. Rev. X}\ }\textbf {\bibinfo {volume} {14}},\ \bibinfo {pages} {011057} (\bibinfo {year} {2024})}\BibitemShut {NoStop}%
\bibitem [{\citenamefont {Lapierre}\ \emph {et~al.}()\citenamefont {Lapierre}, \citenamefont {Trifunovic}, \citenamefont {Neupert},\ and\ \citenamefont {Brouwer}}]{Lapierre-24}%
  \BibitemOpen
  \bibfield  {author} {\bibinfo {author} {\bibfnamefont {B.}~\bibnamefont {Lapierre}}, \bibinfo {author} {\bibfnamefont {L.}~\bibnamefont {Trifunovic}}, \bibinfo {author} {\bibfnamefont {T.}~\bibnamefont {Neupert}},\ and\ \bibinfo {author} {\bibfnamefont {P.~W.}\ \bibnamefont {Brouwer}},\ }\bibfield  {title} {\bibinfo {title} {{Topology of ultra-localized insulators and superconductors}},\ }\Eprint {https://arxiv.org/abs/2407.07957} {arXiv:2407.07957} \BibitemShut {NoStop}%
\bibitem [{\citenamefont {Nakamura}\ \emph {et~al.}()\citenamefont {Nakamura}, \citenamefont {Shiozaki}, \citenamefont {Shimomura}, \citenamefont {Sato},\ and\ \citenamefont {Kawabata}}]{NSSSK-24}%
  \BibitemOpen
  \bibfield  {author} {\bibinfo {author} {\bibfnamefont {D.}~\bibnamefont {Nakamura}}, \bibinfo {author} {\bibfnamefont {K.}~\bibnamefont {Shiozaki}}, \bibinfo {author} {\bibfnamefont {K.}~\bibnamefont {Shimomura}}, \bibinfo {author} {\bibfnamefont {M.}~\bibnamefont {Sato}},\ and\ \bibinfo {author} {\bibfnamefont {K.}~\bibnamefont {Kawabata}},\ }\bibfield  {title} {\bibinfo {title} {{Non-Hermitian Origin of Wannier Localizability and Detachable Topological Boundary States}},\ }\Eprint {https://arxiv.org/abs/2407.09458} {arXiv:2407.09458} \BibitemShut {NoStop}%
\bibitem [{\citenamefont {Shiozaki}\ \emph {et~al.}()\citenamefont {Shiozaki}, \citenamefont {Nakamura}, \citenamefont {Shimomura}, \citenamefont {Sato},\ and\ \citenamefont {Kawabata}}]{SNSSK-24}%
  \BibitemOpen
  \bibfield  {author} {\bibinfo {author} {\bibfnamefont {K.}~\bibnamefont {Shiozaki}}, \bibinfo {author} {\bibfnamefont {D.}~\bibnamefont {Nakamura}}, \bibinfo {author} {\bibfnamefont {K.}~\bibnamefont {Shimomura}}, \bibinfo {author} {\bibfnamefont {M.}~\bibnamefont {Sato}},\ and\ \bibinfo {author} {\bibfnamefont {K.}~\bibnamefont {Kawabata}},\ }\bibfield  {title} {\bibinfo {title} {{$K$-theory classification of Wannier localizability and detachable topological boundary states}},\ }\Eprint {https://arxiv.org/abs/2407.18273} {arXiv:2407.18273} \BibitemShut {NoStop}%
\bibitem [{\citenamefont {Yoshida}\ \emph {et~al.}()\citenamefont {Yoshida}, \citenamefont {Bergholtz},\ and\ \citenamefont {Bzdu\v{s}ek}}]{Yoshida-25}%
  \BibitemOpen
  \bibfield  {author} {\bibinfo {author} {\bibfnamefont {T.}~\bibnamefont {Yoshida}}, \bibinfo {author} {\bibfnamefont {E.~J.}\ \bibnamefont {Bergholtz}},\ and\ \bibinfo {author} {\bibfnamefont {T.}~\bibnamefont {Bzdu\v{s}ek}},\ }\bibfield  {title} {\bibinfo {title} {{Hopf Exceptional Points}},\ }\Eprint {https://arxiv.org/abs/2504.13012} {arXiv:2504.13012} \BibitemShut {NoStop}%
\end{thebibliography}%
\let\addcontentsline\oldaddcontentsline

\end{document}